\documentclass[
 amsmath,amssymb,
reprint,
prx,
floatfix,
]{revtex4-2}
\usepackage{graphicx,xcolor}

\definecolor{darkblue}{RGB}{0,0,150}
\definecolor{nightblue}{RGB}{0,0,100}

\usepackage{mathrsfs,dsfont,mathtools}

\usepackage{dcolumn}% Align table columns on decimal point
\usepackage{bm}% bold math
\usepackage[
colorlinks,
citecolor=darkblue,
linkcolor=darkblue,
urlcolor=nightblue]{hyperref}% add hypertext capabilities
\usepackage[english]{babel}
\usepackage[babel,kerning=true,spacing=true]{microtype}
\usepackage{braket}

\newcommand{\bk}{\mathbf{k}}

\definecolor{DarkRed}{RGB}{100,0,0}

\begin{document}

\title{
Twisted photovoltaics at terahertz frequencies from momentum shift current
}

\author{Daniel Kaplan}
\author{Tobias Holder}
\author{Binghai Yan}
\email{binghai.yan@weizmann.ac.il}
\affiliation{Department of Condensed Matter Physics,
Weizmann Institute of Science,
Rehovot 7610001, Israel}

\begin{abstract}

The bulk photovoltaic effect (BPVE) converts light into a coherent dc current at zero bias, through what is
commonly known as the shift current. This current has previously been attributed to the displacement of the
electronic wave function center in real space, when the sample is excited by light. We reveal that materials like
twisted bilayer graphene (TBG) with a flatband dispersion are uniquely suited to maximize the BPVE because
they lead to an enhanced shift in the momentum space, unlike any previously known shift current mechanism.
We identify properties of quantum geometry, which go beyond the quantum geometric tensor, and are unrelated
to Berry charges, as the physical origin of the large BPVE we observe in TBG. Our calculations show that
TBG with a band gap of several meV exhibits a giant BPVE in a range of 0.2–1 THz, which represents the
strongest BPVE reported so far at this frequency in a two-dimensional material and partially persists even a
room temperature. Our paper provides a design principle for shift current generation, which applies to a broad
range of twisted heterostructures with the potential to overcome the so-called “terahertz gap” in THz sensing.
\end{abstract}

\maketitle

\section{Introduction}
The bulk photovaltaic effect (BPVE)~\cite{Belinicher1980,Boyd2003} refers to the generation of a dc current from a homogeneous solid upon irradiation with light. The so-called shift current~\cite{vonBaltz1981,Sipe2000,Young2012} is one of the most important mechanisms for the BPVE which has been appreciated for a long time for its potential in photovoltaic and photonic applications.
So far, the BVPE has been studied predominantly for near-infrared and optical frequencies, but it may prove useful for overcoming the so-called 'terahertz gap' \cite{Knap2009} in the field of terahertz photonics \cite{Tonouchi2007}; that is  the perceived lack of robust, tunable and broad-frequency materials for terahertz detection. 
% For instance, usual semiconductor infrared detectors~\cite{kinch2000fundamental} harvest the photon-excited conduction electrons, which are competing with the thermal excitation across the energy gap, and thus cannot work in the THz range. Some recent developments attempt to detect plasmons~\cite{Arik2017,Muravev2012,Bandurin2018a} stimulated by the THz radiation. 
Since typical excitation energies in twisted bilayer graphene (TBG) are situated in the THz regime~\cite{Sharpe2019,Serlin2020,Lu2019,Chen2019,Jiang2019,Choi2019,Kerelsky2019,Xie2019,Chen2020,Zondiner2020,Wong2020,
Yankowitz2019,Tomarken2019}, we are motivated to study the BPVE effect in this particular material system, as it will involve resonant transitions which occur within this energy window. 

\begin{figure}
    \centering
    \includegraphics[width=.49\textwidth]{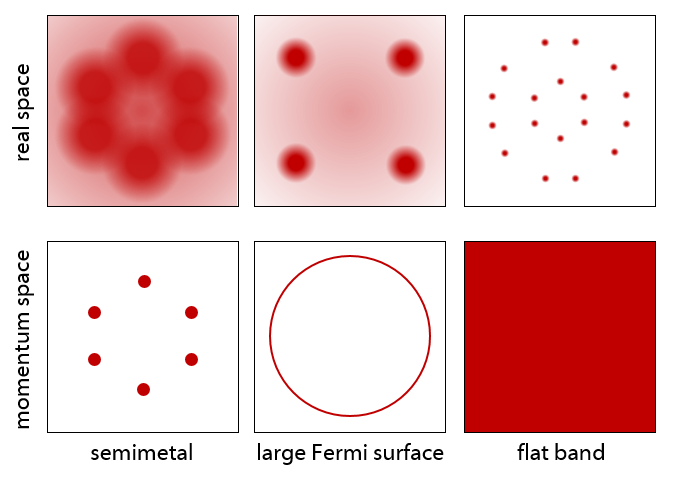}
    \caption{A schematic comparison of the wavefunction in real-space and momentum-space  of electronic states at the Fermi surface for three types of gapless materials. For a Weyl semimetal, states are well localized at few points in the momentum space but quite extended in the real space. In contrast, for a flat band all momenta are equally occupied, leading to sharply localized peaks of the wavefunction in real space. The ordinary metal with a large Fermi surface represents the intermediate region between the Weyl semimetal and the flat-band system.
    If the shift current is viewed as a result of the anomalous acceleration, the Weyl semimetal exhibits a large real-space shift while the flat band displays a major momentum-space shift.
    }
    \label{fig:schematics}
\end{figure}

Our central insight is depicted schematically in Fig.~\ref{fig:schematics}: The shift current appears as the result of a wavefunction displacement in \emph{either} real space or momentum space.
The former interpretation as a real-space shift of wave function centers upon excitation has been proposed since the earliest works~\cite{vonBaltz1981}, 
and it has recently gained renewed interest in studies of Weyl semimetals~\cite{Yan2017,Armitage2018}, 
because the shift is believed to get enhanced by the Berry curvature near Weyl points~\cite{Wu2017,Ma2017,Osterhoudt2019,Morimoto2016,Taguchi2016,Chan2017,deJuan2017,Zhang2018}. 
Very recently, we proposed to instead view the shift current as a result of  anomalous quasiparticle acceleration, which is determined by quantum geometric properties~\cite{Holder2020}. In the latter interpretation, the small effective mass in a Weyl semimetal leads to strong acceleration in the  field of light and hence generates large photocurrent, consistent with the established shift current interpretation~\cite{Morimoto2016}. 
However, materials with a flat dispersion represent the other extreme (large quasiparticle mass) compared to Weyl bands (nearly massless quasiparticles). Using the language of a transient polarization, it is therefore not obvious how a large shift current can emerge for a flat-band system. In contrast, in the new semiclassical acceleration picture, a state can be displaced (i.e. accelerated) in either real-space and momentum-space, (cf. Fig.~\ref{fig:schematics}). Although the shift is indeed small in real-space due to the large quasiparticle mass, for flat bands the acceleration (i.e. the displacement in momentum space) can still be very large. 

% Motivated by the phenomenology of the acceleration point of view, in the following we investigate the shift current of twisted bilayer graphene (TBG).
% TBG exhibits flat bands and narrow band gaps of several meV, and thus provides an ideal platform to investigate the THz BPVE according to all criteria outlined above. 
TBG itself has attracted an immense amount of attention following the recent discovery of correlated insulating and superconducting phases at small twist angle (magic angle) $\theta\sim 1^\circ$~\cite{Cao2018,Cao2018a}.
While the root cause for this exciting behavior is widely accepted to be tied to a highly quenched band structure~\cite{Santos2007,Santos2012,Bistritzer2011}, opinions differ about the mechanisms associated with the various correlated phases which have since been documented in the system~\cite{Sharpe2019,Serlin2020,Lu2019,Chen2019,Jiang2019,Choi2019,Kerelsky2019,Xie2019,Chen2020,Zondiner2020,Wong2020,
Yankowitz2019,Tomarken2019}.
An important stepping stone towards the understanding of these phases is the characterization of the quantum geometric and dispersive features of the flat bands.
Due to inversion symmetry breaking, TBG and similar twisted bilayers are known to exhibit a number of intriguing nonlinear phenomena~\cite{Liu2020a,He2020,Hu2020,Huang2020,Otteneder2020,liu2020b,zhang2020giant,Ikeda2020} such as the BPVE and nonlinear anomalous Hall effect~\cite{Moore2010,Sodemann2015}. These nonlinear probes are an important tool for the investigation of quasiparticle properties in flat bands as they are not suppressed by a vanishing Fermi velocity~\cite{Holder2020}.

In this work, we focus on the shift current generation below $10\mathrm{meV}$ in magic angle TBG across the single particle gaps.
In a numerical analysis, we find a large photocurrent response for light in the low terahertz regime. This is consistent with the acceleration picture outlined above. To further support such a connection, we express the response in terms of the real-space shift current and momentum-space shift current. Due to the flat-band dispersion, the dominant contribution is constituted by the momentum-space shift, which we explicitly show to be dissimilar to an ordinary dispersive material (e.g., MoS$_2$). {We point out that while flat-band materials are expected to be optically active with a large joint density of states (jDOS) \cite{Sato2021}, this alone could not account for the large response, as it exceeds that of other 2D materials with a large jDOS at the band edge}.
Distinct from the nonlinear anomalous Hall effect, the shift current of TBG originates from properties of the quantum geometry of the band structure that are unrelated to the Berry curvature dipole (BCD). The magnitude and tunability of the photocurrent, the broadness of the resonances, and the terahertz frequency range in which they are all observed hold promise of the utilization of TBG devices in terahertz technologies.
We further explore the effect of heterostrain, as it is typically observed in TBG~\cite{Kazmierczak2020}. 
%\looseness=-1

\section{Results}
\subsection{Shift current theory} 

In the clean limit, the Bloch states produce an intrinsic dc-current response to linearly polarized light~\cite{vonBaltz1981}. This nonlinear conductivity $\sigma_{(s)}^{aa;c}$ 
is usually formulated as~\cite{vonBaltz1981,Sipe2000}
\begin{align}
    \sigma^{aa;c}_{(s)}&(0;\omega,-\omega)=\notag\\
    &\frac{\pi e^3}{\hbar^2}\int_{\bm{k}}
    \sum_{mn} f_{mn}
    S^c_{mn}|r_{mn}^a|^2
    \delta(\omega\pm \varepsilon_{mn}),
    \label{eq:aacconductgeneral}
\end{align}
where $m,n$ are the band indices, $\varepsilon_{mn}=\varepsilon_{m}-\varepsilon_{n}$ the band energy difference, $f_{mn}=f(\varepsilon_m)-f(\varepsilon_n)$ the Fermi distribution function difference, and $ r_{mn}^a \equiv \braket{m|r^a|n}$ the dipole transition matrix element,  integration over the Brillouin zone is taken as $\int_\bk = \frac{1}{(2\pi)^2}\int \textrm{d}k_x \textrm{d}k_y$, while $a,c$ represent the light field and current directions, respectively.
$\bm{S}_{mn}$ is the so-called shift vector,
\begin{align} \label{eq:shiftvector}
S^c_{mn}&=(r^c_{mm}-r^c_{nn})+\partial_{k_c}\arg r^c_{mn},
\end{align}
where $(r^c_{mm}-r^c_{nn})$ is the shift of the wave function centers, which is gauge dependent. 
The second term in Eq.~\ref{eq:shiftvector} is the phase derivative of $r^c_{mn}$, which ensures gauge-invariance for the shift vector.
$\bm{S}_{mn}$ has then been interpreted as the real-space shift of the mass center of a quasiparticle upon excitation from band $m$ to $n$~\cite{vonBaltz1981}. 
As we will show now, the phase term in \eqref{eq:shiftvector} gives rise to a shift in momentum space. Therefore, the shift current is actually the result of a shift of the quasiparticle excitation in both real space and momentum space, which is naturally captured using the language of the anomalous quasiparticle acceleration~\cite{Holder2020}.
\begin{figure*}
\includegraphics[width=.85\textwidth]{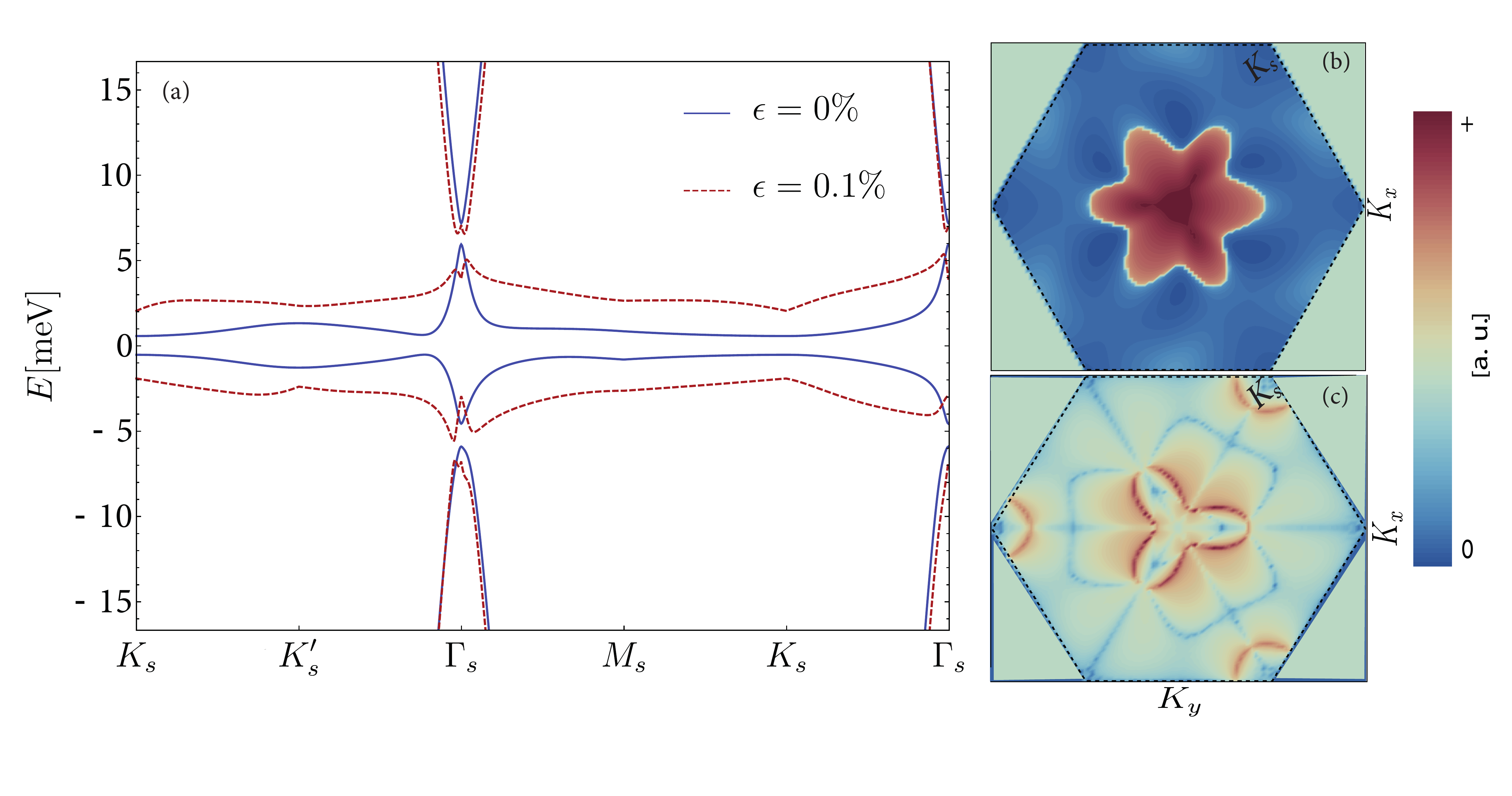} 
\caption{Band structure, Berry curvature and nonlinear conductivity in the mini-Brillouin zone of TBG. 
(a) Band structure of TBG with inversion-breaking at twist angle $\theta=1.05^\circ$. Solid lines show the unstrained case, the dashed lines are at strain $\epsilon=0.001$. Details about the introduction of strain are listed in the SI.
(b) The Berry curvature distribution of all filled bands at half filling for valley $K$. The colorbar is in logarithmic scale. The Berry curvature in each valley is $C_{3z}$ symmetric, which leads to a vanishing Berry curvature dipole in each valley separately.
(c) Modulus of the nonlinear conductivity as given by $\sigma_{(s)}^{yy;x}$ [Eq.~\eqref{eq:shifts2}], in logarithmic scale.
At frequency $\omega = 1.4$ meV, just above the gap, the resonant features are fairly broad in momentum space, a result of the flatness of the dispersion.
}
\label{fig:fig2}
\end{figure*}
To this end, we express the nonlinear conductivity in the form of geometric properties of the band structure,
\begin{widetext} 
\begin{align}
    \sigma^{aa;c}_{(s)}(0;\omega,-\omega)
    =-&\frac{\pi e^3}{\hbar^2} \int_{\bm{k}} R^{aac}_{\mathrm{shift}}+K^{aac}_{\mathrm{shift}}
    \label{eq:shift_new} \\
     R^{aac}_{\mathrm{shift}}  =& \sum_{mn}f_{mn}(r^c_{mm}-r^c_{nn})
    |r^a_{mn}|^2\delta(\omega \pm \varepsilon_{mn}) 
    \label{eq:Rshift}    \\
     K^{aac}_{\mathrm{shift}}  =&\sum_{mn}f_{mn}\Bigl[
    2i r^a_{mn}\lambda^{ac}_{nm}
    -i r^c_{mn}\lambda^{aa}_{nm}    \Bigr]\delta(\omega \pm \varepsilon_{mn})     -\partial_{k_a}\Omega^{ac} |_{\varepsilon_{mn}=\pm\omega}
    \label{eq:Kshift}\\
     \partial_{k_a}\Omega^{ac} |_{\varepsilon_{mn}=\pm\omega} =&
        \sum_{mn} f_{mn}
    \Bigl[
    \tfrac{i}{2}r^a_{mn}\Omega^{ac}_{nm}
    -i r^c_{mn}\lambda^{aa}_{nm}
    +i r^a_{mn}\lambda^{ac}_{nm}
    \Bigr]\delta(\omega\pm \varepsilon_{mn})
    \label{eq:berrydipole}
\end{align}
\end{widetext}
where we use the symmetrized derivative $\lambda^{ab}_{nm} = \frac{1}{2} \left(\partial_{k_a} r^b_{nm} + \partial_{k_b} r^a_{nm}\right)$. {Additionally, we defined the interband Berry curvature, $\Omega^{ac}_{nm} = i\sum_{l \neq (n,m)} (r^a_{nl}r^c_{lm} - r^c_{nl}r^a_{lm})$. For the shift current, this Berry curvature is weighted by the transition element $\delta(\omega \pm \varepsilon_{nm})$, arising from optical selection rules. The derivative of this object gives the interband BCD that appears in Eq.~\eqref{eq:berrydipole} (the derivative does not act on the delta function). } The two pieces $R^{aac}_{\mathrm{shift}}$ and $K^{aac}_{\mathrm{shift}}$ correspond to the contributions from the first term and the second term, respectively, of the shift vector in Eq.~\eqref{eq:shiftvector}.
The second term in $K^{aac}_{\mathrm{shift}}$ is the Berry curvature dipole, and the first term is related to the other geometric quantity, the quantum metric, which characterizes the quantum distance between two states. Thus, $K^{aac}_{\mathrm{shift}}$ originates directly from the quantum geometry of the wave function in $k$-space, and it represents the anomalous acceleration in momentum space~\cite{Holder2020}.
$\lambda^{ab}_{nm}$ encodes the skewness of the acceleration: The Berry curvature dipole involves derivatives of the type $\partial^c(r^a_{mn}r^b_{nm}-r^b_{mn}r^a_{nm})$, 
which does not cover the entire motion in momentum space. The remaining terms involving $\lambda^{ab}$ in Eq.~\eqref{eq:Kshift} can be connected to skew-symmetric derivatives of the structure $(\partial^cr^a_{mn})r^b_{nm}-r^a_{mn}(\partial^cr^b_{nm})
+(\partial^cr^b_{mn})r^a_{nm}-r^b_{mn}(\partial^cr^a_{nm})$. 
Note that in the special case of a two-band Dirac dispersion, it can be shown that the general expressions for the shift current present here can be connected by a pull-back mapping to the Christoffel symbols of the geometric connection on the generalized Bloch sphere~\cite{Ahn2020}. While this reinforces the association of the shift current with a semiclassical acceleration, for the multi-band case discussed here, the more convenient expression is the one presented in Eq.~\eqref{eq:shift_new} in terms of $\lambda^{ab}$ (for another reparametrization cf.~\cite{Ahn2021}).

\subsection{Shift current of TBG}

We model TBG using a modified form of the Bistrizer-Macdonald continuum model \cite{Bistritzer2011, He2020}. We attach two monolayer graphene sheets, with first rotating them with respect to one another with an angle $\theta = 1.05 ^{o}$, and then introducing inter-layer coupling.  By construction, this model is endowed with $C_{3z}$ symmetry, and since both monolayers are inversion symmetric, the model as a whole has inversion symmetry. Inversion symmetry breaking is introduced by a staggered potential $\Delta$, as it typically arises from encapsulation of the bilayer in hBN. On the other hand, $C_{3z}$ is not necessarily broken because it requires some uniaxial strain of magnitude $\varepsilon$ in the bottom layer. The parameters used here all correspond to the original values used in the Bistrizter-Macdonald construction. A detailed description of the model, as well as the results when corrugation effects are included, are given in the supplementary material.
We emphasize that such a non-interacting band structure will of course not capture the important effects of many-body correlations in TBG~\cite{DaLiao2021}. However, since the BVPE across the gap relies on photon energies which greatly exceed the onset temperature of the ordered phases that form in TBG, the qualitative features of the shift current reported here can therefore be expected to also apply for a wide range of temperatures as long as the interacting system remains in the normal state.
The resulting band structure in the mini-Brillouin zone of valley $K$ is shown in Fig.~\ref{fig:fig2}a, both for the unstrained and strained TBG. The flat bands near half filling ($\mu=0\,\mathrm{meV}$) are gapped at $\Gamma$ and $K$ due to inversion symmetry breaking. Fig.~\ref{fig:fig2}b shows the Berry curvature in the mini-Brillouin zone in unstrained TBG, which is $C_3$-symmetric. We note that the Berry curvature near the $K$ and $K'$-points in the original dispersion have opposite signs and cancel.

While the shift vector $\bm{S}$ in Eq.~\eqref{eq:shiftvector} itself is not expected to be further decomposable into gauge invariant pieces, the expression that enters the shift current can indeed be decomposed. Writing $\sigma_{(s)}  = \sigma_{(s1)} + \sigma_{(s2)}$, with
\begin{align}
    \sigma_{(s1)}^{aa;c}&=
    \frac{\pi e^3}{\hbar^2}\int_\bk 
    \partial_{k_a}\Omega^{ac} |_{\varepsilon_{mn}=\omega}
    \label{eq:shifts1}\\
    \sigma_{(s2)}^{aa;c}&= -\frac{\pi e^3}{\hbar^2}\int_\bk 
    (R^{aac}_{\mathrm{shift}}+K^{aac}_{\mathrm{shift}})
    -\sigma_{(s1)}^{aa;c}.
    \label{eq:shifts2}
\end{align}
\begin{figure*}
    \centering
    \includegraphics[width=.95\textwidth]{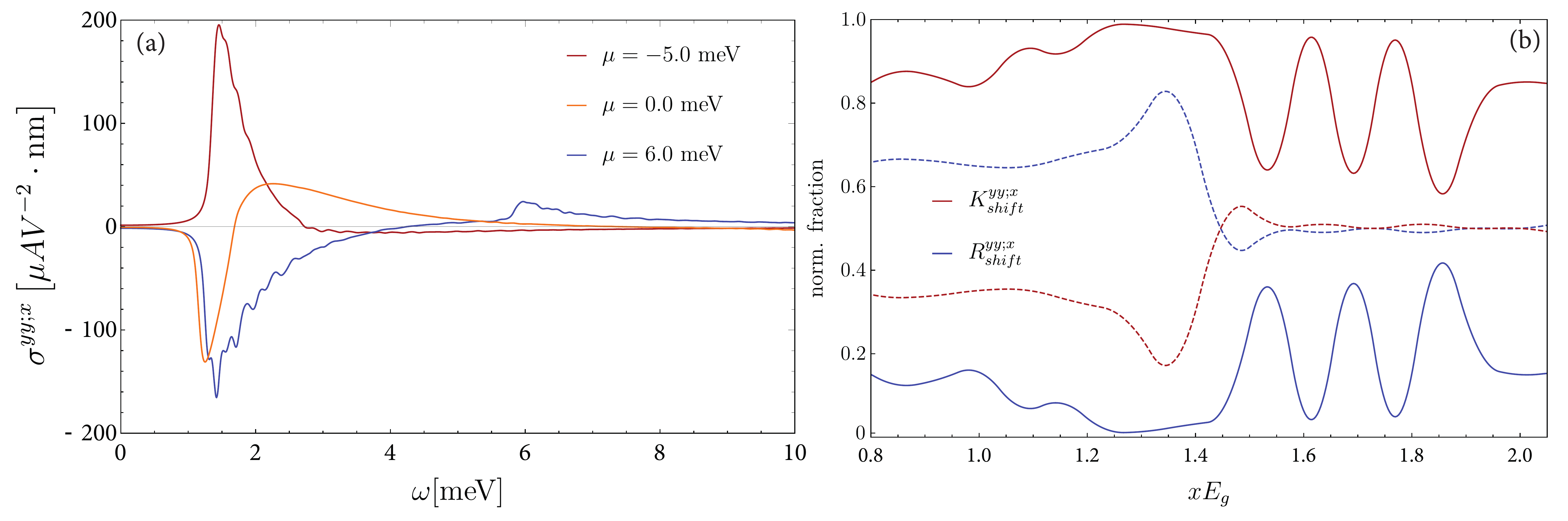}
    \caption{Zero strain nonlinear conductivity of TBG and breakdown of contributions to the conductivity (vs ordinary MoS\textsubscript{2}).
    (a) $\sigma^{yy;x}$ of TBG with $\varepsilon = 0\%$, with $C_3$ symmetry intact. The conductivity is plotted for 3 values of $\mu = -5, 0, 6$ meV, respectively. 
    (b) Decomposition of Eq.~\eqref{eq:shift_new} into the $R^{yy;x}_{\mathrm{shift}}$ and $K^{yy;x}_{\mathrm{shift}}$ pieces, respectively. Solid lines depict TBG, while the dashed lines show MoS\textsubscript{2}, for comparison.
    The shift current is essentially zero for $x<1$, has a large resonance around $x\approx1$ and then decays quickly for values above that. The normalized fraction does not show qualitative changes at $x\approx 1$ but consistently conforms with the semiclassical intuition that the shift receives contributions from both the dispersive acceleration and the anomalous acceleration, where the latter is dominant for a flat-band dispersion. The maximal conductivity for  $\mu = -5.0\textrm{meV}$ is attained at $\omega = 1.4 \textrm{meV}$, with value $\sigma^{yy;x} = +198 \mu \mathrm{A V^{-2} nm}$, and for $\mu = 6.0\textrm{meV}$ it is $-168 \mu \mathrm{A V^{-2} nm}$.}
    \label{fig:fig3}
\end{figure*}
Here, Eq.~\eqref{eq:shifts1} contains the contributions from the {Berry curvature dipole $\partial^a \Omega^{ac}$, which vanishes in the presence of $C_{3z}$ symmetry}. This symmetry renders only two components of $\sigma^{ab;c}$ independent. For simplicity we focus in the main text on $\sigma^{yy;x}$, which encodes the transverse nonlinear conductivity response. The other independent component is $\sigma^{xx;y}$ (Supplementary Figs.~2 and 3); the symmetry analysis can be found in the SI.
In Fig.~\ref{fig:fig3}a we present the total shift current in the unstrained case, for three values of the chemical potential which reside within the three gaps that are opened by the staggered potential. Within our numerical accuracy, the current is indeed found to {have no contributions from the Berry curvature dipole}, i.e. $\sigma^{yy;x}_{(s2)}=\sigma^{yy;x}_{(s)}$. Irrespective of this, for frequencies which cross the single-particle gap the shift current reaches giant values nearly $200 ~\mu \mathrm{A V^{-2} nm}$, far exceeding predicted values for other non-magnetic materials~\cite{Rangel2017,Cook2017}.

As we pointed out, from the quasiparticle shift it is not immediately obvious where such a giant nonlinear response could originate from. For this reason, we  examine the different contributions in Eq.~\eqref{eq:shifts2}.
Fig.~\ref{fig:fig2}c shows the momentum-space structure of the integrand in Eq. \eqref{eq:shifts2}
at $\mu = 0, \omega = 1.4 \textrm{meV}$, i.e. just above the band gap of the flat bands. The largest contributions are from the residual dispersion around $\Gamma$ and from the flat parts of the band structure around $K_s$, both of which are fairly broad in momentum space. 
Fig. \ref{fig:fig3}b illustrates the relative contribution of the pieces $K^{aa;c}_{\mathrm{shift}}$ and  $R^{aa;c}_{\mathrm{shift}}$ in Eq.~\eqref{eq:shifts2}, as a function of the distance from the gap at $0$. Over a large range of frequencies, it holds that $K^{aa;c}_{\mathrm{shift}}\gg R^{aa;c}_{\mathrm{shift}}$, meaning that the shift current is produced mostly by the phase of the Berry connection and not the shift of wavefunction centers. 
For comparison, we performed the same breakdown for the much more dispersive, gapped material MoS\textsubscript{2}, which has very similar symmetry properties [Fig. \ref{fig:fig3}b]. In this latter case, $K^{aa;c}_{\mathrm{shift}}\ll R^{aa;c}_{\mathrm{shift}}$ near the band edge, which supports the notion that the shift current is resulting from a displacement in both real space and momentum space, with the latter being greatly enhanced for a flatband dispersion. As we mentioned in the beginning, these statements might seem questionable upon regauging, a possible shortcoming on which we comment in the discussion.

\subsection{Effect of strain}
\begin{figure*}
    \centering
\includegraphics[width=\textwidth]{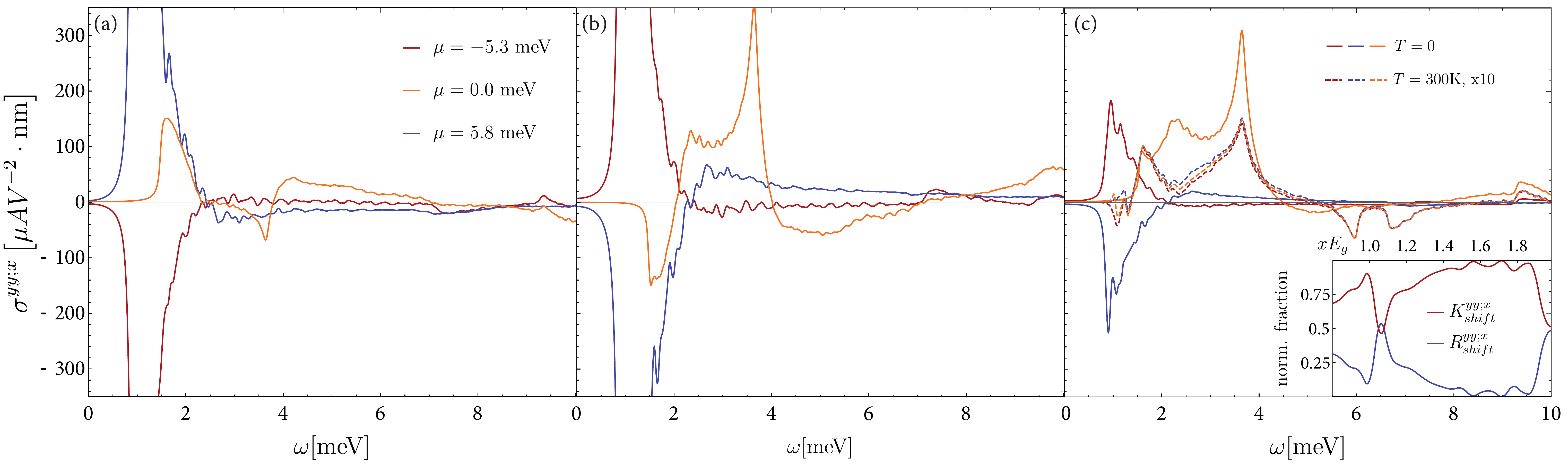}
    \caption{Shift current response of strained TBG, $\sigma^{yy;x}$, for 3 chemical potentials within the gaps created by the staggered potential, and uniaxial strain $\varepsilon = 0.1\%$. (a) Berry curvature dipole contribution, Eq.~\eqref{eq:shifts1}. The largest values are clipped to preserve readability for the total conductivity. (b) Momentum and position shifts contribution, Eq.~\eqref{eq:shifts2} (c) Total conductivity $\sigma^{yy;x}$ at $T=0\mathrm{K}$ (solid lines) and for $T=300\mathrm{K}$ (dashed lines, shown x10 for clarity). 
    For chemical potential $\mu = 0$ the shift peaks at $\sigma^{yy;x} = 315 \mathrm{\mu A nm V^{-2}}$ ($T=0\mathrm{K}$) and $15 \mathrm{\mu A nm V^{-2}}$ ($T=300\mathrm{K}$), respectively.
    Inset: relative size of $R^{yy;x}_{\mathrm{shift}}$ and $K^{yy;x}_{\mathrm{shift}}$ contributing to the total conductivity as a function of the distance from the flat-band band gap, $x E_g$ (here $\mu =0$), cf. Fig. \ref{fig:fig3}b.
    }
    \label{fig:fig4}
\end{figure*}

In the presence of uniaxial strain, the symmetry group of TBG is reduced to $C_1$. All components of the conductivity are now independent, but for clarity we continue to examine only the transverse contribution $\sigma^{yy;x}$. Since the Berry curvature dipole contribution as given by Eq~\eqref{eq:shifts1} is no longer zero, we present both its contribution (Fig. \ref{fig:fig4}a) and the remaining terms in Eq.~\ref{eq:shifts2} for $\sigma^{aa;c}_{(s2)}$ (Fig. \ref{fig:fig4}b). 
The total shift current is depicted in Fig.~\ref{fig:fig4}c, with the relative sizes of $R^{aa;c}_{\mathrm{shift}}$ and $K^{aa;c}_{\mathrm{shift}}$ shown in the inset.

At moderate uniaxial strain of size $\epsilon=0.1\%$, the conductivity $\sigma_{(s1)}$ due to the Berry curvature dipole becomes comparable in size to $\sigma_{(s2)}$, but it has consistently the opposite sign. 
Thus, while the total conductivity unsuprisingly increases due to the reduced symmetry in the system, it has less accentuated resonances for the transitions at the band edges, with values up to $300~\mu \mathrm{A V^{-1} nm}$ across the flatband gap. 
More unexpectedly, the shift current does not seem to profit from {a net Berry curvature dipole}, as this contribution either subtracts from the remaining current, or is almost negligible for transitions across the flatbands.
Our results establish that a large anomalous acceleration in the quasiparticle motion can arise even if $\partial_{k_a}\Omega^{ac}=0$, i.e. its existence does not rely on the presence of {a finite Berry curvature dipole in the system}.
This is an important distinction between the bulk photovoltaic effect and the anomalous Hall effect in TBG.
We further observe that in all cases the shift current at frequencies $\omega\gg 2~\mathrm{meV}$, corresponding to transitions between dispersive bands, is negligible small compared to the resonances around the band edges. 
In other words, the giant shift current shown in Fig.~\ref{fig:fig3} is not tied to the topological properties of the flat bands but rather to their non-dispersive nature.
We note that the maximal shift current obtained for TBG is larger than previously reported values for comparable, non-magnetic two-dimensional materials by a factor of 5 or more~\cite{Tan2016,Cook2017,Schankler2021}.
Figure~\ref{fig:fig3}c also shows the shift current evaluated at room temperature, $T=300\mathrm{K}$ (dashed lines). While the transitions across the gaps of the dispersive bands are strongly suppressed, the large density of states in the flat bands supports a strong signal for $\hbar\omega=4\mathrm{meV}$, with an amplitude of still 5\% of its value at $T=0$. 
We note that these conclusions contain the effect of disorder broadening through the inclusion of a finite quasiparticle relaxation rate~\cite{Monteverde2010}. 
\subsection{Real and momentum space displacements}
\label{sec:realmomdisp}
In introducing $R_{\mathrm{shift}}$ and $K_{\mathrm{shift}}$, we are able to distinguish the sources of displacement that the quasiparticle suffers. These quantities are, however, not measurable observables of the system. In the following we explain why such a decomposition is nonetheless insightful. 
For this, first recall that all types of topological bands (including flat bands) have no uniquely defined center-of-mass coordinate within the unit cell, because the topological nature of the bands prevents such an assignment. However, from this it does not follow that the momentum space integral of $R_{\mathrm{shift}}$ can take arbitrary values, because it contains much more specific information about the \emph{relative} positional difference between two bands, summed for all momenta. Indeed, it was already pointed out a long time ago~\cite{Sipe1999,Sipe2000} that for an arbitrary band structure one cannot generally expect to find a gauge such that $r^a_{mm}-r^a_{nn}$ consistently vanishes for all momenta and all bands.

Drawing from these observations, we therefore suggest that a useful indicator for band flatness is that the integrated positional difference $r^a_{mm}-r^a_{nn}$ between Bloch wavefunctions can be made substantially smaller than the integrated phase contribution $\partial_{k_a}\arg \bm{r}_{mn}$. 
We further conjecture that for highly dispersive bands a similar statement should hold about the smallness of the integrated phase contribution.
A paradigmatic example in the latter case is a two-band semimetal with one band crossing. There, a mostly smooth gauge is at the same time periodic (i.e. without phase jump at the Brillouin zone boundaries), thus completely eliminating the phase contribution from the integrated shift vector. 
This is the expected result for a quasiparticle with vanishing effective mass which is changing position in an applied electric field.
We remark that the difficulties in separating real-space and momentum-space effects of the acceleration into gauge invariant pieces are intrinsic to the more complicated semiclassical motion arising at second order in the applied field.
In particular, the analogous splitting of the quasiparticle velocity into the regular (dispersive) and anomalous velocity has the important distinction that these two components of the velocity are orthogonal to each other, making them linearly independent. Such a decomposition is not straightforward for the acceleration, because it describes the changes to both regular and anomalous velocity components in both normal and perpendicular direction, thus mixing them. 
This being said, we believe that the conclusions outlined above can be made more rigorous by deriving a lower bound for $R_{\mathrm{shift}}$ and $K_{\mathrm{shift}}$, which can then serve as useful indicator for mechanism of shift current generation in a given system - either by a displacement in real space or one in momentum space. This will be the subject of a future work. {Recent advances in ab-initio modelling of TBG \cite{Carr2019,Carr2019b}, coupled with observations of the importance in optical properties of the off-diagonal components of the position operator in the Wannier basis \cite{Ibanez2019}, suggest that a fully ab-initio approach might alter the relative magnitude of $R_{\textrm{shift}}$. We stress that this is not the case for TBG near the magic angle. As Ref. \cite{Ibanez2019} found, for atomically localized Wannier orbitals, the contributions to the off-diagonal parts of the position operator decay quickly beyond nearest-neighbor (NN) coupling. This occurs at the scale $a_0$, where $a_0$ is the monolayer lattice constant, while the continuum model's position operator scales with $L_m$, $L_m$ being the Moir\'e unit cell length. Consequently, all such corrections, near the magic angle -- where $L_m \gg a_0$ -- are expected to be negligible and will not affect our results.
}

\section{Discussion}
The shift vector $\bm{S}$ has previously been connected to the real-space shift of the center-of-mass coordinate between two eigenstates upon excitation from the conduction band $m$ into the valence band $n$. 
To give some intuition, we expand the real-space representation of the periodic eigenfunctions $|u_{n\bm{k}}\rangle$ in terms of local Wannier orbitals $|w_{n\bm{R}}\rangle$ with center coordinate $\bm{R}$~\cite{Vanderbiltbook}
\begin{align}
    \langle\bm{r}|u_{n\bm{k}}\rangle
    &=
    \sum_{\bm{R}}
    e^{-i\bm{k}(\bm{R}-\bm{r})}\langle\bm{r}|w_{n\bm{R}}\rangle.
\end{align}
Then, in momentum space the Berry connection is given by
\begin{align}
    r^a_{mn}&=
    \sum_{\bm{R}\bm{R}'}
    e^{i\bm{k}(\bm{R}-\bm{R}')}
    \int_{cell} d V 
    \langle w_{n\bm{R'}} |r^a|w_{n\bm{R}}\rangle.
    \label{eq:wannierexp}
\end{align}
Evaluating the shift vector $S^c_{mn}=r^c_{mm}-r^c_{nn}+\partial_{k_c}\arg r^c_{mn}$ based on this representation yields $r^a_{mm}-r^a_{nn}=R^a_{mm}-R^a_{nn}$ for the direct difference. {$R^a_{mm}$ refers to the component of the center of the $m$-th Wannier function, in the $a$-direction.}
This is supplemented by the phase derivative $\partial_{k_a}\arg \bm{r}_{mn}$, whose integral over the Brillouin zone is a multiple of $2\pi$. 
If only two Wannier orbitals have a significant overlap, the modulus $|\bm{S}|$ is clearly bounded by $|R^a_{mm}-R^a_{nn}|<a$, with lattice constant $a$. If several orbitals overlap, the phase factors in the sum Eq.~\eqref{eq:wannierexp} become important, with slope of growth in momentum space being at most $a$. 
Then, the phase derivative is expected to contribute similarly at $\mathcal{O}(a)$ to the shift vector. 
This already indicates that interpretation of the shift current as a result of the wavefunction shift is narrow to some extent.
For ease of illustration, imagine a set of Landau levels in symmetric gauge. Their center-of-mass coordinate $\bm{R}$ can be moved around freely in exchange for acquiring an additional phase factor. 
Indeed, for generic flat bands it is to be expected that there is a gauge choice which makes the shift $\bm{S}$ only depend on the phase, because the center of the Wannier functions can be repositioned with an appropriate gauge transformation. 
This is inconsistent with the interpretation of the shift current as the real space shift of the wavefunction center upon absorption of a photon, as the wavefunction only suffers a phase shift.

If the shift current is instead viewed as the anomalous acceleration that a quasiparticle undergoes due to the interaction with the electric field at second order, the photogalvanic response follows as a straightforward generalization of the linear response formalism involving the anomalous velocity~\cite{Xiao2010,Holder2020}, thus removing the direct inference of a current from a real-space displacement. 
Instead, both $R_{\mathrm{shift}}$ and $K_{\mathrm{shift}}$ appear as the result of the same acceleration that changes both the position and the wavevector of the quasiparticle.
As shown in the last section, this is consistent with our numerical findings for $R_{\mathrm{shift}}$ and $K_{\mathrm{shift}}$ using Bloch wavefunctions for a mostly smooth gauge choice within each band [cf. Fig.~\ref{fig:fig3}b]. 

While one might object against inspecting gauge-dependent quantities, we emphasize that $R_{\mathrm{shift}}$ and $K_{\mathrm{shift}}$ can still contain valuable information about the quasiparticle dynamics in the sense that while they are not unique, this does not at all imply that they are arbitrary. 
%We elaborate on this point in Sec. \ref{sec:realmomdisp}. 
Most importantly, based on our results we conjecture there exist nonzero lower bounds for both $R_{\mathrm{shift}}$ and $K_{\mathrm{shift}}$ which allow to uniquely identify the shift in terms of a real space or momentum space displacement. {These bounds generalize previous constraints on the shift vector (viewed as the real space displacement of the quasiparticle wavepacket) \cite{Tan2016} and extend to include the momentum-space shift introduced in this work.} We also remark that the general principles outlined here, inferred from the interplay of $K_{\mathrm{shift}}$ and $R_{\mathrm{shift}}$, as for the magnitude and resonant features of the shift current, are valid even when extrinsic effects are included, such as corrugation and strain (see Appendix). This is because the very nature of the effect relies on flat bands, which survive beyond the original Bistrizer-Macdonald parametrization. 

Finally, we comment on the stability of the shift current signal under experimental conditions. While disorder broadening and thermal broadening present the dominant limitations for the quasiparticle lifetime, more elaborate disorder effects like skew scattering~\cite{Xiao2019,Isobe2020} might in principle affect the shift current. 
In TBG we do not expect skew scattering to play an important role because it is enhanced only in fairly clean systems which additionally feature an asymmetric dispersion near the band edges. Both are conditions which are not met in TBG.
Another important source of disorder are variations in the twist angle, which are known to be present in TBG devices~\cite{Uri2020}. Experimental results for the linear conductivity do not indicate qualitative changes as a function of the twist angle ~\cite{Polshyn2019}, indicating that angle disorder has only a limited influence on transport. 
Additionally, if time-reversal is broken, for example by some magnetically ordered state or by the presence of additional relaxation channels, ballistic currents may appear~\cite{Sturman1992}.
For all these reasons, and based on the line shape, an exciting application of the strong nonlinear signal could be to employ the shift current spectrum to determine the direct band gap in TBG in the normal state. 

We note that for circular polarized light, TBG exhibits a chiral photogalvanic effect~\cite{Gao2020a}. Also, in a calculation including magnetic order at filling 3/4, Ref.~\cite{Liu2020a} has reported a large photogalvanic current at for frequencies of $30\,\mathrm{meV}$ and above. This result, however, should be read in the context of time-reversal breaking. As previously shown \cite{Zhang2019,Holder2020,Fei2020}, a system with broken time-reversal symmetry will generate a large injection current, that will scale inversely as $\gamma^{-1}$, where $\gamma$ is the carrier relaxation rate. Conversely, the shift current discussed here is actually lifetime independent ($\gamma^{0}$), so both effects are experimentally distinct. The decomposition of the nonlinear current at 3/4 filling is discussed separately in App. \ref{sec:threequart}, where the shift current is much smaller than the injection current.

Regarding the possible application of our results for THz sensing, we note that in contrast to existing proposals for Terahertz devices, our proposal relies on the BPVE, and can thus circumvent several issues exhibited for example by  Schottky diodes \cite{Song2015} and Bolometric devices \cite{Zhang2019c}. While the former has been known to generate a broadband response in the THz-range, the conversion efficiency is low and of the order of a few percent, requiring amplification and complicated electrical circuitry. The internal p-n junction used in Schottky diodes requires sensitive doping, and is not as tunable as van der Waals systems (such as TBG) and operates  under a bias field, which is highly sensitive to temperature. Bolometric devices add the additional complication of requiring thermal or mechanical junctions which typically have sub-optimal noise characteristics, thus reducing the applicability of such devices. We stress that setups which rely on the BPVE do not require external biases, amplification, thermal or mechanical junctions, but rather produce a current through resonant absorption of light.

In summary, we report a giant photogalvanic current for TBG which is irradiated by linear polarized light in the terahertz range. 
The magnitude of this shift current exceeds any previous reported numbers in comparable two-dimensional materials~\cite{Tan2016,Cook2017,Schankler2021}.{ Our mechanism for the giant response, which is due to a momentum-space shift and acceleration of quasiparticles is a new facet of nonlinear current generation. This goes beyond ordinary mechanisms for photoconductivity enhancement, such as a large jDOS}.
The resonance profile we observe in both the strained and unstrained cases suggest that TBG is a promising candidate for THz detection and circuits, even at room temperature.
Since the shift current is robust against thermally excited carriers, because it is a coherent bulk effect driven by the quantum geometry~\cite{Morimoto2016,Tan2016,Ahn2020}, we believe it may substantially improve detection capabilities for terahertz radiation. It might also increase the photovoltaic efficiency for energy harvesting, and have further applications in medical imaging, single-photon circuits and novel electronic devices~\cite{Lewis2019,Guillet2014}. 
As the root cause behind the large response we identified the anomalous acceleration due to the skew symmetric properties of the quantum geometry of the band structure as encoded in $\lambda_{mn}^{ab}$, which always appear in the shift current, but are greatly amplified in TBG due to the flat-band dispersion. 
The latter also turned out to be particularly important for retaining a large shift conductivity at higher temperatures.
Our findings present a new design principle for shift current generation which is particularly suitable for twisted heterostructures.
We expect the transverse dc-current reported here to be accessible using current samples and measurement techniques~\cite{Burger2019}. 
The line of reasoning developed in this work can potentially shed light on the quantum geometry of the band structure in similarly twisted van-der-Waals materials with nearly flat bands, for example MoTe\textsubscript{2} or WSe\textsubscript{2}~\cite{Wu2020a}.
\\
\noindent\textit{Acknowledgements} 
We thank 
J.~S.~Hofmann, 
R.~M.~Ribeiro, and 
R.~Queiroz 
for useful discussions.
B.Y. acknowledges the financial support by the European Research Council (ERC Consolidator Grant No. 815869, ``NonlinearTopo'') and Israel Science Foundation (ISF No. 2932/21).

\appendix

\section{Continuum Model construction}
\label{sec:contmod}
\begin{figure}
    \centering
    \includegraphics[width=.45\textwidth]{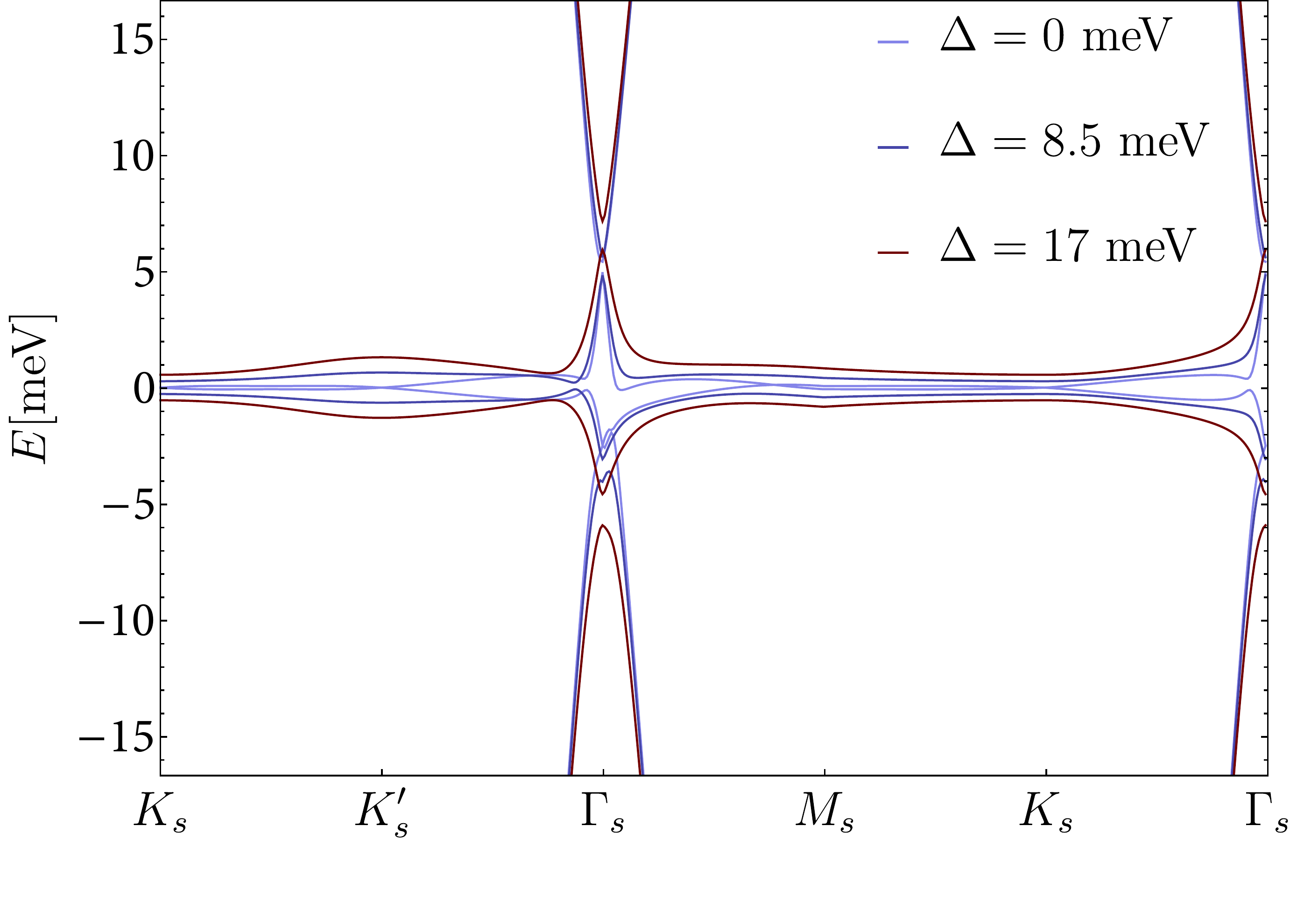}
    \caption{Evolution of the band structure with an applied staggered potential for hBN alignment. Only one layer (the bottom layer) is aligned. The burgundy curve indicates the value used in the main text, $\Delta = 17 \textrm{meV}$.
    }
\label{fig:fig9}
\end{figure}

The continuum model for TBG~\cite{Bistritzer2011,He2020} is well known and only repeated here for convenience of the reader. It is constructed by joining together two monolayer graphene layers at zero effective separation between them.
We choose the following real space unit vectors, for each graphene layer,
\begin{align}
    \mathbf{a}_1 = \sqrt{3}d\left(\frac{1}{2},\frac{\sqrt{3}}{2}\right), ~ \mathbf{a}_2 =\sqrt{3}d\left(\frac{1}{2},-\frac{\sqrt{3}}{2}\right)
\end{align}
In this description, the $A$ and $B$ sublattices are located, respectively, at $v_A = (0,0)$, $v_B = d(0,1)$. The reciprocal lattice vectors are,
\begin{align}
    \mathbf{b}_1 = \frac{4\pi}{3d} \left(\frac{\sqrt{3}}{2}, \frac{1}{2}\right), ~~ \mathbf{b}_2 = \frac{4\pi}{3d} \left(-\frac{\sqrt{3}}{2}, \frac{1}{2}\right),
\end{align}
and the Brillouin zone corners hosting the low energy states are at $K_{u} = \pm \frac{4\pi}{3d}\left(\frac{\sqrt{3}}{2},\frac{1}{2}\right)$. Within the BM model, the bilayer system is symmetric under $C_3$ by construction, and inversion and time-reversal (when both $K_u$ valleys of the original graphene monolayers are included). In order to break inversion symmetry, we introduce a coupling to a substrate (for example, hBN), which lifts inversion symmetry but leaves $C_3$ symmetry intact. This allows for a finite shift current which is entirely independent of the Berry curvature dipole, because the latter is set to zero by $C_3$ symmetry.
The  Hamiltonian of the bilayer system is therefore given by,
\begin{align}
    H = H_{\mathrm{t}} + H_{\mathrm{b}} + H_{\mathrm{tb}},
    \label{eq:total_ham}
\end{align}
where $\textrm{t,b,tb}$ denote the top, bottom layer, and interlayer hopping respectively. The top layer has the following continuum Hamiltonian, for a given momentum $\mathbf{q}$,
\begin{align}
    H_{\textrm{t,u}}(\mathbf{q})= \hbar v_f \sum_{s} u a_{t,s,u}^\dagger(\mathbf{q}) \bm{R_{+}} \mathbf{q}\cdot\bm{\sigma} a_{t,s,u}(\bm{q}),
\end{align}
where $s,u$ designate the spin and valley degrees of freedom; i.e., $s=\uparrow, \downarrow$, $u=\pm 1$, for the $K, K'$ valleys of the original graphene monolayers. $a_{\textrm{t/b},s,u}$ is the annihilation operator for an electron with spin s, and valley u, on the A/B sub-lattices of the top/bottom layers. $\bm{R}_{\pm}$ is the rotation matrix for the top/bottom layers, given by: $\bm{R_{\pm}} = \bm{R}\left(\pm \frac{\theta}{2}\right) = \cos(\frac{\theta}{2})\mp i\sigma_y \sin(\frac{\theta}{2})$, acting on the sub-lattice space, with $\theta \approx 1.05^{\circ}$ denoting the twist angle. The Fermi velocity is $\hbar v_f = 0.596\mathrm{eV\,nm}$~\cite{CastroNeto2009}, and $\bm{\sigma} = (\sigma_x, \sigma_y, \sigma_z)$ are Pauli matrices. The momentum $\mathbf{q}$ is measured relative to the valley centered at $K_u$.
Inversion symmetry breaking and uniaxial strain are introduced in the bottom layer. Throughout this work, the strain is applied along the zigzag direction of the bottom graphene sheet. The Hamiltonian of the bottom layer then takes the form,
\begin{align}
     H_{\textrm{b,u}}(\mathbf{q}) 
     &= \hbar v_f \sum_{s} a_{t,s,u}^\dagger(\mathbf{q}) \bigl(\bm{R_{-}} (1+\bm{\epsilon})(\mathbf{q}+u\mathcal{A})\cdot\bm{\sigma}
     \notag\\&\quad
     +\Delta \sigma_z \bigr)a_{t,s,u}(\mathbf{q}),
\end{align}
Here, $\bm{\epsilon}$ is the uniaxial strain matrix, which has the form $\bm{\epsilon} =\epsilon \left(\begin{smallmatrix}
-1 & 0 \\ 
0 & \nu
\end{smallmatrix}\right)$. $\mathcal{A}$ is the pseudo-gauge field resultant from the application of strain~\cite{Pereira2009,Guinea2010}. $\Delta = 17$ meV, is the staggered potential generated by alignment with an hBN layer, which is applied to the bottom layer only, mimicking realistic symmetry breaking in experiments. The staggered potential term is chosen in such a way as to break inversion symmetry ($\mathcal{P}$), and $C_{2x}$ \cite{Fernandes2020}. 
The evolution of the band structure with the staggered potential for $\theta = 1.05^{o}$ is presented in Fig. \ref{fig:fig9}. 
Strain is introduced through the parameter $\epsilon = 0, 0.1\%$, representing the strain-less and strained cases, respectively. With strain $H_{\textrm{tb}}$ is changed accordingly, as discussed in Sec. \ref{sec:strain}

\section{Continuum model under strain}
\label{sec:strain}
The Bistritzer-MacDonald continuum model contains an inter-layer coupling term, which couples two momenta $\bm{q},\bm{q'}$, if 
$\bm{q}-\bm{q}' = \left\lbrace \bm{q}_{1,u}, 
\bm{q}_{2,u}, \bm{q}_{3,u}\right\rbrace$, 
where $\bm{q}_{1,u} = |q_0| \left(0,-1\right)$, 
$\bm{q}_{2,u} = |q_0| \left(\frac{\sqrt{3}}{2},\frac{1}{2}\right)$,
$\bm{q}_{3,u} = |q_0| \left(-\frac{\sqrt{3}}{2},\frac{1}{2}\right)$, are the Moir\'{e} lattice vectors with $q_0 = \frac{8 \pi \sin(\frac{\theta}{2})}{3\sqrt{3}d}$, and $d = 1.42 \text{\AA}$ is the carbon-carbon bond length in graphene \cite{Bistritzer2011}.
In this framework, the interlayer coupling is included via the $H_{\textrm{tb}}$ term in Eq. \ref{eq:total_ham} as,
\begin{align}
    H_{\textrm{tb}} = \sum_{\bm{q},\bm{q}', s, u} a^\dagger_{t,s,u} (\bm{q})( T_{1,u}(\bm{q},\bm{q'}) + \\ \notag T_{2,u}(\bm{q},\bm{q'})+T_{3,u}(\bm{q},\bm{q'}))a_{b,s,u}(\bm{q'}).
\end{align}
In the presence of the strain defined in the main text, the coupling matrices (acting on the valley index) become,
\begin{align}
    T_{1,u} &= \frac{t}{3} \left(\begin{matrix}
    1 & 1 \\
    1 & 1
    \end{matrix} \right) \delta_{\bm{q}-\bm{q'},\bm{q}_{1,u}} \\ 
     T_{2,u} &= \frac{t}{3} \left(\begin{matrix}
    1 & e^{-i u \frac{2\pi}{3} \left(1-\epsilon^2 \nu^2\right)}   \\ 
    e^{i u \frac{2\pi}{3}\left(1-\epsilon^2 \nu^2\right)} & 1
    \end{matrix} \right) \delta_{\bm{q}-\bm{q'},\bm{q}_{2,u}}  \\ 
    T_{3,u} &= \frac{t}{3} \left(\begin{matrix}
     1 & e^{i u \frac{2\pi}{3} \left(1-\epsilon^2 \nu^2\right)}\\  
    e^{-i u \frac{2\pi}{3}\left(1-\epsilon^2 \nu^2\right)} & 1
    \end{matrix} \right) \delta_{\bm{q}-\bm{q'},\bm{q}_{3,u}}.
    \label{eq:interlayer}
\end{align}
Throughout, we take $t = 0.33$ eV. Recent work on relaxation of twisted graphene bilayers \cite{Koshino2018} suggests that interlayer hopping, $t$ is modified due to resultant corrugation of the graphene layers. The effect of this on the response is minor, as shown in the SM Sec. \ref{sec:corr}.
Accordingly, the lattice vectors of the Moir\'{e} superlattice are deformed in the presence of strain. These have the form,
\begin{align}
    \bm{q}_{1,u} = 
  u \frac{4\pi}{3\sqrt{3} d} \left( \epsilon \cos\tfrac{\theta}{2}, (2+\epsilon)\sin\tfrac{\theta}{2}\right) \\
  \bm{q}_{2,u} = u \frac{2\pi}{9 d} \left(\sqrt{3} \epsilon  \cos \tfrac{\theta}{2} -3(2- \epsilon \nu) \sin \tfrac{\theta}{2}, \right. \\ \notag
  \left. 3\epsilon \nu \cos\tfrac{\theta}{2}-\sqrt{3}(2+\epsilon)\sin\tfrac{\theta}{2}\right) \\
  \bm{q}_{3,u} = -u \frac{2 \pi}{9d} \left(3 \sin \tfrac{\theta}{2} (2- \nu  \epsilon ) -  \sqrt{3} \epsilon  \cos \tfrac{\theta}{2}, \right. \\ \notag \left. 3 \nu  \epsilon  \cos \tfrac{\theta}{2}+\sqrt{3} (2+\epsilon ) \sin \tfrac{\theta}{2}\right)
    \end{align}
\begin{figure*}
    \centering
    \includegraphics[width=.95\textwidth]{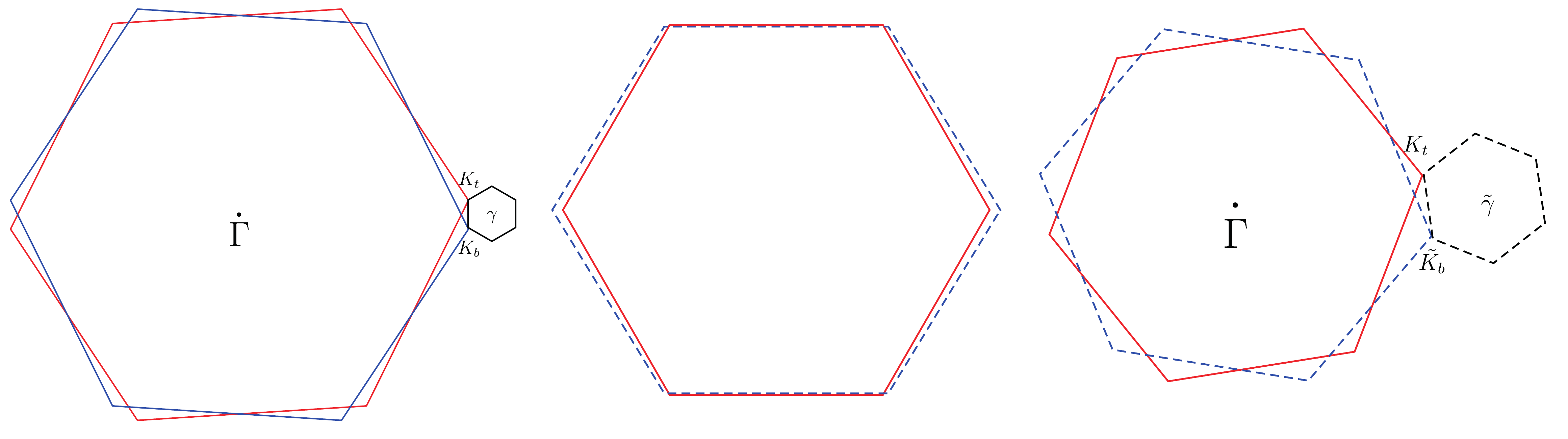}
    \caption{The Brillouin zone of the top (red) and bottom (blue) graphene sheets, with and without strain. 
    Left: Two undistorted ($\varepsilon = 0$) grahpene layers are rotated one with respect to the other, forming a folded mini Brillouin zone (mBZ), when rotated by an angle $\pm \theta/2$. The $K$ point of each layer shifts to $K_{t}$ and $K_{b}$ for the top and bottom layers respectively. The separation between them is denoted by $\bm{q}_0 = \bm{R}_{+}\bm{K}_u - \bm{R}_{-}\bm{K}_u$. When the mBZ is refolded onto the center of the original Brillouin zone, $K_{t/b} = |q_0| \left(\frac{\sqrt{3}}{2}, \pm \frac{1}{2}\right)$. Middle: Upon introduction of uniaxial strain on the bottom layer, the Brillouin zone deformes by expanding one, and contracting in the other direction. The $K$ points transform according to Eq. \ref{eq:kpoint_strain}. 
    Right: when the strained Brillouin zone is rotated with respect to an unstrained one, a deformed mBZ is formed, as shown here. Consequently, tunneling vectors which depend on the positions of $K_{t/b}$ in the mBZ are modified, as shown in Sec. \ref{sec:contmod}. The angle formed between the unstrained $\bf{q}_0$ and the strained vector $\bf{q}'_0$ is given by $\theta = \cos^{-1}\left(\frac{|q|_{0}}{|q'|_0}\right) \approx 3.16^{o}$.}
\label{fig:fig5}
\end{figure*}
The Dirac points transform under strain as,
\begin{align}
    \bar{\bf{K}}_u  = (1-\mathbf{\varepsilon}^{T}) \mathbf{K}_u - u \mathcal{A}
    \label{eq:kpoint_strain}
\end{align}
We introduce a pseudo-gauge field which stems from the underlying two-center approximation for the tunneling matrix \cite{Nam2017}. It is given by,
\begin{align}
    \mathcal{A} = -\frac{\beta \epsilon}{d}(1+\nu,0)
\end{align}
with $\nu = 0.165$, $\beta=1.57$ obtained from monolayer graphene.
Finally, the diagonalization of the Hamiltonian $H$ in Eq. \ref{eq:total_ham} is accomplished by recasting it in the form,
\begin{align}
    H = \sum_{\bm{q},s,u} A^\dagger_{s,u}(\bm{q}) h_u (\bm{q}) A_{s,u},
\end{align}
where now $A_{s,u}(\bm{q}) = [a_{b,s,u}(\bm{q}),a_{t,s,u}(\bm{q}+\bm{q}_{1,u}),a_{t,s,u}(\bm{q}+\bm{q}_{2,u}),a_{t,s,u}(\bm{q}+\bm{q}_{3,u})]^T$ is the infinite-component operator vector satisfying the constraints on momentum transfer. $h_u$ has the following truncated structure, after applying the momentum transfer relations ensuring non-zero $T$ tunnelling,
\begin{align}
    h_u(\bm{q}) = \left(
    \begin{matrix}
    H_{b,u}(\bm{q}) & T_{1,u} & T_{2,u} & T_{3,u} \\  
    T_{1,u}^\dagger & 
    H_{t,u}^{(1)} & 0 & 0 \\ 
    T_{2,u}^\dagger & 0 & 
    H_{t,u}^{(2)} & 0 \\ 
    T_{3,u}^\dagger & 0 & 0 & 
    H_{t,u}^{(3)} \\ 
    \end{matrix}
    \right),
\end{align}
with $H_{t,u}^{(i)}=H_{t,u}(\bm{q}+\bm{q_{i,u}})$.
In this work, we used 81 sites in reciprocal space for the construction of the hamiltonian, which results in a Hamiltonian which is $324 \times 324$. The integrals appearing in Eqs.~\eqref{eq:shifts1},\eqref{eq:shifts2} are computed using a discretized grid of $600 \times 600$ in the $(k_x, k_y)$ plane of the mini Brillouin zone. $10^3$ frequency point samplings in $\omega$ are carried out uniformly. Convergence is checked against the case with $C_3$ symmetry, where verification is done by comparing $\sigma_{xx;x}$ with  $-\sigma_{yy;x}$; and $\sigma_{yy;y}$ with $-\sigma_{xx;y}$. All equalities were verified to within $5\%$.
We stress that both original graphene valleys $K_{\pm}$ are included in every calculation (see SI for model construction details), and, since the expressions appearing in Eqs. 3-6 are time-reversal symmetric, the effect of evaluating both valleys is to double the overall result. This was also verified numerically to the stated accuracy.
The delta functions of Eqs.~\eqref{eq:shifts1},\eqref{eq:shifts2} are broadened with a width $\Gamma = 0.02\mathrm{meV}$  for $T=0\mathrm{K}$ and $\Gamma = 0.1 \textrm{meV}$ at $T=300\mathrm{K}$. This corresponds to transport lifetimes observed in bilayer graphene in the clean limit \cite{Monteverde2010}. 
\section{Symmetry properties of the response}
In the main text, we observe that the response tensor $\sigma^{ab;c}$ has only two independent components, and that the Berry curvature dipole, $\partial^a \Omega^{bc}$ vanishes. We proceed to prove this. The generator of $C_{3z}$ is given by \cite{Tinkham2003},
\begin{align}
    \bf{M} = \left(
    \begin{matrix}
        -\frac{1}{2} & -\frac{\sqrt{3}}{2} & 0
        \\ \frac{\sqrt{3}}{2} & -\frac{1}{2} & 0 \\
         0 & 0 & 1
    \end{matrix}
    \right).
\end{align}
The Berry curvature dipole (BCD), $D^{abc}$ is a gauge invariant material property of the system, and has the form $D^{abc} = \partial_{k_a} \Omega^{bc}$, making it a rank-3 pseudotensor. Firstly, we observe that this quantity is anti-symmetric in $(b,c)$. Applying Neumann's principle, we enforce $D^{abc} = \sum_{\alpha \beta \gamma} M_{a\alpha}M_{b\beta}M_{c\gamma}D^{\alpha\beta\gamma}$. Using the anti-symmetry of $D^{abc} = -D^{acb}$, we focus only on non-trivial components, $D^{axy} = \sum_{\alpha \beta \gamma} M_{a\alpha}M_{x\beta}M_{y\gamma}D^{\alpha\beta\gamma} = \sum_{a\alpha} M_{a\alpha}\left( \frac{1}{4}D^{\alpha xy} - \frac{3}{4} D^{\alpha yx}\right) = \sum_{a \alpha} M_{a \alpha} D^{\alpha xy}$. Since $a = x,y$, we obtain the following set of equations, $D^{xxy} = -\frac{1}{2} D^{xxy} - \frac{\sqrt{3}}{2} D^{yxy}, ~~ D^{yxy} = \frac{\sqrt{3}}{2} D^{xxy}-\frac{1}{2}D^{yxy}$. This set admits only the solution $D^{xxy} = D^{yxy} = 0$, as required, demonstrating that the BCD is zero, under $C_{3z}$. For the general rank-3 symmetric conductivity tensor $\sigma^{ab;c}$, we derive analogous symmetry constraints under $C_{3z}$. This results in two independent components overall,
\begin{align}
    \sigma^{xx;x} = -\sigma^{yy;x} = -\sigma^{xy;y} = -\sigma^{yx;y} \\
    \sigma^{yy;y} = -\sigma^{xx;y} = -\sigma^{yx;x} = -\sigma^{xy;x}.
\end{align}
We further note that under linear-polarized light, the conductivity tensor exhibits a special permutation symmetry, $\sigma^{ab;c} = \sigma^{ba;c}$.

\section{Additional data for the transverse components}
For completeness, we provide the remaining transverse component $\sigma^{xx;y}$ of the shift current, which agrees qualitatively and in part quantitatively with the component $\sigma^{yy;x}$ discussed in the main text. 
We recall that in general, with $C_{3z}$ symmetry the conductivity tensor has only 2 independent components, therefore the longitudinal components can be deduced straightforwardly from the data presented here.

\begin{figure}
    \centering
    \includegraphics[width=.45\textwidth]{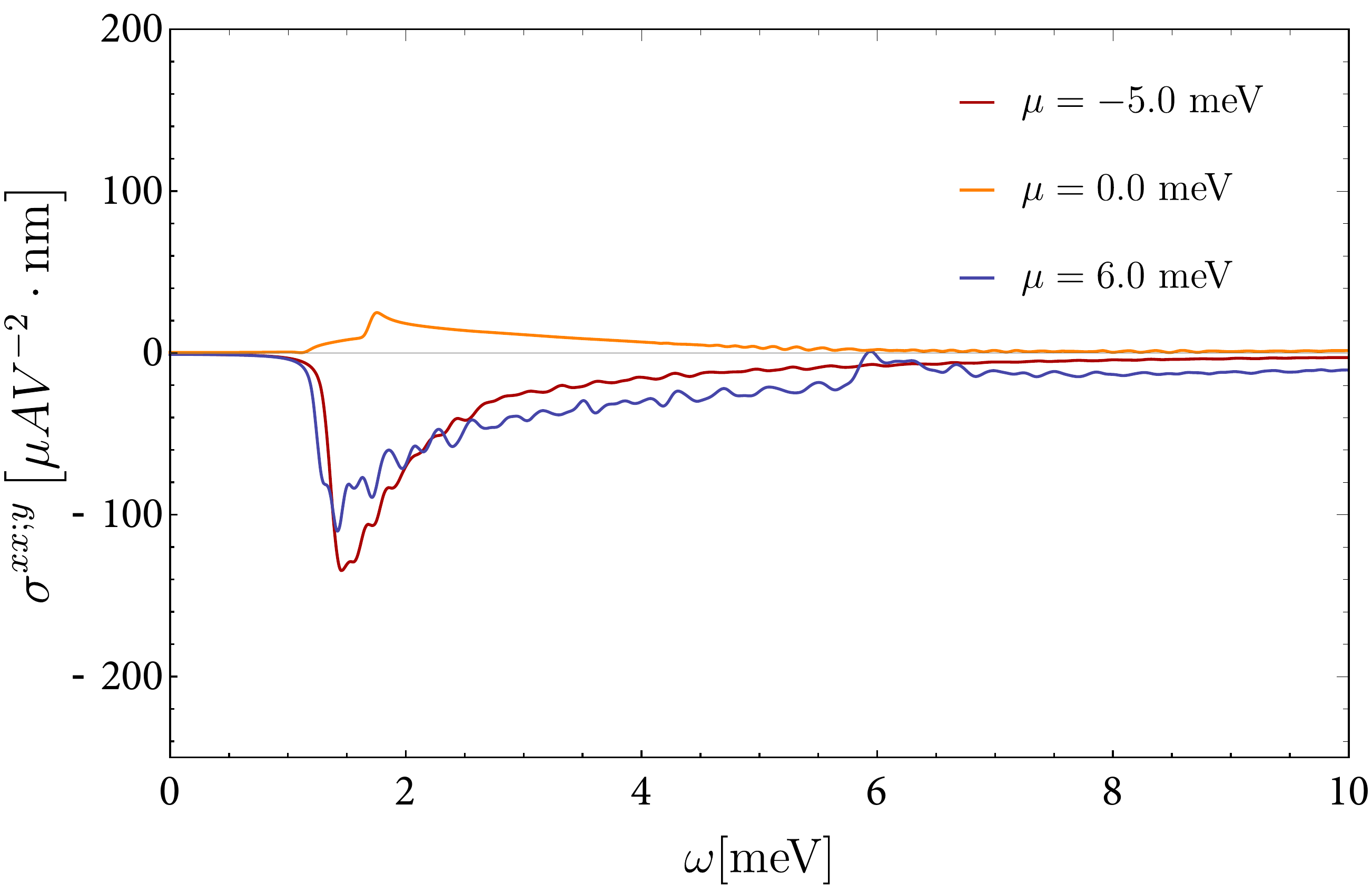}
    \caption{Conductivity $\sigma^{xx;y}$ as a function of frequency, with $\varepsilon = 0$ for 3 values of the chemical potential. Compare with $\sigma^{yy;x}$ in Fig. \ref{fig:fig3}.
    Here, $\sigma^{xx;y}$ is negative for both $\mu = 6.0, -5.0 \mathrm{meV}$, and has the peak values $\sigma^{xx;y} = -108, -121 \mathrm{\mu A nm V^{-2}}$, respectively. For $\mu = 0$, $\sigma^{xx;y} = 28 \mathrm{\mu A nm V^{-2}}$. 
    A sizeable conductivity, $|\sigma^{xx;y}| > 50 \mathrm{\mu A nm V^{-2}}$ is obtained for a wide range of frequencies between $\omega = 1.5- 4.5 \mathrm{meV}$.}
    \label{fig:fig6}
\end{figure}

%\subsection{Without strain}
With $\varepsilon = 0$, the remaining independent component is $\sigma^{xx;y}$, also a transverse component. This is presented in Fig. \ref{fig:fig6}, for 3 chemical potential values. Note that the contribution of Eq.~\eqref{eq:shifts1} is zero, due to $C_{3z}$ symmetry. 

With finite strain, all terms in Eq.~\eqref{eq:shift_new} of the main text contribute to the current. For consistency, we again present the transverse component, $\sigma^{xx;y}$, for 3 chemical potential values, in the same way as in Fig. \ref{fig:fig3} of the main text. 
\begin{figure*}
    \centering
    \includegraphics[width=1.0\textwidth]{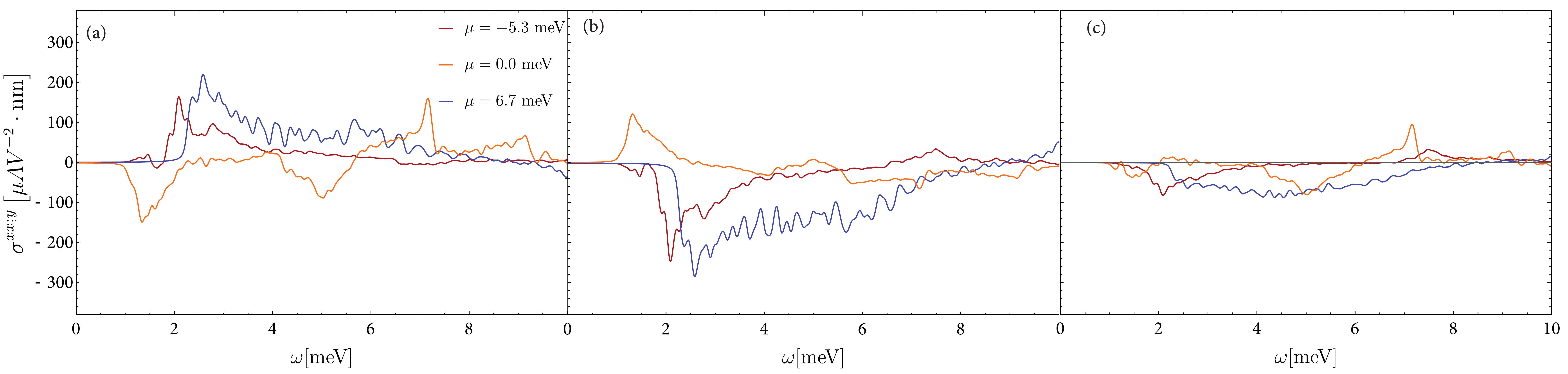}
    \caption{Contributions to the conductivity $\sigma^{xx;y}$, with strain $\varepsilon=0.001$, the same value used in Fig. \ref{fig:fig4} of the main text.
    (a) Berry curvature dipole, Eq.~\eqref{eq:shifts1}. (b) $R_{\mathrm{shift}} + K_{\mathrm{shi ft}}$, as in Eq. ~\eqref{eq:shifts2}. (c) Total conductivity. While the introduction of strain produces a giant Berry curvature dipole, the magnitude of Eq.~\eqref{eq:shifts2} also increases albeit with opposite sign. Consequently, the total conductivity remains comparable to the unstrained case. For $\mu = -5.3, 0.0, 6.7 ~~ \mathrm{meV}$, the maximal values obtained for the conductivity are $\sigma^{xx;y} = -96,100, -102 ~~ \mathrm{\mu A nm V^{-2}}$, respectively. We note that for $\mu = 6.7 \mathrm{meV} $, the response profile is exceptionally broad, and is the conductivity is almost constant for the range $\omega = 2-6 \mathrm{meV}$.
    \label{fig:fig7}
    }
\end{figure*}
With the introduction of strain, the Berry curvature dipole contribution is no longer zero (Fig.~\ref{fig:fig7}a). Although the net conductivity is not substantially enhanced by the introduction of strain (cf. Fig.~\ref{fig:fig7}c), a broad resonance in $\sigma^{xx;y}$ appears at chemical potential $\mu = 6.7 \mathrm{meV}$. 
Note that while the introduction of strain induces a large Berry curvature dipole, it is still smaller than the remaining contributions according to Eq.~\eqref{eq:shifts2}, meaning that the sign of the total conductivity is determined by the latter part of the response. 

\begin{figure*}
    \centering
    \includegraphics[width=1.0\textwidth]{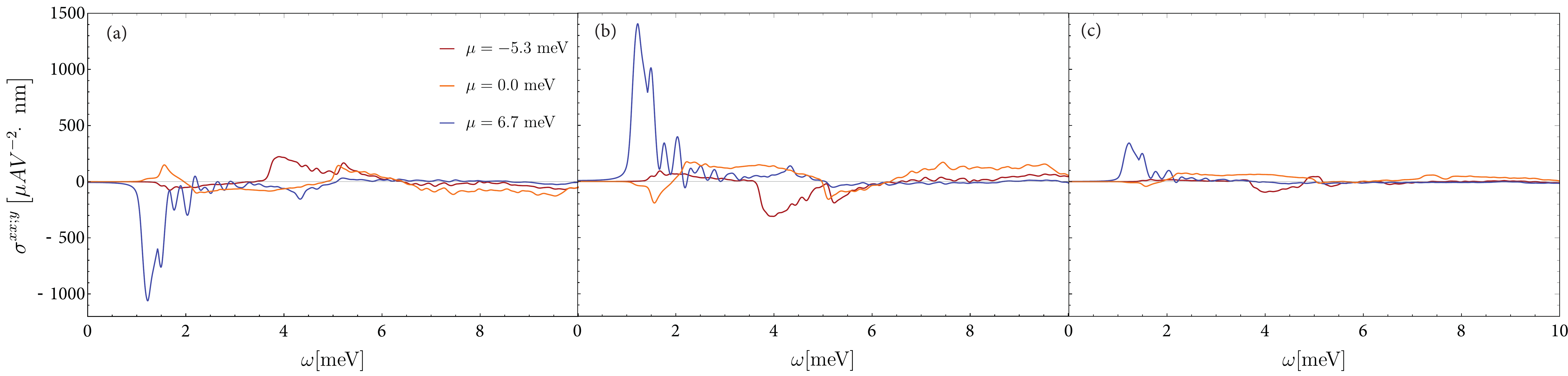}
    \caption{Contributions to the conductivity $\sigma^{xx;y}$ with $\varepsilon = -0.001$, for 3 chemical potential values. (a) BCD term. (b) $K_{\mathrm{shift}}+R_{\mathrm{shift}}$ term. (c) Total conductivity.  }.
    \label{fig:fig8}
\end{figure*}

To examine whether our results depend of the sign of the applied strain (i.e., whether the strain is compressive or tensile), we show in Fig.~\ref{fig:fig8} the conductivity $\sigma^{yy;x}$ for strain with negative (i.e., compressive) magnitude, $\epsilon = -0.001$. While the detailed frequency dependence is indeed sensitively dependent on the strain, both the magnitude of the response and ts resonance structure are very comparable.
One may also be tempted to compare the magnitude of the shift current of $\sim 200 \mathrm{\mu A nm V^{-2}}$ with the nonlinear anomalous Hall signal, which is expected to be around 10x larger~\cite{zhang2020giant} for the same value of strain. However, we emphasize that the latter appears in response to a static electric field, rendering such a comparison moot; the mechanisms for the nonlinear Hall effect and for shift current generation are unrelated.
\section{Effects of corrugation}
\label{sec:corr}
The original work of B-M considered frozen graphene layers, without effects due to relaxation and strain. Recently, reparametrizations of the continuum model with ab-initio methods \cite{Koshino2018, Carr2019} confirmed that the formation of triangular domains of AA/AB regions in the bilayer system may strongly affect interlayer tunneling. For this reason, these works suggested a different interpolation of the interlayer tunneling matrix. We rewrite Eq.~\eqref{eq:interlayer} in a different form,
\begin{align}
    T_{1,u} = \frac{t}{3} \left(\begin{matrix}
    u & 1 \\
    1 & u
    \end{matrix} \right)
\end{align}
Where $u$ is the ratio of $\textrm{AA}$ to $\textrm{AB}$ domain tunneling. The case of $u=1$ reproduces the original B-M parameters, while $u=0$ would correspond to the so-called ``chiral limit" hypothesized to occur under certain conditions in TBG. We stress that the latter has not been observed experimentally.  In what follows, we consider the parameters suggested by Koshino \textit{et al.} \cite{Koshino2018}, and adopt $u = 0.81$, as the value extracted from ab-initio calculations. The substitution of $u$ is repeated for all $T_{1,u},T_{2,u},T_{3,u}$ matrices. The coupling to the hBN substrate is unchanged and remains at $\Delta = 17 \textrm{meV}$.
\begin{figure*}[h!]
    \centering
    \includegraphics[width=\textwidth]{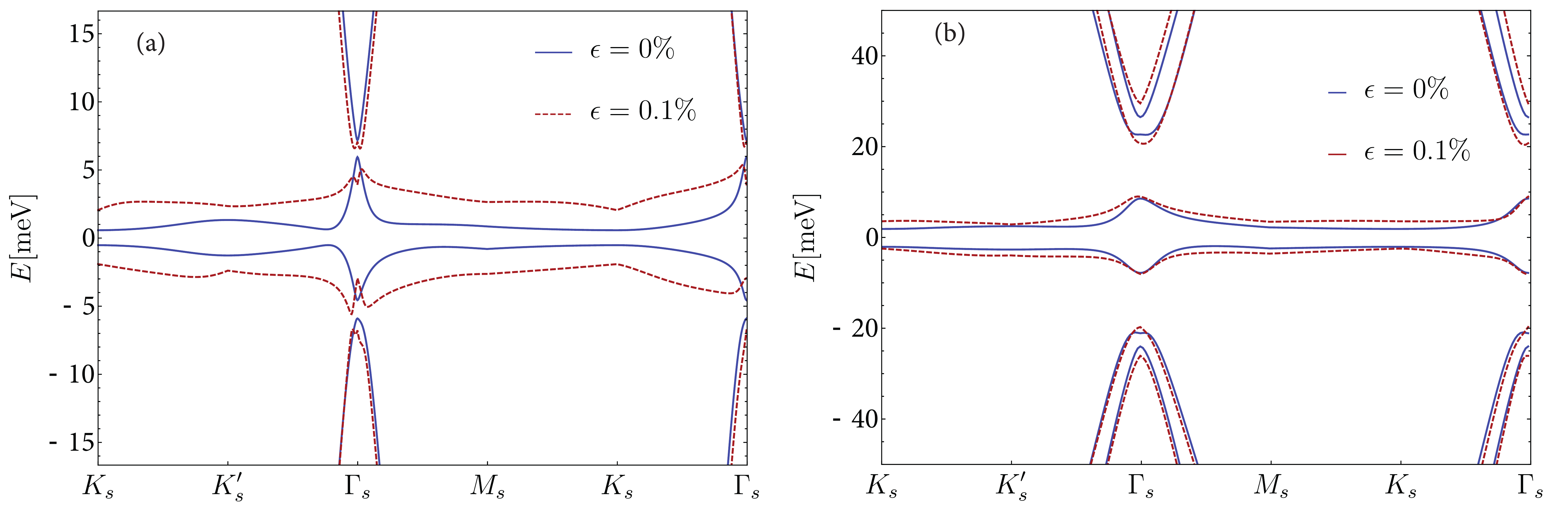}
    \caption{Band structures of TBG with different values of $u$. (a) Original parameterization of B-M with $u=1$. (b) Band structure with $u=0.81$ after Koshino \textit{et al.} \cite{Koshino2018}. }
    \label{fig:corr1}
\end{figure*}
In Fig. \ref{fig:corr1} (b) we plot the band structure with the modified band structure with strain $(\varepsilon = 0.1 \%)$, and without strain $(\varepsilon = 0 \%)$. For completeness, we enclose alongside it, in Fig. \ref{fig:corr1}(a) the band structure with the original B-M parameterization. While certain differences are discernible (such as a shift in the energy distance to the dispersive bands), flat bands appear as they did for the original B-M model. 
\begin{figure*}[h!]
    \centering
    \includegraphics[width=\textwidth]{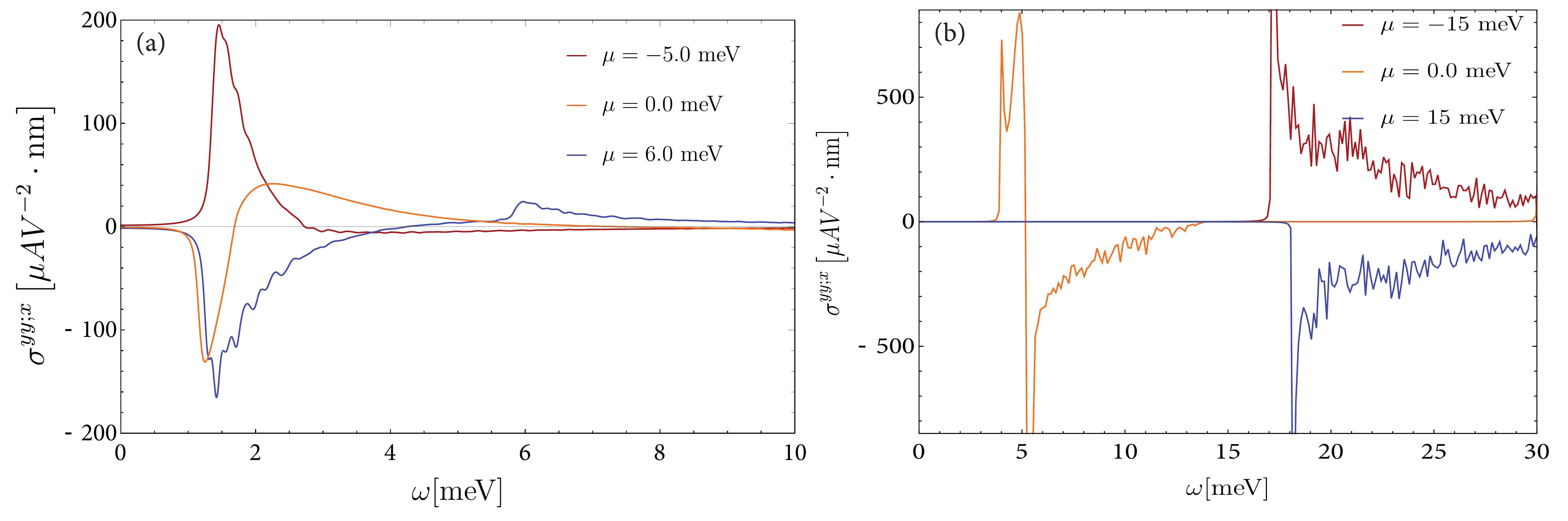}
    \caption{Shift current for different values of $u$, at $\varepsilon = 0 $ strain. (a) Original parameterization of B-M with $u=1$. (b) Shift current with $u=0.81$ after Koshino \textit{et al.} \cite{Koshino2018}. }
    \label{fig:corr2}
\end{figure*}
Following the differences observed in the band structure, we calculate the shift current with the different value of $u$. Clearly, the flat band contribution (seen at $\mu = 0$) produces a large shift-current response in the region $\omega < 10 \textrm{meV}$, as seen indeed Fig. \ref{fig:corr2} (b); compare this with the original B-M parameters \ref{fig:corr2} (a), which similarly show this, albeit at a slightly different frequency $\omega$. The dispersive bands enter at higher frequencies as expected -- $\omega > 15 \textrm{meV}$ -- but their shape resembles the one found in Fig. \ref{fig:fig3} (and shown in Fig. \ref{fig:corr2}(a)). This precisely underscores the point made in the main text that the parameters of the continuum model affect the results qualitatively, but the salient features of the shift current response and our suggested design principle remain unaffected. For completeness, we include the results with strain, in Fig. \ref{fig:corr2}. Here once more we observe that the main effect of the altered parameters is manifest in the magnitude of the response, but not in the main conclusions we provided in the main text, namely, the robustness of the shift current response stemming from flat bands, which are preserved even in the presence of strain. We find an enhancement of the conclusions we have derived the main text: while in the presence of strain the dispersive bands contribute to the shift current response mainly via the presence of a non-vanishing Berry curvature dipole, the flat band response is once again dominated solely by the momentum-space shift current $K_\textrm{shift}$ we have put forward as the mechanism behind shift current generation. 
\begin{figure*}[h!]
    \centering
    \includegraphics[width=\textwidth]{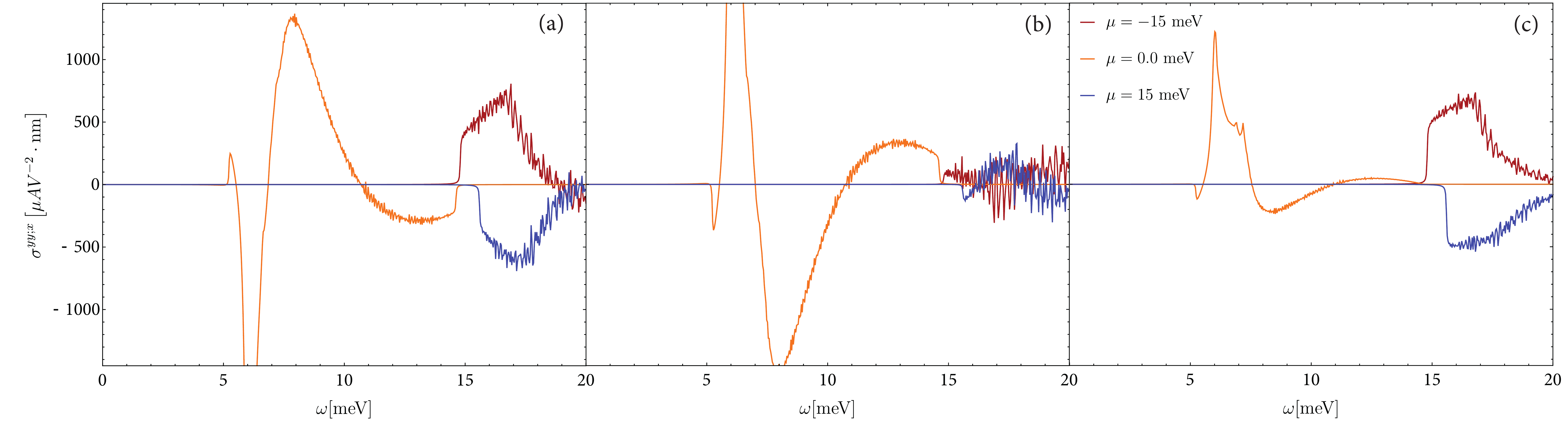}
    \caption{Shift current response for the $\sigma^{yy;x}$ conductivity, with $u = 0.81$ and 3 chemical potential values, with strain $\varepsilon = 0.001$. (a) Berry curvature dipole contribution. (b) Momentum space shift current contribution. (c) Total shift current.}
    \label{fig:corr3}
\end{figure*}
In summary, the tuning of interlayer tunneling does not substantially affect the quantitative results we have observed in the main text. Reducing $u$ to $u=0.81$ (as suggested by ab-initio studies on the properties of graphene bilayers) does not modify our conclusions regarding the robustness of the shift current stemming from flat bands, and the importance of the momentum space picture for understanding the source of the giant response in TBG. We have shown that this is independent of the precise parameterization of the continuum model for the TBG, and therefore, could serve as a vital design principle in a new generation of photovoltaic devices. 
\section{Nonlinear conductivity at 3/4 filling}
\label{sec:threequart}
 \begin{figure}
    \centering
    \includegraphics[width=.45\textwidth]{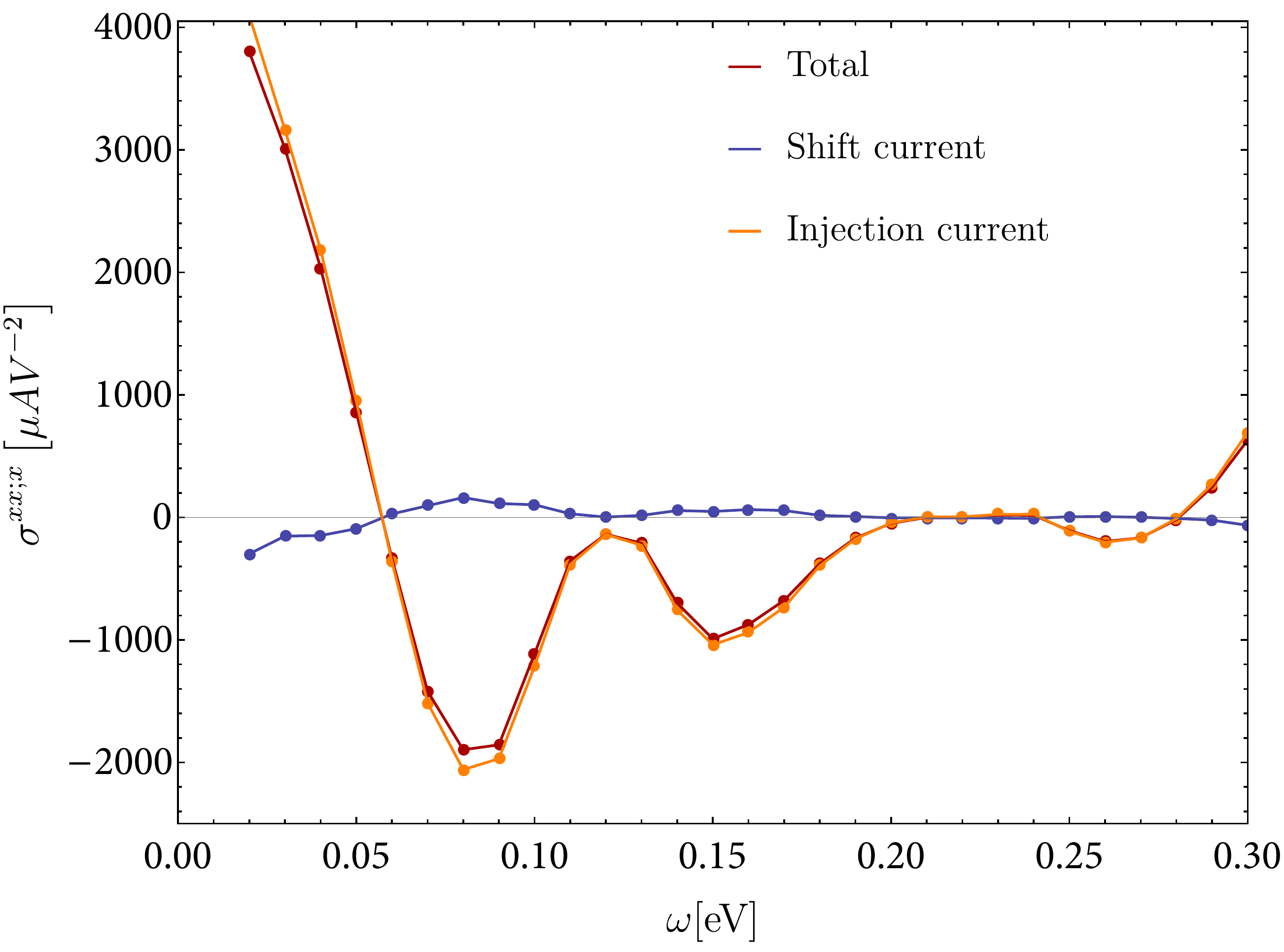}
    \caption{Photoconductivity for TBG at 3/4 filling. The red curve indicates the total photoconductivity (shift and injection). The blue curve denotes the shift current contribution. The yellow curve shows in the injection current contribution. The response is overwhelmingly determined by the injection current. The value taken for the intraband relaxation is {$\gamma = 0.1 \textrm{meV}$. The delta function is broadened by $\Gamma = 0.02 \textrm{meV}$}.}
    \label{fig:decomposition}
\end{figure}
We compare our results to a previous study \cite{Liu2020a} of the nonlinear optical conductivity of TBG focused on a regime in which the ground state breaks time-reversal symmetry (TRS). 
We begin \cite{Holder2020, vonBaltz1981} by listing the dominant contributions from the 3-legged diagrams appearing the perturbative treatment of the light-matter interactions when TRS is broken,
\begin{align}
    \notag \sigma^{aa;c} &= \frac{e^3}{\hbar^2 \omega^2}\int_\bk \sum_{n,m,l} \sum_{\Omega =\pm \omega} f_{nm} \frac{v^{a}_{nm}}{\varepsilon_{nm}+\Omega + i\Gamma} \\  &\left(\frac{v^a_{ml}v^c_{ln}}{\varepsilon_{nl}+i\gamma}-\frac{v^c_{ml}v^a_{ln}}{\varepsilon_{lm}+i\gamma}\right)
    \label{eq:vonbaltz}
 \end{align}
Here we also defined $v^a_{nm} = \langle n(\bk) \left|\frac{\partial H}{\partial k_a}\right| m(\bk) \rangle$. 
In general, Eq.~\eqref{eq:vonbaltz} is insufficient to arrive at all contributions to second-order nonlinear optical response. But for the continuum model of TBG, incorporating only linear dispersion, all higher order vertices vanish since $\frac{\partial^n H}{\partial k_a^n} = 0, n > 1$. Eq.~\eqref{eq:vonbaltz} allows us to isolate divergent contributions, which occur whenever $l=n,m$, since the summation over $l$ is unrestricted. Focusing on these terms,  in the limit $\Gamma \to 0$ one obtains
\begin{align}
    \sigma^{aa;c}_\textrm{inj} = -\frac{\pi e^3}{\hbar^2 \gamma} \sum_{nm} \int_\bk f_{nm}  \frac{|v^a_{nm}|^2}{\varepsilon_{nm}^2} \Delta^c_{mn}\delta(\varepsilon\pm \omega),
\end{align}
where we defined $\Delta^c_{mn} = v^c_{mm} - v^c_{nn}$. Importantly, this part of the conductivity vanishes identically in the presence of TRS. This holds because $\mathcal{T}^{-1} v^c_{nn}(\bk) \mathcal{T} = -v^c_{nn}(-\bk)$, where $\mathcal{T}$ is the time-reversal operator. The case when TRS is preserved is precisely the situation discussed in the present work. However, if TRS is broken, as it is the case for TBG at 3/4 filling, $\sigma^{aa;c}_{\mathrm{inj}}$ survives, and becomes the dominant contribution. Furthermore, this conductivity scales as $\gamma^{-1}$, where $\gamma$ is the carrier relaxation rate. The shift current appears when $l \neq n,m$, which immediately reduces the response by a factor of $\varepsilon_{nl},\varepsilon_{ml} \gg \gamma$, rendering it $\gamma^{0}$. The resulting shift current gives rise to the expressions derived in this work as Eq.~\eqref{eq:shift_new}.
To illustrate this point more clearly, we calculate the total photoconductivity (shift and injection contributions) in the 3/4 filling of TBG. We find that for this filling, the response is entirely dominated by the injection contribution, and scales directly with $\gamma^{-1}$ (cf. Fig.~\ref{fig:decomposition}).
We stress once more that whenever TRS is restored, the injection term vanishes identically and leaves only the shift current. Additionally, the stronger dependence on the carrier relaxation rate means that the injection current is much more effectively suppressed by increasing temperature. For the 3/4 filling case, adding together the shift and injection currents reproduces Ref.~\cite{Liu2020a}.

\bibliography{main.bbl}

\begin{thebibliography}{83}%
\makeatletter
\providecommand \@ifxundefined [1]{%
 \@ifx{#1\undefined}
}%
\providecommand \@ifnum [1]{%
 \ifnum #1\expandafter \@firstoftwo
 \else \expandafter \@secondoftwo
 \fi
}%
\providecommand \@ifx [1]{%
 \ifx #1\expandafter \@firstoftwo
 \else \expandafter \@secondoftwo
 \fi
}%
\providecommand \natexlab [1]{#1}%
\providecommand \enquote  [1]{``#1''}%
\providecommand \bibnamefont  [1]{#1}%
\providecommand \bibfnamefont [1]{#1}%
\providecommand \citenamefont [1]{#1}%
\providecommand \href@noop [0]{\@secondoftwo}%
\providecommand \href [0]{\begingroup \@sanitize@url \@href}%
\providecommand \@href[1]{\@@startlink{#1}\@@href}%
\providecommand \@@href[1]{\endgroup#1\@@endlink}%
\providecommand \@sanitize@url [0]{\catcode `\\12\catcode `\$12\catcode
  `\&12\catcode `\#12\catcode `\^12\catcode `\_12\catcode `\%12\relax}%
\providecommand \@@startlink[1]{}%
\providecommand \@@endlink[0]{}%
\providecommand \url  [0]{\begingroup\@sanitize@url \@url }%
\providecommand \@url [1]{\endgroup\@href {#1}{\urlprefix }}%
\providecommand \urlprefix  [0]{URL }%
\providecommand \Eprint [0]{\href }%
\providecommand \doibase [0]{https://doi.org/}%
\providecommand \selectlanguage [0]{\@gobble}%
\providecommand \bibinfo  [0]{\@secondoftwo}%
\providecommand \bibfield  [0]{\@secondoftwo}%
\providecommand \translation [1]{[#1]}%
\providecommand \BibitemOpen [0]{}%
\providecommand \bibitemStop [0]{}%
\providecommand \bibitemNoStop [0]{.\EOS\space}%
\providecommand \EOS [0]{\spacefactor3000\relax}%
\providecommand \BibitemShut  [1]{\csname bibitem#1\endcsname}%
\let\auto@bib@innerbib\@empty
%</preamble>
\bibitem [{\citenamefont {{Belinicher}}\ and\ \citenamefont
  {{Sturman}}(1980)}]{Belinicher1980}%
  \BibitemOpen
  \bibfield  {author} {\bibinfo {author} {\bibfnamefont {V.~I.}\ \bibnamefont
  {{Belinicher}}}\ and\ \bibinfo {author} {\bibfnamefont {B.~I.}\ \bibnamefont
  {{Sturman}}},\ }\bibfield  {title} {\bibinfo {title} {{The photogalvanic
  effect in media lacking a center of symmetry}},\ }\href
  {https://doi.org/10.1070/PU1980v023n03ABEH004703} {\bibfield  {journal}
  {\bibinfo  {journal} {Sov. Phys. Usp.}\ }\textbf {\bibinfo {volume} {23}},\
  \bibinfo {pages} {199} (\bibinfo {year} {1980})}\BibitemShut {NoStop}%
\bibitem [{\citenamefont {Boyd}(2008)}]{Boyd2003}%
  \BibitemOpen
  \bibfield  {author} {\bibinfo {author} {\bibfnamefont {R.~W.}\ \bibnamefont
  {Boyd}},\ }\href@noop {} {\emph {\bibinfo {title} {Nonlinear Optics (Third
  Edition)}}},\ \bibinfo {edition} {3rd}\ ed.\ (\bibinfo  {publisher} {Academic
  Press},\ \bibinfo {address} {Burlington},\ \bibinfo {year}
  {2008})\BibitemShut {NoStop}%
\bibitem [{\citenamefont {{von Baltz}}\ and\ \citenamefont
  {{Kraut}}(1981)}]{vonBaltz1981}%
  \BibitemOpen
  \bibfield  {author} {\bibinfo {author} {\bibfnamefont {R.}~\bibnamefont {{von
  Baltz}}}\ and\ \bibinfo {author} {\bibfnamefont {W.}~\bibnamefont
  {{Kraut}}},\ }\bibfield  {title} {\bibinfo {title} {{Theory of the bulk
  photovoltaic effect in pure crystals}},\ }\href
  {https://doi.org/10.1103/PhysRevB.23.5590} {\bibfield  {journal} {\bibinfo
  {journal} {Phys. Rev. B}\ }\textbf {\bibinfo {volume} {23}},\ \bibinfo
  {pages} {5590} (\bibinfo {year} {1981})}\BibitemShut {NoStop}%
\bibitem [{\citenamefont {{Sipe}}\ and\ \citenamefont
  {{Shkrebtii}}(2000)}]{Sipe2000}%
  \BibitemOpen
  \bibfield  {author} {\bibinfo {author} {\bibfnamefont {J.~E.}\ \bibnamefont
  {{Sipe}}}\ and\ \bibinfo {author} {\bibfnamefont {A.~I.}\ \bibnamefont
  {{Shkrebtii}}},\ }\bibfield  {title} {\bibinfo {title} {{Second-order optical
  response in semiconductors}},\ }\href
  {https://doi.org/10.1103/PhysRevB.61.5337} {\bibfield  {journal} {\bibinfo
  {journal} {Phys. Rev. B}\ }\textbf {\bibinfo {volume} {61}},\ \bibinfo
  {pages} {5337} (\bibinfo {year} {2000})}\BibitemShut {NoStop}%
\bibitem [{\citenamefont {{Young}}\ and\ \citenamefont
  {{Rappe}}(2012)}]{Young2012}%
  \BibitemOpen
  \bibfield  {author} {\bibinfo {author} {\bibfnamefont {S.~M.}\ \bibnamefont
  {{Young}}}\ and\ \bibinfo {author} {\bibfnamefont {A.~M.}\ \bibnamefont
  {{Rappe}}},\ }\bibfield  {title} {\bibinfo {title} {{First Principles
  Calculation of the Shift Current Photovoltaic Effect in Ferroelectrics}},\
  }\href {https://doi.org/10.1103/PhysRevLett.109.116601} {\bibfield  {journal}
  {\bibinfo  {journal} {Phys. Rev. Lett.}\ }\textbf {\bibinfo {volume} {109}},\
  \bibinfo {eid} {116601} (\bibinfo {year} {2012})},\ \Eprint
  {https://arxiv.org/abs/1202.3168} {arXiv:1202.3168 [cond-mat.mtrl-sci]}
  \BibitemShut {NoStop}%
\bibitem [{\citenamefont {Knap}\ \emph {et~al.}(2009)\citenamefont {Knap},
  \citenamefont {Dyakonov}, \citenamefont {Coquillat}, \citenamefont {Teppe},
  \citenamefont {Dyakonova}, \citenamefont {{\L}usakowski}, \citenamefont
  {Karpierz}, \citenamefont {Sakowicz}, \citenamefont {Valusis}, \citenamefont
  {Seliuta}, \citenamefont {Kasalynas}, \citenamefont {El~Fatimy},
  \citenamefont {Meziani},\ and\ \citenamefont {Otsuji}}]{Knap2009}%
  \BibitemOpen
  \bibfield  {author} {\bibinfo {author} {\bibfnamefont {W.}~\bibnamefont
  {Knap}}, \bibinfo {author} {\bibfnamefont {M.}~\bibnamefont {Dyakonov}},
  \bibinfo {author} {\bibfnamefont {D.}~\bibnamefont {Coquillat}}, \bibinfo
  {author} {\bibfnamefont {F.}~\bibnamefont {Teppe}}, \bibinfo {author}
  {\bibfnamefont {N.}~\bibnamefont {Dyakonova}}, \bibinfo {author}
  {\bibfnamefont {J.}~\bibnamefont {{\L}usakowski}}, \bibinfo {author}
  {\bibfnamefont {K.}~\bibnamefont {Karpierz}}, \bibinfo {author}
  {\bibfnamefont {M.}~\bibnamefont {Sakowicz}}, \bibinfo {author}
  {\bibfnamefont {G.}~\bibnamefont {Valusis}}, \bibinfo {author} {\bibfnamefont
  {D.}~\bibnamefont {Seliuta}}, \bibinfo {author} {\bibfnamefont
  {I.}~\bibnamefont {Kasalynas}}, \bibinfo {author} {\bibfnamefont
  {A.}~\bibnamefont {El~Fatimy}}, \bibinfo {author} {\bibfnamefont {Y.~M.}\
  \bibnamefont {Meziani}},\ and\ \bibinfo {author} {\bibfnamefont
  {T.}~\bibnamefont {Otsuji}},\ }\bibfield  {title} {\bibinfo {title} {Field
  effect transistors for terahertz detection: Physics and first imaging
  applications},\ }\href {https://doi.org/10.1007/s10762-009-9564-9} {\bibfield
   {journal} {\bibinfo  {journal} {Journal of Infrared, Millimeter, and
  Terahertz Waves}\ }\textbf {\bibinfo {volume} {30}},\ \bibinfo {pages} {1319}
  (\bibinfo {year} {2009})}\BibitemShut {NoStop}%
\bibitem [{\citenamefont {Tonouchi}(2007)}]{Tonouchi2007}%
  \BibitemOpen
  \bibfield  {author} {\bibinfo {author} {\bibfnamefont {M.}~\bibnamefont
  {Tonouchi}},\ }\bibfield  {title} {\bibinfo {title} {{Cutting-edge terahertz
  technology}},\ }\href {https://doi.org/10.1038/nphoton.2007.3} {\bibfield
  {journal} {\bibinfo  {journal} {Nat. Photonics}\ }\textbf {\bibinfo {volume}
  {1}},\ \bibinfo {pages} {97} (\bibinfo {year} {2007})}\BibitemShut {NoStop}%
\bibitem [{\citenamefont {{Sharpe}}\ \emph {et~al.}(2019)\citenamefont
  {{Sharpe}}, \citenamefont {{Fox}}, \citenamefont {{Barnard}}, \citenamefont
  {{Finney}}, \citenamefont {{Watanabe}}, \citenamefont {{Taniguchi}},
  \citenamefont {{Kastner}},\ and\ \citenamefont
  {{Goldhaber-Gordon}}}]{Sharpe2019}%
  \BibitemOpen
  \bibfield  {author} {\bibinfo {author} {\bibfnamefont {A.~L.}\ \bibnamefont
  {{Sharpe}}}, \bibinfo {author} {\bibfnamefont {E.~J.}\ \bibnamefont {{Fox}}},
  \bibinfo {author} {\bibfnamefont {A.~W.}\ \bibnamefont {{Barnard}}}, \bibinfo
  {author} {\bibfnamefont {J.}~\bibnamefont {{Finney}}}, \bibinfo {author}
  {\bibfnamefont {K.}~\bibnamefont {{Watanabe}}}, \bibinfo {author}
  {\bibfnamefont {T.}~\bibnamefont {{Taniguchi}}}, \bibinfo {author}
  {\bibfnamefont {M.~A.}\ \bibnamefont {{Kastner}}},\ and\ \bibinfo {author}
  {\bibfnamefont {D.}~\bibnamefont {{Goldhaber-Gordon}}},\ }\bibfield  {title}
  {\bibinfo {title} {{Emergent ferromagnetism near three-quarters filling in
  twisted bilayer graphene}},\ }\href {https://doi.org/10.1126/science.aaw3780}
  {\bibfield  {journal} {\bibinfo  {journal} {Science}\ }\textbf {\bibinfo
  {volume} {365}},\ \bibinfo {pages} {605} (\bibinfo {year} {2019})},\ \Eprint
  {https://arxiv.org/abs/1901.03520} {arXiv:1901.03520 [cond-mat.mes-hall]}
  \BibitemShut {NoStop}%
\bibitem [{\citenamefont {{Serlin}}\ \emph {et~al.}(2020)\citenamefont
  {{Serlin}}, \citenamefont {{Tschirhart}}, \citenamefont {{Polshyn}},
  \citenamefont {{Zhang}}, \citenamefont {{Zhu}}, \citenamefont {{Watanabe}},
  \citenamefont {{Taniguchi}}, \citenamefont {{Balents}},\ and\ \citenamefont
  {{Young}}}]{Serlin2020}%
  \BibitemOpen
  \bibfield  {author} {\bibinfo {author} {\bibfnamefont {M.}~\bibnamefont
  {{Serlin}}}, \bibinfo {author} {\bibfnamefont {C.~L.}\ \bibnamefont
  {{Tschirhart}}}, \bibinfo {author} {\bibfnamefont {H.}~\bibnamefont
  {{Polshyn}}}, \bibinfo {author} {\bibfnamefont {Y.}~\bibnamefont {{Zhang}}},
  \bibinfo {author} {\bibfnamefont {J.}~\bibnamefont {{Zhu}}}, \bibinfo
  {author} {\bibfnamefont {K.}~\bibnamefont {{Watanabe}}}, \bibinfo {author}
  {\bibfnamefont {T.}~\bibnamefont {{Taniguchi}}}, \bibinfo {author}
  {\bibfnamefont {L.}~\bibnamefont {{Balents}}},\ and\ \bibinfo {author}
  {\bibfnamefont {A.~F.}\ \bibnamefont {{Young}}},\ }\bibfield  {title}
  {\bibinfo {title} {{Intrinsic quantized anomalous Hall effect in a moir{\'e}
  heterostructure}},\ }\href {https://doi.org/10.1126/science.aay5533}
  {\bibfield  {journal} {\bibinfo  {journal} {Science}\ }\textbf {\bibinfo
  {volume} {367}},\ \bibinfo {pages} {900} (\bibinfo {year} {2020})},\ \Eprint
  {https://arxiv.org/abs/1907.00261} {arXiv:1907.00261 [cond-mat.str-el]}
  \BibitemShut {NoStop}%
\bibitem [{\citenamefont {{Lu}}\ \emph {et~al.}(2019)\citenamefont {{Lu}},
  \citenamefont {{Stepanov}}, \citenamefont {{Yang}}, \citenamefont {{Xie}},
  \citenamefont {{Aamir}}, \citenamefont {{Das}}, \citenamefont {{Urgell}},
  \citenamefont {{Watanabe}}, \citenamefont {{Taniguchi}}, \citenamefont
  {{Zhang}}, \citenamefont {{Bachtold}}, \citenamefont {{MacDonald}},\ and\
  \citenamefont {{Efetov}}}]{Lu2019}%
  \BibitemOpen
  \bibfield  {author} {\bibinfo {author} {\bibfnamefont {X.}~\bibnamefont
  {{Lu}}}, \bibinfo {author} {\bibfnamefont {P.}~\bibnamefont {{Stepanov}}},
  \bibinfo {author} {\bibfnamefont {W.}~\bibnamefont {{Yang}}}, \bibinfo
  {author} {\bibfnamefont {M.}~\bibnamefont {{Xie}}}, \bibinfo {author}
  {\bibfnamefont {M.~A.}\ \bibnamefont {{Aamir}}}, \bibinfo {author}
  {\bibfnamefont {I.}~\bibnamefont {{Das}}}, \bibinfo {author} {\bibfnamefont
  {C.}~\bibnamefont {{Urgell}}}, \bibinfo {author} {\bibfnamefont
  {K.}~\bibnamefont {{Watanabe}}}, \bibinfo {author} {\bibfnamefont
  {T.}~\bibnamefont {{Taniguchi}}}, \bibinfo {author} {\bibfnamefont
  {G.}~\bibnamefont {{Zhang}}}, \bibinfo {author} {\bibfnamefont
  {A.}~\bibnamefont {{Bachtold}}}, \bibinfo {author} {\bibfnamefont {A.~H.}\
  \bibnamefont {{MacDonald}}},\ and\ \bibinfo {author} {\bibfnamefont {D.~K.}\
  \bibnamefont {{Efetov}}},\ }\bibfield  {title} {\bibinfo {title}
  {{Superconductors, orbital magnets and correlated states in magic-angle
  bilayer graphene}},\ }\href {https://doi.org/10.1038/s41586-019-1695-0}
  {\bibfield  {journal} {\bibinfo  {journal} {Nature}\ }\textbf {\bibinfo
  {volume} {574}},\ \bibinfo {pages} {653} (\bibinfo {year} {2019})},\ \Eprint
  {https://arxiv.org/abs/1903.06513} {arXiv:1903.06513 [cond-mat.str-el]}
  \BibitemShut {NoStop}%
\bibitem [{\citenamefont {{Chen}}\ \emph {et~al.}(2019)\citenamefont {{Chen}},
  \citenamefont {{Jiang}}, \citenamefont {{Wu}}, \citenamefont {{Lyu}},
  \citenamefont {{Li}}, \citenamefont {{Chittari}}, \citenamefont {{Watanabe}},
  \citenamefont {{Taniguchi}}, \citenamefont {{Shi}}, \citenamefont {{Jung}},
  \citenamefont {{Zhang}},\ and\ \citenamefont {{Wang}}}]{Chen2019}%
  \BibitemOpen
  \bibfield  {author} {\bibinfo {author} {\bibfnamefont {G.}~\bibnamefont
  {{Chen}}}, \bibinfo {author} {\bibfnamefont {L.}~\bibnamefont {{Jiang}}},
  \bibinfo {author} {\bibfnamefont {S.}~\bibnamefont {{Wu}}}, \bibinfo {author}
  {\bibfnamefont {B.}~\bibnamefont {{Lyu}}}, \bibinfo {author} {\bibfnamefont
  {H.}~\bibnamefont {{Li}}}, \bibinfo {author} {\bibfnamefont {B.~L.}\
  \bibnamefont {{Chittari}}}, \bibinfo {author} {\bibfnamefont
  {K.}~\bibnamefont {{Watanabe}}}, \bibinfo {author} {\bibfnamefont
  {T.}~\bibnamefont {{Taniguchi}}}, \bibinfo {author} {\bibfnamefont
  {Z.}~\bibnamefont {{Shi}}}, \bibinfo {author} {\bibfnamefont
  {J.}~\bibnamefont {{Jung}}}, \bibinfo {author} {\bibfnamefont
  {Y.}~\bibnamefont {{Zhang}}},\ and\ \bibinfo {author} {\bibfnamefont
  {F.}~\bibnamefont {{Wang}}},\ }\bibfield  {title} {\bibinfo {title}
  {{Evidence of a gate-tunable Mott insulator in a trilayer graphene moir{\'e}
  superlattice}},\ }\href {https://doi.org/10.1038/s41567-018-0387-2}
  {\bibfield  {journal} {\bibinfo  {journal} {Nat. Phys.}\ }\textbf {\bibinfo
  {volume} {15}},\ \bibinfo {pages} {237} (\bibinfo {year} {2019})},\ \Eprint
  {https://arxiv.org/abs/1803.01985} {arXiv:1803.01985 [cond-mat.mes-hall]}
  \BibitemShut {NoStop}%
\bibitem [{\citenamefont {{Jiang}}\ \emph {et~al.}(2019)\citenamefont
  {{Jiang}}, \citenamefont {{Lai}}, \citenamefont {{Watanabe}}, \citenamefont
  {{Taniguchi}}, \citenamefont {{Haule}}, \citenamefont {{Mao}},\ and\
  \citenamefont {{Andrei}}}]{Jiang2019}%
  \BibitemOpen
  \bibfield  {author} {\bibinfo {author} {\bibfnamefont {Y.}~\bibnamefont
  {{Jiang}}}, \bibinfo {author} {\bibfnamefont {X.}~\bibnamefont {{Lai}}},
  \bibinfo {author} {\bibfnamefont {K.}~\bibnamefont {{Watanabe}}}, \bibinfo
  {author} {\bibfnamefont {T.}~\bibnamefont {{Taniguchi}}}, \bibinfo {author}
  {\bibfnamefont {K.}~\bibnamefont {{Haule}}}, \bibinfo {author} {\bibfnamefont
  {J.}~\bibnamefont {{Mao}}},\ and\ \bibinfo {author} {\bibfnamefont {E.~Y.}\
  \bibnamefont {{Andrei}}},\ }\bibfield  {title} {\bibinfo {title} {{Charge
  order and broken rotational symmetry in magic-angle twisted bilayer
  graphene}},\ }\href {https://doi.org/10.1038/s41586-019-1460-4} {\bibfield
  {journal} {\bibinfo  {journal} {Nature}\ }\textbf {\bibinfo {volume} {573}},\
  \bibinfo {pages} {91} (\bibinfo {year} {2019})},\ \Eprint
  {https://arxiv.org/abs/1904.10153} {arXiv:1904.10153 [cond-mat.mes-hall]}
  \BibitemShut {NoStop}%
\bibitem [{\citenamefont {{Choi}}\ \emph {et~al.}(2019)\citenamefont {{Choi}},
  \citenamefont {{Kemmer}}, \citenamefont {{Peng}}, \citenamefont {{Thomson}},
  \citenamefont {{Arora}}, \citenamefont {{Polski}}, \citenamefont {{Zhang}},
  \citenamefont {{Ren}}, \citenamefont {{Alicea}}, \citenamefont {{Refael}},
  \citenamefont {{von Oppen}}, \citenamefont {{Watanabe}}, \citenamefont
  {{Taniguchi}},\ and\ \citenamefont {{Nadj-Perge}}}]{Choi2019}%
  \BibitemOpen
  \bibfield  {author} {\bibinfo {author} {\bibfnamefont {Y.}~\bibnamefont
  {{Choi}}}, \bibinfo {author} {\bibfnamefont {J.}~\bibnamefont {{Kemmer}}},
  \bibinfo {author} {\bibfnamefont {Y.}~\bibnamefont {{Peng}}}, \bibinfo
  {author} {\bibfnamefont {A.}~\bibnamefont {{Thomson}}}, \bibinfo {author}
  {\bibfnamefont {H.}~\bibnamefont {{Arora}}}, \bibinfo {author} {\bibfnamefont
  {R.}~\bibnamefont {{Polski}}}, \bibinfo {author} {\bibfnamefont
  {Y.}~\bibnamefont {{Zhang}}}, \bibinfo {author} {\bibfnamefont
  {H.}~\bibnamefont {{Ren}}}, \bibinfo {author} {\bibfnamefont
  {J.}~\bibnamefont {{Alicea}}}, \bibinfo {author} {\bibfnamefont
  {G.}~\bibnamefont {{Refael}}}, \bibinfo {author} {\bibfnamefont
  {F.}~\bibnamefont {{von Oppen}}}, \bibinfo {author} {\bibfnamefont
  {K.}~\bibnamefont {{Watanabe}}}, \bibinfo {author} {\bibfnamefont
  {T.}~\bibnamefont {{Taniguchi}}},\ and\ \bibinfo {author} {\bibfnamefont
  {S.}~\bibnamefont {{Nadj-Perge}}},\ }\bibfield  {title} {\bibinfo {title}
  {{Electronic correlations in twisted bilayer graphene near the magic
  angle}},\ }\href {https://doi.org/10.1038/s41567-019-0606-5} {\bibfield
  {journal} {\bibinfo  {journal} {Nat. Phys.}\ }\textbf {\bibinfo {volume}
  {15}},\ \bibinfo {pages} {1174} (\bibinfo {year} {2019})},\ \Eprint
  {https://arxiv.org/abs/1901.02997} {arXiv:1901.02997 [cond-mat.mes-hall]}
  \BibitemShut {NoStop}%
\bibitem [{\citenamefont {{Kerelsky}}\ \emph {et~al.}(2019)\citenamefont
  {{Kerelsky}}, \citenamefont {{McGilly}}, \citenamefont {{Kennes}},
  \citenamefont {{Xian}}, \citenamefont {{Yankowitz}}, \citenamefont {{Chen}},
  \citenamefont {{Watanabe}}, \citenamefont {{Taniguchi}}, \citenamefont
  {{Hone}}, \citenamefont {{Dean}}, \citenamefont {{Rubio}},\ and\
  \citenamefont {{Pasupathy}}}]{Kerelsky2019}%
  \BibitemOpen
  \bibfield  {author} {\bibinfo {author} {\bibfnamefont {A.}~\bibnamefont
  {{Kerelsky}}}, \bibinfo {author} {\bibfnamefont {L.~J.}\ \bibnamefont
  {{McGilly}}}, \bibinfo {author} {\bibfnamefont {D.~M.}\ \bibnamefont
  {{Kennes}}}, \bibinfo {author} {\bibfnamefont {L.}~\bibnamefont {{Xian}}},
  \bibinfo {author} {\bibfnamefont {M.}~\bibnamefont {{Yankowitz}}}, \bibinfo
  {author} {\bibfnamefont {S.}~\bibnamefont {{Chen}}}, \bibinfo {author}
  {\bibfnamefont {K.}~\bibnamefont {{Watanabe}}}, \bibinfo {author}
  {\bibfnamefont {T.}~\bibnamefont {{Taniguchi}}}, \bibinfo {author}
  {\bibfnamefont {J.}~\bibnamefont {{Hone}}}, \bibinfo {author} {\bibfnamefont
  {C.}~\bibnamefont {{Dean}}}, \bibinfo {author} {\bibfnamefont
  {A.}~\bibnamefont {{Rubio}}},\ and\ \bibinfo {author} {\bibfnamefont {A.~N.}\
  \bibnamefont {{Pasupathy}}},\ }\bibfield  {title} {\bibinfo {title}
  {{Maximized electron interactions at the magic angle in twisted bilayer
  graphene}},\ }\href {https://doi.org/10.1038/s41586-019-1431-9} {\bibfield
  {journal} {\bibinfo  {journal} {Nature}\ }\textbf {\bibinfo {volume} {572}},\
  \bibinfo {pages} {95} (\bibinfo {year} {2019})}\BibitemShut {NoStop}%
\bibitem [{\citenamefont {{Xie}}\ \emph {et~al.}(2019)\citenamefont {{Xie}},
  \citenamefont {{Lian}}, \citenamefont {{J{\"a}ck}}, \citenamefont {{Liu}},
  \citenamefont {{Chiu}}, \citenamefont {{Watanabe}}, \citenamefont
  {{Taniguchi}}, \citenamefont {{Bernevig}},\ and\ \citenamefont
  {{Yazdani}}}]{Xie2019}%
  \BibitemOpen
  \bibfield  {author} {\bibinfo {author} {\bibfnamefont {Y.}~\bibnamefont
  {{Xie}}}, \bibinfo {author} {\bibfnamefont {B.}~\bibnamefont {{Lian}}},
  \bibinfo {author} {\bibfnamefont {B.}~\bibnamefont {{J{\"a}ck}}}, \bibinfo
  {author} {\bibfnamefont {X.}~\bibnamefont {{Liu}}}, \bibinfo {author}
  {\bibfnamefont {C.-L.}\ \bibnamefont {{Chiu}}}, \bibinfo {author}
  {\bibfnamefont {K.}~\bibnamefont {{Watanabe}}}, \bibinfo {author}
  {\bibfnamefont {T.}~\bibnamefont {{Taniguchi}}}, \bibinfo {author}
  {\bibfnamefont {B.~A.}\ \bibnamefont {{Bernevig}}},\ and\ \bibinfo {author}
  {\bibfnamefont {A.}~\bibnamefont {{Yazdani}}},\ }\bibfield  {title} {\bibinfo
  {title} {{Spectroscopic signatures of many-body correlations in magic-angle
  twisted bilayer graphene}},\ }\href
  {https://doi.org/10.1038/s41586-019-1422-x} {\bibfield  {journal} {\bibinfo
  {journal} {Nature}\ }\textbf {\bibinfo {volume} {572}},\ \bibinfo {pages}
  {101} (\bibinfo {year} {2019})},\ \Eprint {https://arxiv.org/abs/1906.09274}
  {arXiv:1906.09274 [cond-mat.mes-hall]} \BibitemShut {NoStop}%
\bibitem [{\citenamefont {{Chen}}\ \emph {et~al.}(2020)\citenamefont {{Chen}},
  \citenamefont {{Sharpe}}, \citenamefont {{Fox}}, \citenamefont {{Zhang}},
  \citenamefont {{Wang}}, \citenamefont {{Jiang}}, \citenamefont {{Lyu}},
  \citenamefont {{Li}}, \citenamefont {{Watanabe}}, \citenamefont
  {{Taniguchi}}, \citenamefont {{Shi}}, \citenamefont {{Senthil}},
  \citenamefont {{Goldhaber-Gordon}}, \citenamefont {{Zhang}},\ and\
  \citenamefont {{Wang}}}]{Chen2020}%
  \BibitemOpen
  \bibfield  {author} {\bibinfo {author} {\bibfnamefont {G.}~\bibnamefont
  {{Chen}}}, \bibinfo {author} {\bibfnamefont {A.~L.}\ \bibnamefont
  {{Sharpe}}}, \bibinfo {author} {\bibfnamefont {E.~J.}\ \bibnamefont {{Fox}}},
  \bibinfo {author} {\bibfnamefont {Y.-H.}\ \bibnamefont {{Zhang}}}, \bibinfo
  {author} {\bibfnamefont {S.}~\bibnamefont {{Wang}}}, \bibinfo {author}
  {\bibfnamefont {L.}~\bibnamefont {{Jiang}}}, \bibinfo {author} {\bibfnamefont
  {B.}~\bibnamefont {{Lyu}}}, \bibinfo {author} {\bibfnamefont
  {H.}~\bibnamefont {{Li}}}, \bibinfo {author} {\bibfnamefont {K.}~\bibnamefont
  {{Watanabe}}}, \bibinfo {author} {\bibfnamefont {T.}~\bibnamefont
  {{Taniguchi}}}, \bibinfo {author} {\bibfnamefont {Z.}~\bibnamefont {{Shi}}},
  \bibinfo {author} {\bibfnamefont {T.}~\bibnamefont {{Senthil}}}, \bibinfo
  {author} {\bibfnamefont {D.}~\bibnamefont {{Goldhaber-Gordon}}}, \bibinfo
  {author} {\bibfnamefont {Y.}~\bibnamefont {{Zhang}}},\ and\ \bibinfo {author}
  {\bibfnamefont {F.}~\bibnamefont {{Wang}}},\ }\bibfield  {title} {\bibinfo
  {title} {{Tunable correlated Chern insulator and ferromagnetism in a
  moir{\'e} superlattice}},\ }\href {https://doi.org/10.1038/s41586-020-2049-7}
  {\bibfield  {journal} {\bibinfo  {journal} {Nature}\ }\textbf {\bibinfo
  {volume} {579}},\ \bibinfo {pages} {56} (\bibinfo {year} {2020})},\ \Eprint
  {https://arxiv.org/abs/1905.06535} {arXiv:1905.06535 [cond-mat.mes-hall]}
  \BibitemShut {NoStop}%
\bibitem [{\citenamefont {{Zondiner}}\ \emph {et~al.}(2020)\citenamefont
  {{Zondiner}}, \citenamefont {{Rozen}}, \citenamefont {{Rodan-Legrain}},
  \citenamefont {{Cao}}, \citenamefont {{Queiroz}}, \citenamefont
  {{Taniguchi}}, \citenamefont {{Watanabe}}, \citenamefont {{Oreg}},
  \citenamefont {{von Oppen}}, \citenamefont {{Stern}}, \citenamefont {{Berg}},
  \citenamefont {{Jarillo-Herrero}},\ and\ \citenamefont
  {{Ilani}}}]{Zondiner2020}%
  \BibitemOpen
  \bibfield  {author} {\bibinfo {author} {\bibfnamefont {U.}~\bibnamefont
  {{Zondiner}}}, \bibinfo {author} {\bibfnamefont {A.}~\bibnamefont {{Rozen}}},
  \bibinfo {author} {\bibfnamefont {D.}~\bibnamefont {{Rodan-Legrain}}},
  \bibinfo {author} {\bibfnamefont {Y.}~\bibnamefont {{Cao}}}, \bibinfo
  {author} {\bibfnamefont {R.}~\bibnamefont {{Queiroz}}}, \bibinfo {author}
  {\bibfnamefont {T.}~\bibnamefont {{Taniguchi}}}, \bibinfo {author}
  {\bibfnamefont {K.}~\bibnamefont {{Watanabe}}}, \bibinfo {author}
  {\bibfnamefont {Y.}~\bibnamefont {{Oreg}}}, \bibinfo {author} {\bibfnamefont
  {F.}~\bibnamefont {{von Oppen}}}, \bibinfo {author} {\bibfnamefont
  {A.}~\bibnamefont {{Stern}}}, \bibinfo {author} {\bibfnamefont
  {E.}~\bibnamefont {{Berg}}}, \bibinfo {author} {\bibfnamefont
  {P.}~\bibnamefont {{Jarillo-Herrero}}},\ and\ \bibinfo {author}
  {\bibfnamefont {S.}~\bibnamefont {{Ilani}}},\ }\bibfield  {title} {\bibinfo
  {title} {{Cascade of phase transitions and Dirac revivals in magic-angle
  graphene}},\ }\href {https://doi.org/10.1038/s41586-020-2373-y} {\bibfield
  {journal} {\bibinfo  {journal} {Nature}\ }\textbf {\bibinfo {volume} {582}},\
  \bibinfo {pages} {203} (\bibinfo {year} {2020})},\ \Eprint
  {https://arxiv.org/abs/1912.06150} {arXiv:1912.06150 [cond-mat.mes-hall]}
  \BibitemShut {NoStop}%
\bibitem [{\citenamefont {{Wong}}\ \emph {et~al.}(2020)\citenamefont {{Wong}},
  \citenamefont {{Nuckolls}}, \citenamefont {{Oh}}, \citenamefont {{Lian}},
  \citenamefont {{Xie}}, \citenamefont {{Jeon}}, \citenamefont {{Watanabe}},
  \citenamefont {{Taniguchi}}, \citenamefont {{Bernevig}},\ and\ \citenamefont
  {{Yazdani}}}]{Wong2020}%
  \BibitemOpen
  \bibfield  {author} {\bibinfo {author} {\bibfnamefont {D.}~\bibnamefont
  {{Wong}}}, \bibinfo {author} {\bibfnamefont {K.~P.}\ \bibnamefont
  {{Nuckolls}}}, \bibinfo {author} {\bibfnamefont {M.}~\bibnamefont {{Oh}}},
  \bibinfo {author} {\bibfnamefont {B.}~\bibnamefont {{Lian}}}, \bibinfo
  {author} {\bibfnamefont {Y.}~\bibnamefont {{Xie}}}, \bibinfo {author}
  {\bibfnamefont {S.}~\bibnamefont {{Jeon}}}, \bibinfo {author} {\bibfnamefont
  {K.}~\bibnamefont {{Watanabe}}}, \bibinfo {author} {\bibfnamefont
  {T.}~\bibnamefont {{Taniguchi}}}, \bibinfo {author} {\bibfnamefont {B.~A.}\
  \bibnamefont {{Bernevig}}},\ and\ \bibinfo {author} {\bibfnamefont
  {A.}~\bibnamefont {{Yazdani}}},\ }\bibfield  {title} {\bibinfo {title}
  {{Cascade of electronic transitions in magic-angle twisted bilayer
  graphene}},\ }\href {https://doi.org/10.1038/s41586-020-2339-0} {\bibfield
  {journal} {\bibinfo  {journal} {Nature}\ }\textbf {\bibinfo {volume} {582}},\
  \bibinfo {pages} {198} (\bibinfo {year} {2020})},\ \Eprint
  {https://arxiv.org/abs/1912.06145} {arXiv:1912.06145 [cond-mat.mes-hall]}
  \BibitemShut {NoStop}%
\bibitem [{\citenamefont {{Yankowitz}}\ \emph {et~al.}(2019)\citenamefont
  {{Yankowitz}}, \citenamefont {{Chen}}, \citenamefont {{Polshyn}},
  \citenamefont {{Zhang}}, \citenamefont {{Watanabe}}, \citenamefont
  {{Taniguchi}}, \citenamefont {{Graf}}, \citenamefont {{Young}},\ and\
  \citenamefont {{Dean}}}]{Yankowitz2019}%
  \BibitemOpen
  \bibfield  {author} {\bibinfo {author} {\bibfnamefont {M.}~\bibnamefont
  {{Yankowitz}}}, \bibinfo {author} {\bibfnamefont {S.}~\bibnamefont {{Chen}}},
  \bibinfo {author} {\bibfnamefont {H.}~\bibnamefont {{Polshyn}}}, \bibinfo
  {author} {\bibfnamefont {Y.}~\bibnamefont {{Zhang}}}, \bibinfo {author}
  {\bibfnamefont {K.}~\bibnamefont {{Watanabe}}}, \bibinfo {author}
  {\bibfnamefont {T.}~\bibnamefont {{Taniguchi}}}, \bibinfo {author}
  {\bibfnamefont {D.}~\bibnamefont {{Graf}}}, \bibinfo {author} {\bibfnamefont
  {A.~F.}\ \bibnamefont {{Young}}},\ and\ \bibinfo {author} {\bibfnamefont
  {C.~R.}\ \bibnamefont {{Dean}}},\ }\bibfield  {title} {\bibinfo {title}
  {{Tuning superconductivity in twisted bilayer graphene}},\ }\href
  {https://doi.org/10.1126/science.aav1910} {\bibfield  {journal} {\bibinfo
  {journal} {Science}\ }\textbf {\bibinfo {volume} {363}},\ \bibinfo {pages}
  {1059} (\bibinfo {year} {2019})},\ \Eprint {https://arxiv.org/abs/1808.07865}
  {arXiv:1808.07865 [cond-mat.mes-hall]} \BibitemShut {NoStop}%
\bibitem [{\citenamefont {{Tomarken}}\ \emph {et~al.}(2019)\citenamefont
  {{Tomarken}}, \citenamefont {{Cao}}, \citenamefont {{Demir}}, \citenamefont
  {{Watanabe}}, \citenamefont {{Taniguchi}}, \citenamefont
  {{Jarillo-Herrero}},\ and\ \citenamefont {{Ashoori}}}]{Tomarken2019}%
  \BibitemOpen
  \bibfield  {author} {\bibinfo {author} {\bibfnamefont {S.~L.}\ \bibnamefont
  {{Tomarken}}}, \bibinfo {author} {\bibfnamefont {Y.}~\bibnamefont {{Cao}}},
  \bibinfo {author} {\bibfnamefont {A.}~\bibnamefont {{Demir}}}, \bibinfo
  {author} {\bibfnamefont {K.}~\bibnamefont {{Watanabe}}}, \bibinfo {author}
  {\bibfnamefont {T.}~\bibnamefont {{Taniguchi}}}, \bibinfo {author}
  {\bibfnamefont {P.}~\bibnamefont {{Jarillo-Herrero}}},\ and\ \bibinfo
  {author} {\bibfnamefont {R.~C.}\ \bibnamefont {{Ashoori}}},\ }\bibfield
  {title} {\bibinfo {title} {{Electronic Compressibility of Magic-Angle
  Graphene Superlattices}},\ }\href
  {https://doi.org/10.1103/PhysRevLett.123.046601} {\bibfield  {journal}
  {\bibinfo  {journal} {Phys. Rev. Lett.}\ }\textbf {\bibinfo {volume} {123}},\
  \bibinfo {eid} {046601} (\bibinfo {year} {2019})},\ \Eprint
  {https://arxiv.org/abs/1903.10492} {arXiv:1903.10492 [cond-mat.mes-hall]}
  \BibitemShut {NoStop}%
\bibitem [{\citenamefont {{Yan}}\ and\ \citenamefont
  {{Felser}}(2017)}]{Yan2017}%
  \BibitemOpen
  \bibfield  {author} {\bibinfo {author} {\bibfnamefont {B.}~\bibnamefont
  {{Yan}}}\ and\ \bibinfo {author} {\bibfnamefont {C.}~\bibnamefont
  {{Felser}}},\ }\bibfield  {title} {\bibinfo {title} {{Topological Materials:
  Weyl Semimetals}},\ }\href
  {https://doi.org/10.1146/annurev-conmatphys-031016-025458} {\bibfield
  {journal} {\bibinfo  {journal} {Annu. Rev. Condens. Matter Phys.}\ }\textbf
  {\bibinfo {volume} {8}},\ \bibinfo {pages} {337} (\bibinfo {year} {2017})},\
  \Eprint {https://arxiv.org/abs/1611.04182} {arXiv:1611.04182
  [cond-mat.mtrl-sci]} \BibitemShut {NoStop}%
\bibitem [{\citenamefont {{Armitage}}\ \emph {et~al.}(2018)\citenamefont
  {{Armitage}}, \citenamefont {{Mele}},\ and\ \citenamefont
  {{Vishwanath}}}]{Armitage2018}%
  \BibitemOpen
  \bibfield  {author} {\bibinfo {author} {\bibfnamefont {N.~P.}\ \bibnamefont
  {{Armitage}}}, \bibinfo {author} {\bibfnamefont {E.~J.}\ \bibnamefont
  {{Mele}}},\ and\ \bibinfo {author} {\bibfnamefont {A.}~\bibnamefont
  {{Vishwanath}}},\ }\bibfield  {title} {\bibinfo {title} {{Weyl and Dirac
  semimetals in three-dimensional solids}},\ }\href
  {https://doi.org/10.1103/RevModPhys.90.015001} {\bibfield  {journal}
  {\bibinfo  {journal} {Rev. Mod. Phys.}\ }\textbf {\bibinfo {volume} {90}},\
  \bibinfo {pages} {015001} (\bibinfo {year} {2018})},\ \Eprint
  {https://arxiv.org/abs/1705.01111} {arXiv:1705.01111 [cond-mat.str-el]}
  \BibitemShut {NoStop}%
\bibitem [{\citenamefont {{Wu}}\ \emph {et~al.}(2017)\citenamefont {{Wu}},
  \citenamefont {{Patankar}}, \citenamefont {{Morimoto}}, \citenamefont
  {{Nair}}, \citenamefont {{Thewalt}}, \citenamefont {{Little}}, \citenamefont
  {{Analytis}}, \citenamefont {{Moore}},\ and\ \citenamefont
  {{Orenstein}}}]{Wu2017}%
  \BibitemOpen
  \bibfield  {author} {\bibinfo {author} {\bibfnamefont {L.}~\bibnamefont
  {{Wu}}}, \bibinfo {author} {\bibfnamefont {S.}~\bibnamefont {{Patankar}}},
  \bibinfo {author} {\bibfnamefont {T.}~\bibnamefont {{Morimoto}}}, \bibinfo
  {author} {\bibfnamefont {N.~L.}\ \bibnamefont {{Nair}}}, \bibinfo {author}
  {\bibfnamefont {E.}~\bibnamefont {{Thewalt}}}, \bibinfo {author}
  {\bibfnamefont {A.}~\bibnamefont {{Little}}}, \bibinfo {author}
  {\bibfnamefont {J.~G.}\ \bibnamefont {{Analytis}}}, \bibinfo {author}
  {\bibfnamefont {J.~E.}\ \bibnamefont {{Moore}}},\ and\ \bibinfo {author}
  {\bibfnamefont {J.}~\bibnamefont {{Orenstein}}},\ }\bibfield  {title}
  {\bibinfo {title} {{Giant anisotropic nonlinear optical response in
  transition metal monopnictide Weyl semimetals}},\ }\href
  {https://doi.org/10.1038/nphys3969} {\bibfield  {journal} {\bibinfo
  {journal} {Nat. Phys.}\ }\textbf {\bibinfo {volume} {13}},\ \bibinfo {pages}
  {350} (\bibinfo {year} {2017})},\ \Eprint {https://arxiv.org/abs/1609.04894}
  {arXiv:1609.04894 [cond-mat.mtrl-sci]} \BibitemShut {NoStop}%
\bibitem [{\citenamefont {{Ma}}\ \emph {et~al.}(2017)\citenamefont {{Ma}},
  \citenamefont {{Xu}}, \citenamefont {{Chan}}, \citenamefont {{Zhang}},
  \citenamefont {{Chang}}, \citenamefont {{Lin}}, \citenamefont {{Xie}},
  \citenamefont {{Palacios}}, \citenamefont {{Lin}}, \citenamefont {{Jia}},
  \citenamefont {{Lee}}, \citenamefont {{Jarillo-Herrero}},\ and\ \citenamefont
  {{Gedik}}}]{Ma2017}%
  \BibitemOpen
  \bibfield  {author} {\bibinfo {author} {\bibfnamefont {Q.}~\bibnamefont
  {{Ma}}}, \bibinfo {author} {\bibfnamefont {S.-Y.}\ \bibnamefont {{Xu}}},
  \bibinfo {author} {\bibfnamefont {C.-K.}\ \bibnamefont {{Chan}}}, \bibinfo
  {author} {\bibfnamefont {C.-L.}\ \bibnamefont {{Zhang}}}, \bibinfo {author}
  {\bibfnamefont {G.}~\bibnamefont {{Chang}}}, \bibinfo {author} {\bibfnamefont
  {Y.}~\bibnamefont {{Lin}}}, \bibinfo {author} {\bibfnamefont
  {W.}~\bibnamefont {{Xie}}}, \bibinfo {author} {\bibfnamefont
  {T.}~\bibnamefont {{Palacios}}}, \bibinfo {author} {\bibfnamefont
  {H.}~\bibnamefont {{Lin}}}, \bibinfo {author} {\bibfnamefont
  {S.}~\bibnamefont {{Jia}}}, \bibinfo {author} {\bibfnamefont {P.~A.}\
  \bibnamefont {{Lee}}}, \bibinfo {author} {\bibfnamefont {P.}~\bibnamefont
  {{Jarillo-Herrero}}},\ and\ \bibinfo {author} {\bibfnamefont
  {N.}~\bibnamefont {{Gedik}}},\ }\bibfield  {title} {\bibinfo {title} {{Direct
  optical detection of Weyl fermion chirality in a topological semimetal}},\
  }\href {https://doi.org/10.1038/nphys4146} {\bibfield  {journal} {\bibinfo
  {journal} {Nat. Phys.}\ }\textbf {\bibinfo {volume} {13}},\ \bibinfo {pages}
  {842} (\bibinfo {year} {2017})},\ \Eprint {https://arxiv.org/abs/1705.00590}
  {arXiv:1705.00590 [cond-mat.mtrl-sci]} \BibitemShut {NoStop}%
\bibitem [{\citenamefont {{Osterhoudt}}\ \emph {et~al.}(2019)\citenamefont
  {{Osterhoudt}}, \citenamefont {{Diebel}}, \citenamefont {{Gray}},
  \citenamefont {{Yang}}, \citenamefont {{Stanco}}, \citenamefont {{Huang}},
  \citenamefont {{Shen}}, \citenamefont {{Ni}}, \citenamefont {{Moll}},
  \citenamefont {{Ran}},\ and\ \citenamefont {{Burch}}}]{Osterhoudt2019}%
  \BibitemOpen
  \bibfield  {author} {\bibinfo {author} {\bibfnamefont {G.~B.}\ \bibnamefont
  {{Osterhoudt}}}, \bibinfo {author} {\bibfnamefont {L.~K.}\ \bibnamefont
  {{Diebel}}}, \bibinfo {author} {\bibfnamefont {M.~J.}\ \bibnamefont
  {{Gray}}}, \bibinfo {author} {\bibfnamefont {X.}~\bibnamefont {{Yang}}},
  \bibinfo {author} {\bibfnamefont {J.}~\bibnamefont {{Stanco}}}, \bibinfo
  {author} {\bibfnamefont {X.}~\bibnamefont {{Huang}}}, \bibinfo {author}
  {\bibfnamefont {B.}~\bibnamefont {{Shen}}}, \bibinfo {author} {\bibfnamefont
  {N.}~\bibnamefont {{Ni}}}, \bibinfo {author} {\bibfnamefont {P.~J.~W.}\
  \bibnamefont {{Moll}}}, \bibinfo {author} {\bibfnamefont {Y.}~\bibnamefont
  {{Ran}}},\ and\ \bibinfo {author} {\bibfnamefont {K.~S.}\ \bibnamefont
  {{Burch}}},\ }\bibfield  {title} {\bibinfo {title} {{Colossal mid-infrared
  bulk photovoltaic effect in a type-I Weyl semimetal}},\ }\href
  {https://doi.org/10.1038/s41563-019-0297-4} {\bibfield  {journal} {\bibinfo
  {journal} {Nat. Mater.}\ }\textbf {\bibinfo {volume} {18}},\ \bibinfo {pages}
  {471} (\bibinfo {year} {2019})},\ \Eprint {https://arxiv.org/abs/1712.04951}
  {arXiv:1712.04951} \BibitemShut {NoStop}%
\bibitem [{\citenamefont {{Morimoto}}\ and\ \citenamefont
  {{Nagaosa}}(2016)}]{Morimoto2016}%
  \BibitemOpen
  \bibfield  {author} {\bibinfo {author} {\bibfnamefont {T.}~\bibnamefont
  {{Morimoto}}}\ and\ \bibinfo {author} {\bibfnamefont {N.}~\bibnamefont
  {{Nagaosa}}},\ }\bibfield  {title} {\bibinfo {title} {{Topological nature of
  nonlinear optical effects in solids}},\ }\href
  {https://doi.org/10.1126/sciadv.1501524} {\bibfield  {journal} {\bibinfo
  {journal} {Sci. Adv.}\ }\textbf {\bibinfo {volume} {2}},\ \bibinfo {pages}
  {e1501524} (\bibinfo {year} {2016})},\ \Eprint
  {https://arxiv.org/abs/1510.08112} {arXiv:1510.08112 [cond-mat.mes-hall]}
  \BibitemShut {NoStop}%
\bibitem [{\citenamefont {{Taguchi}}\ \emph {et~al.}(2016)\citenamefont
  {{Taguchi}}, \citenamefont {{Imaeda}}, \citenamefont {{Sato}},\ and\
  \citenamefont {{Tanaka}}}]{Taguchi2016}%
  \BibitemOpen
  \bibfield  {author} {\bibinfo {author} {\bibfnamefont {K.}~\bibnamefont
  {{Taguchi}}}, \bibinfo {author} {\bibfnamefont {T.}~\bibnamefont {{Imaeda}}},
  \bibinfo {author} {\bibfnamefont {M.}~\bibnamefont {{Sato}}},\ and\ \bibinfo
  {author} {\bibfnamefont {Y.}~\bibnamefont {{Tanaka}}},\ }\bibfield  {title}
  {\bibinfo {title} {{Photovoltaic chiral magnetic effect in Weyl
  semimetals}},\ }\href {https://doi.org/10.1103/PhysRevB.93.201202} {\bibfield
   {journal} {\bibinfo  {journal} {Phys. Rev. B}\ }\textbf {\bibinfo {volume}
  {93}},\ \bibinfo {eid} {201202} (\bibinfo {year} {2016})}\BibitemShut
  {NoStop}%
\bibitem [{\citenamefont {{Chan}}\ \emph {et~al.}(2017)\citenamefont {{Chan}},
  \citenamefont {{Lindner}}, \citenamefont {{Refael}},\ and\ \citenamefont
  {{Lee}}}]{Chan2017}%
  \BibitemOpen
  \bibfield  {author} {\bibinfo {author} {\bibfnamefont {C.-K.}\ \bibnamefont
  {{Chan}}}, \bibinfo {author} {\bibfnamefont {N.~H.}\ \bibnamefont
  {{Lindner}}}, \bibinfo {author} {\bibfnamefont {G.}~\bibnamefont
  {{Refael}}},\ and\ \bibinfo {author} {\bibfnamefont {P.~A.}\ \bibnamefont
  {{Lee}}},\ }\bibfield  {title} {\bibinfo {title} {{Photocurrents in Weyl
  semimetals}},\ }\href {https://doi.org/10.1103/PhysRevB.95.041104} {\bibfield
   {journal} {\bibinfo  {journal} {Phys. Rev. B}\ }\textbf {\bibinfo {volume}
  {95}},\ \bibinfo {eid} {041104} (\bibinfo {year} {2017})},\ \Eprint
  {https://arxiv.org/abs/1607.07839} {arXiv:1607.07839 [cond-mat.mes-hall]}
  \BibitemShut {NoStop}%
\bibitem [{\citenamefont {{de Juan}}\ \emph {et~al.}(2017)\citenamefont {{de
  Juan}}, \citenamefont {{Grushin}}, \citenamefont {{Morimoto}},\ and\
  \citenamefont {{Moore}}}]{deJuan2017}%
  \BibitemOpen
  \bibfield  {author} {\bibinfo {author} {\bibfnamefont {F.}~\bibnamefont {{de
  Juan}}}, \bibinfo {author} {\bibfnamefont {A.~G.}\ \bibnamefont {{Grushin}}},
  \bibinfo {author} {\bibfnamefont {T.}~\bibnamefont {{Morimoto}}},\ and\
  \bibinfo {author} {\bibfnamefont {J.~E.}\ \bibnamefont {{Moore}}},\
  }\bibfield  {title} {\bibinfo {title} {{Quantized circular photogalvanic
  effect in Weyl semimetals}},\ }\href {https://doi.org/10.1038/ncomms15995}
  {\bibfield  {journal} {\bibinfo  {journal} {Nat. Commun.}\ }\textbf {\bibinfo
  {volume} {8}},\ \bibinfo {eid} {15995} (\bibinfo {year} {2017})},\ \Eprint
  {https://arxiv.org/abs/1611.05887} {arXiv:1611.05887 [cond-mat.str-el]}
  \BibitemShut {NoStop}%
\bibitem [{\citenamefont {{Zhang}}\ \emph {et~al.}(2018)\citenamefont
  {{Zhang}}, \citenamefont {{Ishizuka}}, \citenamefont {{van den Brink}},
  \citenamefont {{Felser}}, \citenamefont {{Yan}},\ and\ \citenamefont
  {{Nagaosa}}}]{Zhang2018}%
  \BibitemOpen
  \bibfield  {author} {\bibinfo {author} {\bibfnamefont {Y.}~\bibnamefont
  {{Zhang}}}, \bibinfo {author} {\bibfnamefont {H.}~\bibnamefont {{Ishizuka}}},
  \bibinfo {author} {\bibfnamefont {J.}~\bibnamefont {{van den Brink}}},
  \bibinfo {author} {\bibfnamefont {C.}~\bibnamefont {{Felser}}}, \bibinfo
  {author} {\bibfnamefont {B.}~\bibnamefont {{Yan}}},\ and\ \bibinfo {author}
  {\bibfnamefont {N.}~\bibnamefont {{Nagaosa}}},\ }\bibfield  {title} {\bibinfo
  {title} {{Photogalvanic effect in Weyl semimetals from first principles}},\
  }\href {https://doi.org/10.1103/PhysRevB.97.241118} {\bibfield  {journal}
  {\bibinfo  {journal} {Phys. Rev. B}\ }\textbf {\bibinfo {volume} {97}},\
  \bibinfo {eid} {241118} (\bibinfo {year} {2018})},\ \Eprint
  {https://arxiv.org/abs/1803.00562} {arXiv:1803.00562 [cond-mat.str-el]}
  \BibitemShut {NoStop}%
\bibitem [{\citenamefont {{Holder}}\ \emph {et~al.}(2020)\citenamefont
  {{Holder}}, \citenamefont {{Kaplan}},\ and\ \citenamefont
  {{Yan}}}]{Holder2020}%
  \BibitemOpen
  \bibfield  {author} {\bibinfo {author} {\bibfnamefont {T.}~\bibnamefont
  {{Holder}}}, \bibinfo {author} {\bibfnamefont {D.}~\bibnamefont {{Kaplan}}},\
  and\ \bibinfo {author} {\bibfnamefont {B.}~\bibnamefont {{Yan}}},\ }\bibfield
   {title} {\bibinfo {title} {{Consequences of time-reversal-symmetry breaking
  in the light-matter interaction: Berry curvature, quantum metric, and
  diabatic motion}},\ }\href {https://doi.org/10.1103/PhysRevResearch.2.033100}
  {\bibfield  {journal} {\bibinfo  {journal} {Phys. Rev. Research}\ }\textbf
  {\bibinfo {volume} {2}},\ \bibinfo {eid} {033100} (\bibinfo {year} {2020})},\
  \Eprint {https://arxiv.org/abs/1911.05667} {arXiv:1911.05667
  [cond-mat.mes-hall]} \BibitemShut {NoStop}%
\bibitem [{\citenamefont {{Cao}}\ \emph
  {et~al.}(2018{\natexlab{a}})\citenamefont {{Cao}}, \citenamefont {{Fatemi}},
  \citenamefont {{Fang}}, \citenamefont {{Watanabe}}, \citenamefont
  {{Taniguchi}}, \citenamefont {{Kaxiras}},\ and\ \citenamefont
  {{Jarillo-Herrero}}}]{Cao2018}%
  \BibitemOpen
  \bibfield  {author} {\bibinfo {author} {\bibfnamefont {Y.}~\bibnamefont
  {{Cao}}}, \bibinfo {author} {\bibfnamefont {V.}~\bibnamefont {{Fatemi}}},
  \bibinfo {author} {\bibfnamefont {S.}~\bibnamefont {{Fang}}}, \bibinfo
  {author} {\bibfnamefont {K.}~\bibnamefont {{Watanabe}}}, \bibinfo {author}
  {\bibfnamefont {T.}~\bibnamefont {{Taniguchi}}}, \bibinfo {author}
  {\bibfnamefont {E.}~\bibnamefont {{Kaxiras}}},\ and\ \bibinfo {author}
  {\bibfnamefont {P.}~\bibnamefont {{Jarillo-Herrero}}},\ }\bibfield  {title}
  {\bibinfo {title} {{Unconventional superconductivity in magic-angle graphene
  superlattices}},\ }\href {https://doi.org/10.1038/nature26160} {\bibfield
  {journal} {\bibinfo  {journal} {Nature}\ }\textbf {\bibinfo {volume} {556}},\
  \bibinfo {pages} {43} (\bibinfo {year} {2018}{\natexlab{a}})},\ \Eprint
  {https://arxiv.org/abs/1803.02342} {arXiv:1803.02342 [cond-mat.mes-hall]}
  \BibitemShut {NoStop}%
\bibitem [{\citenamefont {{Cao}}\ \emph
  {et~al.}(2018{\natexlab{b}})\citenamefont {{Cao}}, \citenamefont {{Fatemi}},
  \citenamefont {{Demir}}, \citenamefont {{Fang}}, \citenamefont {{Tomarken}},
  \citenamefont {{Luo}}, \citenamefont {{Sanchez-Yamagishi}}, \citenamefont
  {{Watanabe}}, \citenamefont {{Taniguchi}}, \citenamefont {{Kaxiras}},
  \citenamefont {{Ashoori}},\ and\ \citenamefont
  {{Jarillo-Herrero}}}]{Cao2018a}%
  \BibitemOpen
  \bibfield  {author} {\bibinfo {author} {\bibfnamefont {Y.}~\bibnamefont
  {{Cao}}}, \bibinfo {author} {\bibfnamefont {V.}~\bibnamefont {{Fatemi}}},
  \bibinfo {author} {\bibfnamefont {A.}~\bibnamefont {{Demir}}}, \bibinfo
  {author} {\bibfnamefont {S.}~\bibnamefont {{Fang}}}, \bibinfo {author}
  {\bibfnamefont {S.~L.}\ \bibnamefont {{Tomarken}}}, \bibinfo {author}
  {\bibfnamefont {J.~Y.}\ \bibnamefont {{Luo}}}, \bibinfo {author}
  {\bibfnamefont {J.~D.}\ \bibnamefont {{Sanchez-Yamagishi}}}, \bibinfo
  {author} {\bibfnamefont {K.}~\bibnamefont {{Watanabe}}}, \bibinfo {author}
  {\bibfnamefont {T.}~\bibnamefont {{Taniguchi}}}, \bibinfo {author}
  {\bibfnamefont {E.}~\bibnamefont {{Kaxiras}}}, \bibinfo {author}
  {\bibfnamefont {R.~C.}\ \bibnamefont {{Ashoori}}},\ and\ \bibinfo {author}
  {\bibfnamefont {P.}~\bibnamefont {{Jarillo-Herrero}}},\ }\bibfield  {title}
  {\bibinfo {title} {{Correlated insulator behaviour at half-filling in
  magic-angle graphene superlattices}},\ }\href
  {https://doi.org/10.1038/nature26154} {\bibfield  {journal} {\bibinfo
  {journal} {Nature}\ }\textbf {\bibinfo {volume} {556}},\ \bibinfo {pages}
  {80} (\bibinfo {year} {2018}{\natexlab{b}})},\ \Eprint
  {https://arxiv.org/abs/1802.00553} {arXiv:1802.00553 [cond-mat.mes-hall]}
  \BibitemShut {NoStop}%
\bibitem [{\citenamefont {{Santos}}\ \emph {et~al.}(2007)\citenamefont
  {{Santos}}, \citenamefont {{Peres}},\ and\ \citenamefont {{Castro
  Neto}}}]{Santos2007}%
  \BibitemOpen
  \bibfield  {author} {\bibinfo {author} {\bibfnamefont {J.~M.~B.~L.}\
  \bibnamefont {{Santos}}}, \bibinfo {author} {\bibfnamefont {N.~M.~R.}\
  \bibnamefont {{Peres}}},\ and\ \bibinfo {author} {\bibfnamefont {A.~H.}\
  \bibnamefont {{Castro Neto}}},\ }\bibfield  {title} {\bibinfo {title}
  {{Graphene Bilayer with a Twist: Electronic Structure}},\ }\href
  {https://doi.org/10.1103/PhysRevLett.99.256802} {\bibfield  {journal}
  {\bibinfo  {journal} {\prl}\ }\textbf {\bibinfo {volume} {99}},\ \bibinfo
  {eid} {256802} (\bibinfo {year} {2007})},\ \Eprint
  {https://arxiv.org/abs/0704.2128} {arXiv:0704.2128 [cond-mat.mtrl-sci]}
  \BibitemShut {NoStop}%
\bibitem [{\citenamefont {{Santos}}\ \emph {et~al.}(2012)\citenamefont
  {{Santos}}, \citenamefont {{Peres}},\ and\ \citenamefont {{Castro
  Neto}}}]{Santos2012}%
  \BibitemOpen
  \bibfield  {author} {\bibinfo {author} {\bibfnamefont {J.~M.~B.~L.}\
  \bibnamefont {{Santos}}}, \bibinfo {author} {\bibfnamefont {N.~M.~R.}\
  \bibnamefont {{Peres}}},\ and\ \bibinfo {author} {\bibfnamefont {A.~H.}\
  \bibnamefont {{Castro Neto}}},\ }\bibfield  {title} {\bibinfo {title}
  {{Continuum model of the twisted graphene bilayer}},\ }\href
  {https://doi.org/10.1103/PhysRevB.86.155449} {\bibfield  {journal} {\bibinfo
  {journal} {\prb}\ }\textbf {\bibinfo {volume} {86}},\ \bibinfo {eid} {155449}
  (\bibinfo {year} {2012})},\ \Eprint {https://arxiv.org/abs/1202.1088}
  {arXiv:1202.1088 [cond-mat.mtrl-sci]} \BibitemShut {NoStop}%
\bibitem [{\citenamefont {{Bistritzer}}\ and\ \citenamefont
  {{MacDonald}}(2011)}]{Bistritzer2011}%
  \BibitemOpen
  \bibfield  {author} {\bibinfo {author} {\bibfnamefont {R.}~\bibnamefont
  {{Bistritzer}}}\ and\ \bibinfo {author} {\bibfnamefont {A.~H.}\ \bibnamefont
  {{MacDonald}}},\ }\bibfield  {title} {\bibinfo {title} {{Moir{\'e} bands in
  twisted double-layer graphene}},\ }\href
  {https://doi.org/10.1073/pnas.1108174108} {\bibfield  {journal} {\bibinfo
  {journal} {PNAS}\ }\textbf {\bibinfo {volume} {108}},\ \bibinfo {pages}
  {12233} (\bibinfo {year} {2011})},\ \Eprint {https://arxiv.org/abs/1009.4203}
  {arXiv:1009.4203 [cond-mat.mes-hall]} \BibitemShut {NoStop}%
\bibitem [{\citenamefont {{Liu}}\ and\ \citenamefont {{Dai}}(2020)}]{Liu2020a}%
  \BibitemOpen
  \bibfield  {author} {\bibinfo {author} {\bibfnamefont {J.}~\bibnamefont
  {{Liu}}}\ and\ \bibinfo {author} {\bibfnamefont {X.}~\bibnamefont {{Dai}}},\
  }\bibfield  {title} {\bibinfo {title} {{Anomalous Hall effect,
  magneto-optical properties, and nonlinear optical properties of twisted
  graphene systems}},\ }\href {https://doi.org/10.1038/s41524-020-0299-4}
  {\bibfield  {journal} {\bibinfo  {journal} {npj Comput. Mater.}\ }\textbf
  {\bibinfo {volume} {6}},\ \bibinfo {eid} {57} (\bibinfo {year} {2020})},\
  \Eprint {https://arxiv.org/abs/1907.08932} {arXiv:1907.08932
  [cond-mat.mes-hall]} \BibitemShut {NoStop}%
\bibitem [{\citenamefont {{He}}\ \emph {et~al.}(2020)\citenamefont {{He}},
  \citenamefont {{Goldhaber-Gordon}},\ and\ \citenamefont {{Law}}}]{He2020}%
  \BibitemOpen
  \bibfield  {author} {\bibinfo {author} {\bibfnamefont {W.-Y.}\ \bibnamefont
  {{He}}}, \bibinfo {author} {\bibfnamefont {D.}~\bibnamefont
  {{Goldhaber-Gordon}}},\ and\ \bibinfo {author} {\bibfnamefont {K.~T.}\
  \bibnamefont {{Law}}},\ }\bibfield  {title} {\bibinfo {title} {{Giant orbital
  magnetoelectric effect and current-induced magnetization switching in twisted
  bilayer graphene}},\ }\href {https://doi.org/10.1038/s41467-020-15473-9}
  {\bibfield  {journal} {\bibinfo  {journal} {Nat. Commun.}\ }\textbf {\bibinfo
  {volume} {11}},\ \bibinfo {eid} {1650} (\bibinfo {year} {2020})}\BibitemShut
  {NoStop}%
\bibitem [{\citenamefont {{Hu}}\ \emph {et~al.}(2020)\citenamefont {{Hu}},
  \citenamefont {{Zhang}}, \citenamefont {{Xie}},\ and\ \citenamefont
  {{Law}}}]{Hu2020}%
  \BibitemOpen
  \bibfield  {author} {\bibinfo {author} {\bibfnamefont {J.-X.}\ \bibnamefont
  {{Hu}}}, \bibinfo {author} {\bibfnamefont {C.-P.}\ \bibnamefont {{Zhang}}},
  \bibinfo {author} {\bibfnamefont {Y.-M.}\ \bibnamefont {{Xie}}},\ and\
  \bibinfo {author} {\bibfnamefont {K.~T.}\ \bibnamefont {{Law}}},\ }\bibfield
  {title} {\bibinfo {title} {{Nonlinear Hall Effects in Strained Twisted
  Bilayer WSe$_2$}},\ }\href@noop {} {\bibfield  {journal} {\bibinfo  {journal}
  {arXiv}\ ,\ \bibinfo {eid} {arXiv:2004.14140}} (\bibinfo {year} {2020})},\
  \Eprint {https://arxiv.org/abs/2004.14140} {arXiv:2004.14140
  [cond-mat.mes-hall]} \BibitemShut {NoStop}%
\bibitem [{\citenamefont {{Huang}}\ \emph {et~al.}(2020)\citenamefont
  {{Huang}}, \citenamefont {{Wu}}, \citenamefont {{Hu}}, \citenamefont {{Cai}},
  \citenamefont {{Li}}, \citenamefont {{An}}, \citenamefont {{Feng}},
  \citenamefont {{Ye}}, \citenamefont {{Lin}}, \citenamefont {{Tuen Law}},\
  and\ \citenamefont {{Wang}}}]{Huang2020}%
  \BibitemOpen
  \bibfield  {author} {\bibinfo {author} {\bibfnamefont {M.}~\bibnamefont
  {{Huang}}}, \bibinfo {author} {\bibfnamefont {Z.}~\bibnamefont {{Wu}}},
  \bibinfo {author} {\bibfnamefont {J.}~\bibnamefont {{Hu}}}, \bibinfo {author}
  {\bibfnamefont {X.}~\bibnamefont {{Cai}}}, \bibinfo {author} {\bibfnamefont
  {E.}~\bibnamefont {{Li}}}, \bibinfo {author} {\bibfnamefont {L.}~\bibnamefont
  {{An}}}, \bibinfo {author} {\bibfnamefont {X.}~\bibnamefont {{Feng}}},
  \bibinfo {author} {\bibfnamefont {Z.}~\bibnamefont {{Ye}}}, \bibinfo {author}
  {\bibfnamefont {N.}~\bibnamefont {{Lin}}}, \bibinfo {author} {\bibfnamefont
  {K.}~\bibnamefont {{Tuen Law}}},\ and\ \bibinfo {author} {\bibfnamefont
  {N.}~\bibnamefont {{Wang}}},\ }\bibfield  {title} {\bibinfo {title} {{Giant
  nonlinear Hall effect in twisted WSe$_2$}},\ }\href@noop {} {\bibfield
  {journal} {\bibinfo  {journal} {arXiv}\ ,\ \bibinfo {eid} {arXiv:2006.05615}}
  (\bibinfo {year} {2020})},\ \Eprint {https://arxiv.org/abs/2006.05615}
  {arXiv:2006.05615 [cond-mat.mes-hall]} \BibitemShut {NoStop}%
\bibitem [{\citenamefont {Otteneder}\ \emph {et~al.}(2020)\citenamefont
  {Otteneder}, \citenamefont {Hubmann}, \citenamefont {Lu}, \citenamefont
  {Kozlov}, \citenamefont {Golub}, \citenamefont {Watanabe}, \citenamefont
  {Taniguchi}, \citenamefont {Efetov},\ and\ \citenamefont
  {Ganichev}}]{Otteneder2020}%
  \BibitemOpen
  \bibfield  {author} {\bibinfo {author} {\bibfnamefont {M.}~\bibnamefont
  {Otteneder}}, \bibinfo {author} {\bibfnamefont {S.}~\bibnamefont {Hubmann}},
  \bibinfo {author} {\bibfnamefont {X.}~\bibnamefont {Lu}}, \bibinfo {author}
  {\bibfnamefont {D.~A.}\ \bibnamefont {Kozlov}}, \bibinfo {author}
  {\bibfnamefont {L.~E.}\ \bibnamefont {Golub}}, \bibinfo {author}
  {\bibfnamefont {K.}~\bibnamefont {Watanabe}}, \bibinfo {author}
  {\bibfnamefont {T.}~\bibnamefont {Taniguchi}}, \bibinfo {author}
  {\bibfnamefont {D.~K.}\ \bibnamefont {Efetov}},\ and\ \bibinfo {author}
  {\bibfnamefont {S.~D.}\ \bibnamefont {Ganichev}},\ }\bibfield  {title}
  {\bibinfo {title} {{Terahertz photogalvanics in twisted bilayer graphene
  close to the second magic angle}},\ }\href
  {https://doi.org/10.1021/acs.nanolett.0c02474} {\bibfield  {journal}
  {\bibinfo  {journal} {Nano Lett.}\ }\textbf {\bibinfo {volume} {20}},\
  \bibinfo {pages} {7152–7158} (\bibinfo {year} {2020})}\BibitemShut
  {NoStop}%
\bibitem [{\citenamefont {{Liu}}\ \emph {et~al.}(2021)\citenamefont {{Liu}},
  \citenamefont {{Holder}},\ and\ \citenamefont {{Yan}}}]{liu2020b}%
  \BibitemOpen
  \bibfield  {author} {\bibinfo {author} {\bibfnamefont {Y.}~\bibnamefont
  {{Liu}}}, \bibinfo {author} {\bibfnamefont {T.}~\bibnamefont {{Holder}}},\
  and\ \bibinfo {author} {\bibfnamefont {B.}~\bibnamefont {{Yan}}},\ }\bibfield
   {title} {\bibinfo {title} {{Chirality-Induced Giant Unidirectional
  Magnetoresistance in Twisted Bilayer Graphene}},\ }\href
  {https://doi.org/10.1016/j.xinn.2021.100085} {\bibfield  {journal} {\bibinfo
  {journal} {The Innovation}\ }\textbf {\bibinfo {volume} {2}},\ \bibinfo
  {pages} {100085} (\bibinfo {year} {2021})},\ \Eprint
  {https://arxiv.org/abs/2010.08385} {arXiv:2010.08385 [cond-mat.mes-hall]}
  \BibitemShut {NoStop}%
\bibitem [{\citenamefont {{Zhang}}\ \emph {et~al.}(2020)\citenamefont
  {{Zhang}}, \citenamefont {{Xiao}}, \citenamefont {{Zhou}}, \citenamefont
  {{Hu}}, \citenamefont {{Xie}}, \citenamefont {{Yan}},\ and\ \citenamefont
  {{Law}}}]{zhang2020giant}%
  \BibitemOpen
  \bibfield  {author} {\bibinfo {author} {\bibfnamefont {C.-P.}\ \bibnamefont
  {{Zhang}}}, \bibinfo {author} {\bibfnamefont {J.}~\bibnamefont {{Xiao}}},
  \bibinfo {author} {\bibfnamefont {B.~T.}\ \bibnamefont {{Zhou}}}, \bibinfo
  {author} {\bibfnamefont {J.-X.}\ \bibnamefont {{Hu}}}, \bibinfo {author}
  {\bibfnamefont {Y.-M.}\ \bibnamefont {{Xie}}}, \bibinfo {author}
  {\bibfnamefont {B.}~\bibnamefont {{Yan}}},\ and\ \bibinfo {author}
  {\bibfnamefont {K.~T.}\ \bibnamefont {{Law}}},\ }\bibfield  {title} {\bibinfo
  {title} {{Giant nonlinear Hall effect in strained twisted bilayer
  graphene}},\ }\href@noop {} {\bibfield  {journal} {\bibinfo  {journal} {arXiv
  e-prints}\ ,\ \bibinfo {eid} {arXiv:2010.08333}} (\bibinfo {year} {2020})},\
  \Eprint {https://arxiv.org/abs/2010.08333} {arXiv:2010.08333
  [cond-mat.mes-hall]} \BibitemShut {NoStop}%
\bibitem [{\citenamefont {Ikeda}(2020)}]{Ikeda2020}%
  \BibitemOpen
  \bibfield  {author} {\bibinfo {author} {\bibfnamefont {T.~N.}\ \bibnamefont
  {Ikeda}},\ }\bibfield  {title} {\bibinfo {title} {High-order nonlinear
  optical response of a twisted bilayer graphene},\ }\href
  {https://doi.org/10.1103/PhysRevResearch.2.032015} {\bibfield  {journal}
  {\bibinfo  {journal} {Phys. Rev. Research}\ }\textbf {\bibinfo {volume}
  {2}},\ \bibinfo {pages} {032015} (\bibinfo {year} {2020})},\ \Eprint
  {https://arxiv.org/abs/2005.01718} {arXiv:2005.01718 [cond-mat.str-el]}
  \BibitemShut {NoStop}%
\bibitem [{\citenamefont {{Moore}}\ and\ \citenamefont
  {{Orenstein}}(2010)}]{Moore2010}%
  \BibitemOpen
  \bibfield  {author} {\bibinfo {author} {\bibfnamefont {J.~E.}\ \bibnamefont
  {{Moore}}}\ and\ \bibinfo {author} {\bibfnamefont {J.}~\bibnamefont
  {{Orenstein}}},\ }\bibfield  {title} {\bibinfo {title} {{Confinement-Induced
  Berry Phase and Helicity-Dependent Photocurrents}},\ }\href
  {https://doi.org/10.1103/PhysRevLett.105.026805} {\bibfield  {journal}
  {\bibinfo  {journal} {\prl}\ }\textbf {\bibinfo {volume} {105}},\ \bibinfo
  {eid} {026805} (\bibinfo {year} {2010})},\ \Eprint
  {https://arxiv.org/abs/0911.3630} {arXiv:0911.3630 [cond-mat.mes-hall]}
  \BibitemShut {NoStop}%
\bibitem [{\citenamefont {{Sodemann}}\ and\ \citenamefont
  {{Fu}}(2015)}]{Sodemann2015}%
  \BibitemOpen
  \bibfield  {author} {\bibinfo {author} {\bibfnamefont {I.}~\bibnamefont
  {{Sodemann}}}\ and\ \bibinfo {author} {\bibfnamefont {L.}~\bibnamefont
  {{Fu}}},\ }\bibfield  {title} {\bibinfo {title} {{Quantum Nonlinear Hall
  Effect Induced by Berry Curvature Dipole in Time-Reversal Invariant
  Materials}},\ }\href {https://doi.org/10.1103/PhysRevLett.115.216806}
  {\bibfield  {journal} {\bibinfo  {journal} {Phys. Rev. Lett.}\ }\textbf
  {\bibinfo {volume} {115}},\ \bibinfo {eid} {216806} (\bibinfo {year}
  {2015})},\ \Eprint {https://arxiv.org/abs/1508.00571} {arXiv:1508.00571
  [cond-mat.mes-hall]} \BibitemShut {NoStop}%
\bibitem [{\citenamefont {{Sato}}\ \emph {et~al.}(2021)\citenamefont {{Sato}},
  \citenamefont {{Hayashi}}, \citenamefont {{Ito}}, \citenamefont {{Masago}},
  \citenamefont {{Takamura}}, \citenamefont {{Morimoto}}, \citenamefont
  {{Maekawa}}, \citenamefont {{Lee}}, \citenamefont {{Qiao}}, \citenamefont
  {{Kim}}, \citenamefont {{Nakagahara}}, \citenamefont {{Wakabayashi}},
  \citenamefont {{Hibino}},\ and\ \citenamefont {{Norimatsu}}}]{Sato2021}%
  \BibitemOpen
  \bibfield  {author} {\bibinfo {author} {\bibfnamefont {K.}~\bibnamefont
  {{Sato}}}, \bibinfo {author} {\bibfnamefont {N.}~\bibnamefont {{Hayashi}}},
  \bibinfo {author} {\bibfnamefont {T.}~\bibnamefont {{Ito}}}, \bibinfo
  {author} {\bibfnamefont {N.}~\bibnamefont {{Masago}}}, \bibinfo {author}
  {\bibfnamefont {M.}~\bibnamefont {{Takamura}}}, \bibinfo {author}
  {\bibfnamefont {M.}~\bibnamefont {{Morimoto}}}, \bibinfo {author}
  {\bibfnamefont {T.}~\bibnamefont {{Maekawa}}}, \bibinfo {author}
  {\bibfnamefont {D.}~\bibnamefont {{Lee}}}, \bibinfo {author} {\bibfnamefont
  {K.}~\bibnamefont {{Qiao}}}, \bibinfo {author} {\bibfnamefont
  {J.}~\bibnamefont {{Kim}}}, \bibinfo {author} {\bibfnamefont
  {K.}~\bibnamefont {{Nakagahara}}}, \bibinfo {author} {\bibfnamefont
  {K.}~\bibnamefont {{Wakabayashi}}}, \bibinfo {author} {\bibfnamefont
  {H.}~\bibnamefont {{Hibino}}},\ and\ \bibinfo {author} {\bibfnamefont
  {W.}~\bibnamefont {{Norimatsu}}},\ }\bibfield  {title} {\bibinfo {title}
  {{Observation of a flat band and bandgap in millimeter-scale twisted bilayer
  graphene}},\ }\href {https://doi.org/10.1038/s43246-021-00221-3} {\bibfield
  {journal} {\bibinfo  {journal} {Commun. Mater.}\ }\textbf {\bibinfo {volume}
  {2}},\ \bibinfo {eid} {117} (\bibinfo {year} {2021})}\BibitemShut {NoStop}%
\bibitem [{\citenamefont {Kazmierczak}\ \emph {et~al.}(2021)\citenamefont
  {Kazmierczak}, \citenamefont {{Van Winkle}}, \citenamefont {Ophus},
  \citenamefont {Bustillo}, \citenamefont {Brown}, \citenamefont {Carr},
  \citenamefont {Ciston}, \citenamefont {Taniguchi}, \citenamefont {Watanabe},\
  and\ \citenamefont {Bediako}}]{Kazmierczak2020}%
  \BibitemOpen
  \bibfield  {author} {\bibinfo {author} {\bibfnamefont {N.~P.}\ \bibnamefont
  {Kazmierczak}}, \bibinfo {author} {\bibfnamefont {M.}~\bibnamefont {{Van
  Winkle}}}, \bibinfo {author} {\bibfnamefont {C.}~\bibnamefont {Ophus}},
  \bibinfo {author} {\bibfnamefont {K.~C.}\ \bibnamefont {Bustillo}}, \bibinfo
  {author} {\bibfnamefont {H.~G.}\ \bibnamefont {Brown}}, \bibinfo {author}
  {\bibfnamefont {S.}~\bibnamefont {Carr}}, \bibinfo {author} {\bibfnamefont
  {J.}~\bibnamefont {Ciston}}, \bibinfo {author} {\bibfnamefont
  {T.}~\bibnamefont {Taniguchi}}, \bibinfo {author} {\bibfnamefont
  {K.}~\bibnamefont {Watanabe}},\ and\ \bibinfo {author} {\bibfnamefont
  {D.~K.}\ \bibnamefont {Bediako}},\ }\bibfield  {title} {\bibinfo {title}
  {{Strain fields in twisted bilayer graphene}},\ }\href
  {https://doi.org/10.1038/s41563-021-00973-w} {\bibfield  {journal} {\bibinfo
  {journal} {Nat. Mater.}\ ,\ \bibinfo {pages} {1}} (\bibinfo {year} {2021})},\
  \Eprint {https://arxiv.org/abs/2008.09761} {arXiv:2008.09761} \BibitemShut
  {NoStop}%
\bibitem [{\citenamefont {{Ahn}}\ \emph {et~al.}(2020)\citenamefont {{Ahn}},
  \citenamefont {{Guo}},\ and\ \citenamefont {{Nagaosa}}}]{Ahn2020}%
  \BibitemOpen
  \bibfield  {author} {\bibinfo {author} {\bibfnamefont {J.}~\bibnamefont
  {{Ahn}}}, \bibinfo {author} {\bibfnamefont {G.-Y.}\ \bibnamefont {{Guo}}},\
  and\ \bibinfo {author} {\bibfnamefont {N.}~\bibnamefont {{Nagaosa}}},\
  }\bibfield  {title} {\bibinfo {title} {{Low-Frequency Divergence and Quantum
  Geometry of the Bulk Photovoltaic Effect in Topological Semimetals}},\ }\href
  {https://doi.org/10.1103/PhysRevX.10.041041} {\bibfield  {journal} {\bibinfo
  {journal} {Physical Review X}\ }\textbf {\bibinfo {volume} {10}},\ \bibinfo
  {eid} {041041} (\bibinfo {year} {2020})},\ \Eprint
  {https://arxiv.org/abs/2006.06709} {arXiv:2006.06709 [cond-mat.mes-hall]}
  \BibitemShut {NoStop}%
\bibitem [{\citenamefont {{Ahn}}\ \emph {et~al.}(2021)\citenamefont {{Ahn}},
  \citenamefont {{Guo}}, \citenamefont {{Nagaosa}},\ and\ \citenamefont
  {{Vishwanath}}}]{Ahn2021}%
  \BibitemOpen
  \bibfield  {author} {\bibinfo {author} {\bibfnamefont {J.}~\bibnamefont
  {{Ahn}}}, \bibinfo {author} {\bibfnamefont {G.-Y.}\ \bibnamefont {{Guo}}},
  \bibinfo {author} {\bibfnamefont {N.}~\bibnamefont {{Nagaosa}}},\ and\
  \bibinfo {author} {\bibfnamefont {A.}~\bibnamefont {{Vishwanath}}},\
  }\bibfield  {title} {\bibinfo {title} {{Riemannian Geometry of Resonant
  Optical Responses}},\ }\href@noop {} {\bibfield  {journal} {\bibinfo
  {journal} {arXiv}\ ,\ \bibinfo {eid} {arXiv:2103.01241}} (\bibinfo {year}
  {2021})},\ \Eprint {https://arxiv.org/abs/2103.01241} {arXiv:2103.01241
  [cond-mat.mes-hall]} \BibitemShut {NoStop}%
\bibitem [{\citenamefont {{Da Liao}}\ \emph {et~al.}(2021)\citenamefont {{Da
  Liao}}, \citenamefont {{Kang}}, \citenamefont {{Brei{\o}}}, \citenamefont
  {{Xu}}, \citenamefont {{Wu}}, \citenamefont {{Andersen}}, \citenamefont
  {{Fernandes}},\ and\ \citenamefont {{Meng}}}]{DaLiao2021}%
  \BibitemOpen
  \bibfield  {author} {\bibinfo {author} {\bibfnamefont {Y.}~\bibnamefont {{Da
  Liao}}}, \bibinfo {author} {\bibfnamefont {J.}~\bibnamefont {{Kang}}},
  \bibinfo {author} {\bibfnamefont {C.~N.}\ \bibnamefont {{Brei{\o}}}},
  \bibinfo {author} {\bibfnamefont {X.~Y.}\ \bibnamefont {{Xu}}}, \bibinfo
  {author} {\bibfnamefont {H.-Q.}\ \bibnamefont {{Wu}}}, \bibinfo {author}
  {\bibfnamefont {B.~M.}\ \bibnamefont {{Andersen}}}, \bibinfo {author}
  {\bibfnamefont {R.~M.}\ \bibnamefont {{Fernandes}}},\ and\ \bibinfo {author}
  {\bibfnamefont {Z.~Y.}\ \bibnamefont {{Meng}}},\ }\bibfield  {title}
  {\bibinfo {title} {{Correlation-Induced Insulating Topological Phases at
  Charge Neutrality in Twisted Bilayer Graphene}},\ }\href
  {https://doi.org/10.1103/PhysRevX.11.011014} {\bibfield  {journal} {\bibinfo
  {journal} {Phys. Rev. X}\ }\textbf {\bibinfo {volume} {11}},\ \bibinfo {eid}
  {011014} (\bibinfo {year} {2021})},\ \Eprint
  {https://arxiv.org/abs/2004.12536} {arXiv:2004.12536 [cond-mat.str-el]}
  \BibitemShut {NoStop}%
\bibitem [{\citenamefont {Rangel}\ \emph {et~al.}(2017)\citenamefont {Rangel},
  \citenamefont {Fregoso}, \citenamefont {Mendoza}, \citenamefont {Morimoto},
  \citenamefont {Moore},\ and\ \citenamefont {Neaton}}]{Rangel2017}%
  \BibitemOpen
  \bibfield  {author} {\bibinfo {author} {\bibfnamefont {T.}~\bibnamefont
  {Rangel}}, \bibinfo {author} {\bibfnamefont {B.~M.}\ \bibnamefont {Fregoso}},
  \bibinfo {author} {\bibfnamefont {B.~S.}\ \bibnamefont {Mendoza}}, \bibinfo
  {author} {\bibfnamefont {T.}~\bibnamefont {Morimoto}}, \bibinfo {author}
  {\bibfnamefont {J.~E.}\ \bibnamefont {Moore}},\ and\ \bibinfo {author}
  {\bibfnamefont {J.~B.}\ \bibnamefont {Neaton}},\ }\bibfield  {title}
  {\bibinfo {title} {Large bulk photovoltaic effect and spontaneous
  polarization of single-layer monochalcogenides},\ }\href
  {https://doi.org/10.1103/PhysRevLett.119.067402} {\bibfield  {journal}
  {\bibinfo  {journal} {Phys. Rev. Lett.}\ }\textbf {\bibinfo {volume} {119}},\
  \bibinfo {pages} {067402} (\bibinfo {year} {2017})}\BibitemShut {NoStop}%
\bibitem [{\citenamefont {{Cook}}\ \emph {et~al.}(2017)\citenamefont {{Cook}},
  \citenamefont {{M.~Fregoso}}, \citenamefont {{de Juan}}, \citenamefont
  {{Coh}},\ and\ \citenamefont {{Moore}}}]{Cook2017}%
  \BibitemOpen
  \bibfield  {author} {\bibinfo {author} {\bibfnamefont {A.~M.}\ \bibnamefont
  {{Cook}}}, \bibinfo {author} {\bibfnamefont {B.}~\bibnamefont
  {{M.~Fregoso}}}, \bibinfo {author} {\bibfnamefont {F.}~\bibnamefont {{de
  Juan}}}, \bibinfo {author} {\bibfnamefont {S.}~\bibnamefont {{Coh}}},\ and\
  \bibinfo {author} {\bibfnamefont {J.~E.}\ \bibnamefont {{Moore}}},\
  }\bibfield  {title} {\bibinfo {title} {{Design principles for shift current
  photovoltaics}},\ }\href {https://doi.org/10.1038/ncomms14176} {\bibfield
  {journal} {\bibinfo  {journal} {Nat. Commun.}\ }\textbf {\bibinfo {volume}
  {8}},\ \bibinfo {pages} {14176} (\bibinfo {year} {2017})}\BibitemShut
  {NoStop}%
\bibitem [{\citenamefont {{Tan}}\ \emph {et~al.}(2016)\citenamefont {{Tan}},
  \citenamefont {{Zheng}}, \citenamefont {{Young}}, \citenamefont {{Wang}},
  \citenamefont {{Liu}},\ and\ \citenamefont {{Rappe}}}]{Tan2016}%
  \BibitemOpen
  \bibfield  {author} {\bibinfo {author} {\bibfnamefont {L.~Z.}\ \bibnamefont
  {{Tan}}}, \bibinfo {author} {\bibfnamefont {F.}~\bibnamefont {{Zheng}}},
  \bibinfo {author} {\bibfnamefont {S.~M.}\ \bibnamefont {{Young}}}, \bibinfo
  {author} {\bibfnamefont {F.}~\bibnamefont {{Wang}}}, \bibinfo {author}
  {\bibfnamefont {S.}~\bibnamefont {{Liu}}},\ and\ \bibinfo {author}
  {\bibfnamefont {A.~M.}\ \bibnamefont {{Rappe}}},\ }\bibfield  {title}
  {\bibinfo {title} {{Shift current bulk photovoltaic effect in polar
  materials{\textemdash}hybrid and oxide perovskites and beyond}},\ }\href
  {https://doi.org/10.1038/npjcompumats.2016.26} {\bibfield  {journal}
  {\bibinfo  {journal} {npj Comput. Mater.}\ }\textbf {\bibinfo {volume} {2}},\
  \bibinfo {eid} {16026} (\bibinfo {year} {2016})}\BibitemShut {NoStop}%
\bibitem [{\citenamefont {{Schankler}}\ \emph {et~al.}(2021)\citenamefont
  {{Schankler}}, \citenamefont {{Gao}},\ and\ \citenamefont
  {{Rappe}}}]{Schankler2021}%
  \BibitemOpen
  \bibfield  {author} {\bibinfo {author} {\bibfnamefont {A.~M.}\ \bibnamefont
  {{Schankler}}}, \bibinfo {author} {\bibfnamefont {L.}~\bibnamefont {{Gao}}},\
  and\ \bibinfo {author} {\bibfnamefont {A.~M.}\ \bibnamefont {{Rappe}}},\
  }\bibfield  {title} {\bibinfo {title} {{Large Bulk Piezophotovoltaic Effect
  of Monolayer 2H-MoS$_2$}},\ }\href
  {https://doi.org/10.1021/acs.jpclett.0c03503} {\bibfield  {journal} {\bibinfo
   {journal} {J. Phys. Chem. Lett.}\ }\textbf {\bibinfo {volume} {12}},\
  \bibinfo {pages} {1244} (\bibinfo {year} {2021})}\BibitemShut {NoStop}%
\bibitem [{\citenamefont {Monteverde}\ \emph {et~al.}(2010)\citenamefont
  {Monteverde}, \citenamefont {Ojeda-Aristizabal}, \citenamefont {Weil},
  \citenamefont {Bennaceur}, \citenamefont {Ferrier}, \citenamefont
  {Gu{\'{e}}ron}, \citenamefont {Glattli}, \citenamefont {Bouchiat},
  \citenamefont {Fuchs},\ and\ \citenamefont {Maslov}}]{Monteverde2010}%
  \BibitemOpen
  \bibfield  {author} {\bibinfo {author} {\bibfnamefont {M.}~\bibnamefont
  {Monteverde}}, \bibinfo {author} {\bibfnamefont {C.}~\bibnamefont
  {Ojeda-Aristizabal}}, \bibinfo {author} {\bibfnamefont {R.}~\bibnamefont
  {Weil}}, \bibinfo {author} {\bibfnamefont {K.}~\bibnamefont {Bennaceur}},
  \bibinfo {author} {\bibfnamefont {M.}~\bibnamefont {Ferrier}}, \bibinfo
  {author} {\bibfnamefont {S.}~\bibnamefont {Gu{\'{e}}ron}}, \bibinfo {author}
  {\bibfnamefont {C.}~\bibnamefont {Glattli}}, \bibinfo {author} {\bibfnamefont
  {H.}~\bibnamefont {Bouchiat}}, \bibinfo {author} {\bibfnamefont {J.~N.}\
  \bibnamefont {Fuchs}},\ and\ \bibinfo {author} {\bibfnamefont {D.~L.}\
  \bibnamefont {Maslov}},\ }\bibfield  {title} {\bibinfo {title} {{Transport
  and Elastic Scattering Times as Probes of the Nature of Impurity Scattering
  in Single-Layer and Bilayer Graphene}},\ }\href
  {https://doi.org/10.1103/PhysRevLett.104.126801} {\bibfield  {journal}
  {\bibinfo  {journal} {Phys. Rev. Lett.}\ }\textbf {\bibinfo {volume} {104}},\
  \bibinfo {pages} {126801} (\bibinfo {year} {2010})}\BibitemShut {NoStop}%
\bibitem [{\citenamefont {Sipe}\ and\ \citenamefont {Zak}(1999)}]{Sipe1999}%
  \BibitemOpen
  \bibfield  {author} {\bibinfo {author} {\bibfnamefont {J.~E.}\ \bibnamefont
  {Sipe}}\ and\ \bibinfo {author} {\bibfnamefont {J.}~\bibnamefont {Zak}},\
  }\bibfield  {title} {\bibinfo {title} {{Geometric phase for electric
  polarization along 'rational' directions in crystals}},\ }\href
  {https://doi.org/10.1016/S0375-9601(99)00308-4} {\bibfield  {journal}
  {\bibinfo  {journal} {Phys. Lett. A}\ }\textbf {\bibinfo {volume} {258}},\
  \bibinfo {pages} {406} (\bibinfo {year} {1999})}\BibitemShut {NoStop}%
\bibitem [{\citenamefont {{Carr}}\ \emph
  {et~al.}(2019{\natexlab{a}})\citenamefont {{Carr}}, \citenamefont {{Fang}},
  \citenamefont {{Zhu}},\ and\ \citenamefont {{Kaxiras}}}]{Carr2019}%
  \BibitemOpen
  \bibfield  {author} {\bibinfo {author} {\bibfnamefont {S.}~\bibnamefont
  {{Carr}}}, \bibinfo {author} {\bibfnamefont {S.}~\bibnamefont {{Fang}}},
  \bibinfo {author} {\bibfnamefont {Z.}~\bibnamefont {{Zhu}}},\ and\ \bibinfo
  {author} {\bibfnamefont {E.}~\bibnamefont {{Kaxiras}}},\ }\bibfield  {title}
  {\bibinfo {title} {{Exact continuum model for low-energy electronic states of
  twisted bilayer graphene}},\ }\href
  {https://doi.org/10.1103/PhysRevResearch.1.013001} {\bibfield  {journal}
  {\bibinfo  {journal} {Phys. Rev. Research}\ }\textbf {\bibinfo {volume}
  {1}},\ \bibinfo {eid} {013001} (\bibinfo {year} {2019}{\natexlab{a}})},\
  \Eprint {https://arxiv.org/abs/1901.03420} {arXiv:1901.03420
  [cond-mat.mes-hall]} \BibitemShut {NoStop}%
\bibitem [{\citenamefont {{Carr}}\ \emph
  {et~al.}(2019{\natexlab{b}})\citenamefont {{Carr}}, \citenamefont {{Fang}},
  \citenamefont {{Po}}, \citenamefont {{Vishwanath}},\ and\ \citenamefont
  {{Kaxiras}}}]{Carr2019b}%
  \BibitemOpen
  \bibfield  {author} {\bibinfo {author} {\bibfnamefont {S.}~\bibnamefont
  {{Carr}}}, \bibinfo {author} {\bibfnamefont {S.}~\bibnamefont {{Fang}}},
  \bibinfo {author} {\bibfnamefont {H.~C.}\ \bibnamefont {{Po}}}, \bibinfo
  {author} {\bibfnamefont {A.}~\bibnamefont {{Vishwanath}}},\ and\ \bibinfo
  {author} {\bibfnamefont {E.}~\bibnamefont {{Kaxiras}}},\ }\bibfield  {title}
  {\bibinfo {title} {{Derivation of Wannier orbitals and minimal-basis
  tight-binding Hamiltonians for twisted bilayer graphene: First-principles
  approach}},\ }\href {https://doi.org/10.1103/PhysRevResearch.1.033072}
  {\bibfield  {journal} {\bibinfo  {journal} {Physical Review Research}\
  }\textbf {\bibinfo {volume} {1}},\ \bibinfo {eid} {033072} (\bibinfo {year}
  {2019}{\natexlab{b}})},\ \Eprint {https://arxiv.org/abs/1907.06282}
  {arXiv:1907.06282 [cond-mat.str-el]} \BibitemShut {NoStop}%
\bibitem [{\citenamefont {{Iba{\~n}ez-Azpiroz}}\ \emph
  {et~al.}(2019)\citenamefont {{Iba{\~n}ez-Azpiroz}}, \citenamefont {{de
  Juan}},\ and\ \citenamefont {{Souza}}}]{Ibanez2019}%
  \BibitemOpen
  \bibfield  {author} {\bibinfo {author} {\bibfnamefont {J.}~\bibnamefont
  {{Iba{\~n}ez-Azpiroz}}}, \bibinfo {author} {\bibfnamefont {F.}~\bibnamefont
  {{de Juan}}},\ and\ \bibinfo {author} {\bibfnamefont {I.}~\bibnamefont
  {{Souza}}},\ }\bibfield  {title} {\bibinfo {title} {{Assessing the role of
  interatomic position matrix elements in tight-binding calculations of optical
  properties}},\ }\href@noop {} {\bibfield  {journal} {\bibinfo  {journal}
  {arXiv e-prints}\ ,\ \bibinfo {eid} {arXiv:1910.06172}} (\bibinfo {year}
  {2019})},\ \Eprint {https://arxiv.org/abs/1910.06172} {arXiv:1910.06172
  [cond-mat.mes-hall]} \BibitemShut {NoStop}%
\bibitem [{\citenamefont {Vanderbilt}(2018)}]{Vanderbiltbook}%
  \BibitemOpen
  \bibfield  {author} {\bibinfo {author} {\bibfnamefont {D.}~\bibnamefont
  {Vanderbilt}},\ }\href {https://doi.org/10.1017/9781316662205} {\emph
  {\bibinfo {title} {Berry Phases in Electronic Structure Theory: Electric
  Polarization, Orbital Magnetization and Topological Insulators}}}\ (\bibinfo
  {publisher} {Cambridge University Press},\ \bibinfo {year}
  {2018})\BibitemShut {NoStop}%
\bibitem [{\citenamefont {{Xiao}}\ \emph {et~al.}(2010)\citenamefont {{Xiao}},
  \citenamefont {{Chang}},\ and\ \citenamefont {{Niu}}}]{Xiao2010}%
  \BibitemOpen
  \bibfield  {author} {\bibinfo {author} {\bibfnamefont {D.}~\bibnamefont
  {{Xiao}}}, \bibinfo {author} {\bibfnamefont {M.-C.}\ \bibnamefont
  {{Chang}}},\ and\ \bibinfo {author} {\bibfnamefont {Q.}~\bibnamefont
  {{Niu}}},\ }\bibfield  {title} {\bibinfo {title} {{Berry phase effects on
  electronic properties}},\ }\href {https://doi.org/10.1103/RevModPhys.82.1959}
  {\bibfield  {journal} {\bibinfo  {journal} {Rev. Mod. Phys.}\ }\textbf
  {\bibinfo {volume} {82}},\ \bibinfo {pages} {1959} (\bibinfo {year}
  {2010})},\ \Eprint {https://arxiv.org/abs/0907.2021} {arXiv:0907.2021
  [cond-mat.mes-hall]} \BibitemShut {NoStop}%
\bibitem [{\citenamefont {{Xiao}}\ \emph {et~al.}(2019)\citenamefont {{Xiao}},
  \citenamefont {{Du}},\ and\ \citenamefont {{Niu}}}]{Xiao2019}%
  \BibitemOpen
  \bibfield  {author} {\bibinfo {author} {\bibfnamefont {C.}~\bibnamefont
  {{Xiao}}}, \bibinfo {author} {\bibfnamefont {Z.~Z.}\ \bibnamefont {{Du}}},\
  and\ \bibinfo {author} {\bibfnamefont {Q.}~\bibnamefont {{Niu}}},\ }\bibfield
   {title} {\bibinfo {title} {{Theory of nonlinear Hall effects: Modified
  semiclassics from quantum kinetics}},\ }\href
  {https://doi.org/10.1103/PhysRevB.100.165422} {\bibfield  {journal} {\bibinfo
   {journal} {Phys. Rev. B}\ }\textbf {\bibinfo {volume} {100}},\ \bibinfo
  {eid} {165422} (\bibinfo {year} {2019})},\ \Eprint
  {https://arxiv.org/abs/1907.00577} {arXiv:1907.00577 [cond-mat.mes-hall]}
  \BibitemShut {NoStop}%
\bibitem [{\citenamefont {{Isobe}}\ \emph {et~al.}(2020)\citenamefont
  {{Isobe}}, \citenamefont {{Xu}},\ and\ \citenamefont {{Fu}}}]{Isobe2020}%
  \BibitemOpen
  \bibfield  {author} {\bibinfo {author} {\bibfnamefont {H.}~\bibnamefont
  {{Isobe}}}, \bibinfo {author} {\bibfnamefont {S.-Y.}\ \bibnamefont {{Xu}}},\
  and\ \bibinfo {author} {\bibfnamefont {L.}~\bibnamefont {{Fu}}},\ }\bibfield
  {title} {\bibinfo {title} {{High-frequency rectification via chiral Bloch
  electrons}},\ }\href {https://doi.org/10.1126/sciadv.aay2497} {\bibfield
  {journal} {\bibinfo  {journal} {Sci. Adv.}\ }\textbf {\bibinfo {volume}
  {6}},\ \bibinfo {pages} {eaay2497} (\bibinfo {year} {2020})}\BibitemShut
  {NoStop}%
\bibitem [{\citenamefont {{Uri}}\ \emph {et~al.}(2020)\citenamefont {{Uri}},
  \citenamefont {{Grover}}, \citenamefont {{Cao}}, \citenamefont {{Crosse}},
  \citenamefont {{Bagani}}, \citenamefont {{Rodan-Legrain}}, \citenamefont
  {{Myasoedov}}, \citenamefont {{Watanabe}}, \citenamefont {{Taniguchi}},
  \citenamefont {{Moon}}, \citenamefont {{Koshino}}, \citenamefont
  {{Jarillo-Herrero}},\ and\ \citenamefont {{Zeldov}}}]{Uri2020}%
  \BibitemOpen
  \bibfield  {author} {\bibinfo {author} {\bibfnamefont {A.}~\bibnamefont
  {{Uri}}}, \bibinfo {author} {\bibfnamefont {S.}~\bibnamefont {{Grover}}},
  \bibinfo {author} {\bibfnamefont {Y.}~\bibnamefont {{Cao}}}, \bibinfo
  {author} {\bibfnamefont {J.~{\^A}.~A.}\ \bibnamefont {{Crosse}}}, \bibinfo
  {author} {\bibfnamefont {K.}~\bibnamefont {{Bagani}}}, \bibinfo {author}
  {\bibfnamefont {D.}~\bibnamefont {{Rodan-Legrain}}}, \bibinfo {author}
  {\bibfnamefont {Y.}~\bibnamefont {{Myasoedov}}}, \bibinfo {author}
  {\bibfnamefont {K.}~\bibnamefont {{Watanabe}}}, \bibinfo {author}
  {\bibfnamefont {T.}~\bibnamefont {{Taniguchi}}}, \bibinfo {author}
  {\bibfnamefont {P.}~\bibnamefont {{Moon}}}, \bibinfo {author} {\bibfnamefont
  {M.}~\bibnamefont {{Koshino}}}, \bibinfo {author} {\bibfnamefont
  {P.}~\bibnamefont {{Jarillo-Herrero}}},\ and\ \bibinfo {author}
  {\bibfnamefont {E.}~\bibnamefont {{Zeldov}}},\ }\bibfield  {title} {\bibinfo
  {title} {{Mapping the twist-angle disorder and Landau levels in magic-angle
  graphene}},\ }\href {https://doi.org/10.1038/s41586-020-2255-3} {\bibfield
  {journal} {\bibinfo  {journal} {\nat}\ }\textbf {\bibinfo {volume} {581}},\
  \bibinfo {pages} {47} (\bibinfo {year} {2020})},\ \Eprint
  {https://arxiv.org/abs/1908.04595} {arXiv:1908.04595 [cond-mat.mes-hall]}
  \BibitemShut {NoStop}%
\bibitem [{\citenamefont {{Polshyn}}\ \emph {et~al.}(2019)\citenamefont
  {{Polshyn}}, \citenamefont {{Yankowitz}}, \citenamefont {{Chen}},
  \citenamefont {{Zhang}}, \citenamefont {{Watanabe}}, \citenamefont
  {{Taniguchi}}, \citenamefont {{Dean}},\ and\ \citenamefont
  {{Young}}}]{Polshyn2019}%
  \BibitemOpen
  \bibfield  {author} {\bibinfo {author} {\bibfnamefont {H.}~\bibnamefont
  {{Polshyn}}}, \bibinfo {author} {\bibfnamefont {M.}~\bibnamefont
  {{Yankowitz}}}, \bibinfo {author} {\bibfnamefont {S.}~\bibnamefont {{Chen}}},
  \bibinfo {author} {\bibfnamefont {Y.}~\bibnamefont {{Zhang}}}, \bibinfo
  {author} {\bibfnamefont {K.}~\bibnamefont {{Watanabe}}}, \bibinfo {author}
  {\bibfnamefont {T.}~\bibnamefont {{Taniguchi}}}, \bibinfo {author}
  {\bibfnamefont {C.~R.}\ \bibnamefont {{Dean}}},\ and\ \bibinfo {author}
  {\bibfnamefont {A.~F.}\ \bibnamefont {{Young}}},\ }\bibfield  {title}
  {\bibinfo {title} {{Large linear-in-temperature resistivity in twisted
  bilayer graphene}},\ }\href {https://doi.org/10.1038/s41567-019-0596-3}
  {\bibfield  {journal} {\bibinfo  {journal} {Nat. Phys.}\ }\textbf {\bibinfo
  {volume} {15}},\ \bibinfo {pages} {1011} (\bibinfo {year}
  {2019})}\BibitemShut {NoStop}%
\bibitem [{\citenamefont {{Sturman}}\ and\ \citenamefont
  {{Fridkin}}(1992)}]{Sturman1992}%
  \BibitemOpen
  \bibfield  {author} {\bibinfo {author} {\bibfnamefont {B.~I.}\ \bibnamefont
  {{Sturman}}}\ and\ \bibinfo {author} {\bibfnamefont {V.~M.}\ \bibnamefont
  {{Fridkin}}},\ }\href@noop {} {\emph {\bibinfo {title} {The Photovoltaic and
  Photorefractive Effects in Noncentrosymmetric Materials}}}\ (\bibinfo
  {publisher} {Gordon and Breach Philadelphia},\ \bibinfo {year}
  {1992})\BibitemShut {NoStop}%
\bibitem [{\citenamefont {{Gao}}\ \emph {et~al.}(2020)\citenamefont {{Gao}},
  \citenamefont {{Zhang}},\ and\ \citenamefont {{Xiao}}}]{Gao2020a}%
  \BibitemOpen
  \bibfield  {author} {\bibinfo {author} {\bibfnamefont {Y.}~\bibnamefont
  {{Gao}}}, \bibinfo {author} {\bibfnamefont {Y.}~\bibnamefont {{Zhang}}},\
  and\ \bibinfo {author} {\bibfnamefont {D.}~\bibnamefont {{Xiao}}},\
  }\bibfield  {title} {\bibinfo {title} {{Tunable Layer Circular Photogalvanic
  Effect in Twisted Bilayers}},\ }\href
  {https://doi.org/10.1103/PhysRevLett.124.077401} {\bibfield  {journal}
  {\bibinfo  {journal} {Phys. Rev. Lett.}\ }\textbf {\bibinfo {volume} {124}},\
  \bibinfo {eid} {077401} (\bibinfo {year} {2020})},\ \Eprint
  {https://arxiv.org/abs/1911.04049} {arXiv:1911.04049 [cond-mat.mes-hall]}
  \BibitemShut {NoStop}%
\bibitem [{\citenamefont {{Zhang}}\ \emph {et~al.}(2019)\citenamefont
  {{Zhang}}, \citenamefont {{Holder}}, \citenamefont {{Ishizuka}},
  \citenamefont {{de Juan}}, \citenamefont {{Nagaosa}}, \citenamefont
  {{Felser}},\ and\ \citenamefont {{Yan}}}]{Zhang2019}%
  \BibitemOpen
  \bibfield  {author} {\bibinfo {author} {\bibfnamefont {Y.}~\bibnamefont
  {{Zhang}}}, \bibinfo {author} {\bibfnamefont {T.}~\bibnamefont {{Holder}}},
  \bibinfo {author} {\bibfnamefont {H.}~\bibnamefont {{Ishizuka}}}, \bibinfo
  {author} {\bibfnamefont {F.}~\bibnamefont {{de Juan}}}, \bibinfo {author}
  {\bibfnamefont {N.}~\bibnamefont {{Nagaosa}}}, \bibinfo {author}
  {\bibfnamefont {C.}~\bibnamefont {{Felser}}},\ and\ \bibinfo {author}
  {\bibfnamefont {B.}~\bibnamefont {{Yan}}},\ }\bibfield  {title} {\bibinfo
  {title} {{Switchable magnetic bulk photovoltaic effect in the two-dimensional
  magnet CrI$_{3}$}},\ }\href {https://doi.org/10.1038/s41467-019-11832-3}
  {\bibfield  {journal} {\bibinfo  {journal} {Nat. Commun.}\ }\textbf {\bibinfo
  {volume} {10}},\ \bibinfo {eid} {3783} (\bibinfo {year} {2019})},\ \Eprint
  {https://arxiv.org/abs/1903.06264} {arXiv:1903.06264 [cond-mat.mes-hall]}
  \BibitemShut {NoStop}%
\bibitem [{\citenamefont {{Fei}}\ \emph {et~al.}(2020)\citenamefont {{Fei}},
  \citenamefont {{Song}},\ and\ \citenamefont {{Yang}}}]{Fei2020}%
  \BibitemOpen
  \bibfield  {author} {\bibinfo {author} {\bibfnamefont {R.}~\bibnamefont
  {{Fei}}}, \bibinfo {author} {\bibfnamefont {W.}~\bibnamefont {{Song}}},\ and\
  \bibinfo {author} {\bibfnamefont {L.}~\bibnamefont {{Yang}}},\ }\bibfield
  {title} {\bibinfo {title} {{Giant photogalvanic effect and second-harmonic
  generation in magnetic axion insulators}},\ }\href
  {https://doi.org/10.1103/PhysRevB.102.035440} {\bibfield  {journal} {\bibinfo
   {journal} {Phys. Rev. B}\ }\textbf {\bibinfo {volume} {102}},\ \bibinfo
  {eid} {035440} (\bibinfo {year} {2020})},\ \Eprint
  {https://arxiv.org/abs/2003.01576} {arXiv:2003.01576 [cond-mat.mtrl-sci]}
  \BibitemShut {NoStop}%
\bibitem [{\citenamefont {Song}\ and\ \citenamefont
  {Nagatsuma}(2015)}]{Song2015}%
  \BibitemOpen
  \bibfield  {author} {\bibinfo {author} {\bibfnamefont {H.~J.}\ \bibnamefont
  {Song}}\ and\ \bibinfo {author} {\bibfnamefont {T.}~\bibnamefont
  {Nagatsuma}},\ }\href
  {https://www.google.com/books?hl=en&lr=&id=rHh3CAAAQBAJ&oi=fnd&pg=PP1&ots=_ZQ6ynE-We&sig=1AqYYCrzeJx7B9BwyobWioUPLvM}
  {\emph {\bibinfo {title} {Handbook of Terahertz Technologies: Devices and
  Applications}}}\ (\bibinfo  {publisher} {Taylor \& Francis},\ \bibinfo {year}
  {2015})\ pp.\ \bibinfo {pages} {1--594}\BibitemShut {NoStop}%
\bibitem [{\citenamefont {Zhang}\ \emph {et~al.}(2019)\citenamefont {Zhang},
  \citenamefont {Hosono}, \citenamefont {Nagai}, \citenamefont {Song},\ and\
  \citenamefont {Hirakawa}}]{Zhang2019c}%
  \BibitemOpen
  \bibfield  {author} {\bibinfo {author} {\bibfnamefont {Y.}~\bibnamefont
  {Zhang}}, \bibinfo {author} {\bibfnamefont {S.}~\bibnamefont {Hosono}},
  \bibinfo {author} {\bibfnamefont {N.}~\bibnamefont {Nagai}}, \bibinfo
  {author} {\bibfnamefont {S.~H.}\ \bibnamefont {Song}},\ and\ \bibinfo
  {author} {\bibfnamefont {K.}~\bibnamefont {Hirakawa}},\ }\bibfield  {title}
  {\bibinfo {title} {{Fast and sensitive bolometric terahertz detection at room
  temperature through thermomechanical transduction}},\ }\href
  {https://doi.org/10.1063/1.5045256} {\bibfield  {journal} {\bibinfo
  {journal} {J. Appl. Phys}\ }\textbf {\bibinfo {volume} {125}},\ \bibinfo
  {pages} {151602} (\bibinfo {year} {2019})}\BibitemShut {NoStop}%
\bibitem [{\citenamefont {Lewis}(2019)}]{Lewis2019}%
  \BibitemOpen
  \bibfield  {author} {\bibinfo {author} {\bibfnamefont {R.~A.}\ \bibnamefont
  {Lewis}},\ }\bibfield  {title} {\bibinfo {title} {{A review of terahertz
  detectors}},\ }\href {https://doi.org/10.1088/1361-6463/ab31d5} {\bibfield
  {journal} {\bibinfo  {journal} {J. Phys. D}\ }\textbf {\bibinfo {volume}
  {52}},\ \bibinfo {pages} {433001} (\bibinfo {year} {2019})}\BibitemShut
  {NoStop}%
\bibitem [{\citenamefont {Guillet}\ \emph {et~al.}(2014)\citenamefont
  {Guillet}, \citenamefont {Recur}, \citenamefont {Frederique}, \citenamefont
  {Bousquet}, \citenamefont {Canioni}, \citenamefont {Manek-H{\"{o}}nninger},
  \citenamefont {Desbarats},\ and\ \citenamefont {Mounaix}}]{Guillet2014}%
  \BibitemOpen
  \bibfield  {author} {\bibinfo {author} {\bibfnamefont {J.~P.}\ \bibnamefont
  {Guillet}}, \bibinfo {author} {\bibfnamefont {B.}~\bibnamefont {Recur}},
  \bibinfo {author} {\bibfnamefont {L.}~\bibnamefont {Frederique}}, \bibinfo
  {author} {\bibfnamefont {B.}~\bibnamefont {Bousquet}}, \bibinfo {author}
  {\bibfnamefont {L.}~\bibnamefont {Canioni}}, \bibinfo {author} {\bibfnamefont
  {I.}~\bibnamefont {Manek-H{\"{o}}nninger}}, \bibinfo {author} {\bibfnamefont
  {P.}~\bibnamefont {Desbarats}},\ and\ \bibinfo {author} {\bibfnamefont
  {P.}~\bibnamefont {Mounaix}},\ }\bibfield  {title} {\bibinfo {title} {{Review
  of terahertz tomography techniques}},\ }\href
  {https://doi.org/10.1007/s10762-014-0057-0} {\bibfield  {journal} {\bibinfo
  {journal} {Journal of Infrared, Millimeter, and Terahertz Waves}\ }\textbf
  {\bibinfo {volume} {35}},\ \bibinfo {pages} {382} (\bibinfo {year}
  {2014})}\BibitemShut {NoStop}%
\bibitem [{\citenamefont {{Burger}}\ \emph {et~al.}(2019)\citenamefont
  {{Burger}}, \citenamefont {{Agarwal}}, \citenamefont {{Aprelev}},
  \citenamefont {{Schruba}}, \citenamefont {{Gutierrez-Perez}}, \citenamefont
  {{Fridkin}},\ and\ \citenamefont {{Spanier}}}]{Burger2019}%
  \BibitemOpen
  \bibfield  {author} {\bibinfo {author} {\bibfnamefont {A.~M.}\ \bibnamefont
  {{Burger}}}, \bibinfo {author} {\bibfnamefont {R.}~\bibnamefont {{Agarwal}}},
  \bibinfo {author} {\bibfnamefont {A.}~\bibnamefont {{Aprelev}}}, \bibinfo
  {author} {\bibfnamefont {E.}~\bibnamefont {{Schruba}}}, \bibinfo {author}
  {\bibfnamefont {A.}~\bibnamefont {{Gutierrez-Perez}}}, \bibinfo {author}
  {\bibfnamefont {V.~M.}\ \bibnamefont {{Fridkin}}},\ and\ \bibinfo {author}
  {\bibfnamefont {J.~E.}\ \bibnamefont {{Spanier}}},\ }\bibfield  {title}
  {\bibinfo {title} {{Direct observation of shift and ballistic photovoltaic
  currents}},\ }\href {https://doi.org/10.1126/sciadv.aau5588} {\bibfield
  {journal} {\bibinfo  {journal} {Sci. Adv.}\ }\textbf {\bibinfo {volume}
  {5}},\ \bibinfo {pages} {eaau5588} (\bibinfo {year} {2019})}\BibitemShut
  {NoStop}%
\bibitem [{\citenamefont {Wu}\ \emph {et~al.}(2018)\citenamefont {Wu},
  \citenamefont {Lovorn}, \citenamefont {Tutuc},\ and\ \citenamefont
  {MacDonald}}]{Wu2020a}%
  \BibitemOpen
  \bibfield  {author} {\bibinfo {author} {\bibfnamefont {F.}~\bibnamefont
  {Wu}}, \bibinfo {author} {\bibfnamefont {T.}~\bibnamefont {Lovorn}}, \bibinfo
  {author} {\bibfnamefont {E.}~\bibnamefont {Tutuc}},\ and\ \bibinfo {author}
  {\bibfnamefont {A.~H.}\ \bibnamefont {MacDonald}},\ }\bibfield  {title}
  {\bibinfo {title} {Hubbard model physics in transition metal dichalcogenide
  moir\'e bands},\ }\href {https://doi.org/10.1103/PhysRevLett.121.026402}
  {\bibfield  {journal} {\bibinfo  {journal} {Phys. Rev. Lett.}\ }\textbf
  {\bibinfo {volume} {121}},\ \bibinfo {pages} {026402} (\bibinfo {year}
  {2018})}\BibitemShut {NoStop}%
\bibitem [{\citenamefont {{Castro Neto}}\ \emph {et~al.}(2009)\citenamefont
  {{Castro Neto}}, \citenamefont {{Guinea}}, \citenamefont {{Peres}},
  \citenamefont {{Novoselov}},\ and\ \citenamefont {{Geim}}}]{CastroNeto2009}%
  \BibitemOpen
  \bibfield  {author} {\bibinfo {author} {\bibfnamefont {A.~H.}\ \bibnamefont
  {{Castro Neto}}}, \bibinfo {author} {\bibfnamefont {F.}~\bibnamefont
  {{Guinea}}}, \bibinfo {author} {\bibfnamefont {N.~M.~R.}\ \bibnamefont
  {{Peres}}}, \bibinfo {author} {\bibfnamefont {K.~S.}\ \bibnamefont
  {{Novoselov}}},\ and\ \bibinfo {author} {\bibfnamefont {A.~K.}\ \bibnamefont
  {{Geim}}},\ }\bibfield  {title} {\bibinfo {title} {{The electronic properties
  of graphene}},\ }\href {https://doi.org/10.1103/RevModPhys.81.109} {\bibfield
   {journal} {\bibinfo  {journal} {Rev. Mod. Phys.}\ }\textbf {\bibinfo
  {volume} {81}},\ \bibinfo {pages} {109} (\bibinfo {year} {2009})},\ \Eprint
  {https://arxiv.org/abs/0709.1163} {arXiv:0709.1163} \BibitemShut {NoStop}%
\bibitem [{\citenamefont {{Pereira}}\ \emph {et~al.}(2009)\citenamefont
  {{Pereira}}, \citenamefont {{Castro Neto}},\ and\ \citenamefont
  {{Peres}}}]{Pereira2009}%
  \BibitemOpen
  \bibfield  {author} {\bibinfo {author} {\bibfnamefont {V.~M.}\ \bibnamefont
  {{Pereira}}}, \bibinfo {author} {\bibfnamefont {A.~H.}\ \bibnamefont {{Castro
  Neto}}},\ and\ \bibinfo {author} {\bibfnamefont {N.~M.~R.}\ \bibnamefont
  {{Peres}}},\ }\bibfield  {title} {\bibinfo {title} {{Tight-binding approach
  to uniaxial strain in graphene}},\ }\href
  {https://doi.org/10.1103/PhysRevB.80.045401} {\bibfield  {journal} {\bibinfo
  {journal} {Phys. Rev. B}\ }\textbf {\bibinfo {volume} {80}},\ \bibinfo {eid}
  {045401} (\bibinfo {year} {2009})}\BibitemShut {NoStop}%
\bibitem [{\citenamefont {{Guinea}}\ \emph {et~al.}(2010)\citenamefont
  {{Guinea}}, \citenamefont {{Katsnelson}},\ and\ \citenamefont
  {{Geim}}}]{Guinea2010}%
  \BibitemOpen
  \bibfield  {author} {\bibinfo {author} {\bibfnamefont {F.}~\bibnamefont
  {{Guinea}}}, \bibinfo {author} {\bibfnamefont {M.~I.}\ \bibnamefont
  {{Katsnelson}}},\ and\ \bibinfo {author} {\bibfnamefont {A.~K.}\ \bibnamefont
  {{Geim}}},\ }\bibfield  {title} {\bibinfo {title} {{Energy gaps and a
  zero-field quantum Hall effect in graphene by strain engineering}},\ }\href
  {https://doi.org/10.1038/nphys1420} {\bibfield  {journal} {\bibinfo
  {journal} {Nat. Phys.}\ }\textbf {\bibinfo {volume} {6}},\ \bibinfo {pages}
  {30} (\bibinfo {year} {2010})},\ \Eprint {https://arxiv.org/abs/0909.1787}
  {arXiv:0909.1787 [cond-mat.mes-hall]} \BibitemShut {NoStop}%
\bibitem [{\citenamefont {Fernandes}\ and\ \citenamefont
  {Venderbos}(2020)}]{Fernandes2020}%
  \BibitemOpen
  \bibfield  {author} {\bibinfo {author} {\bibfnamefont {R.~M.}\ \bibnamefont
  {Fernandes}}\ and\ \bibinfo {author} {\bibfnamefont {J.~W.}\ \bibnamefont
  {Venderbos}},\ }\bibfield  {title} {\bibinfo {title} {{Nematicity with a
  twist: Rotational symmetry breaking in a moir{\'{e}} superlattice}},\ }\href
  {https://doi.org/10.1126/sciadv.aba8834} {\bibfield  {journal} {\bibinfo
  {journal} {Science Advances}\ }\textbf {\bibinfo {volume} {6}},\ \bibinfo
  {pages} {eaba8834} (\bibinfo {year} {2020})},\ \Eprint
  {https://arxiv.org/abs/1911.11367} {arXiv:1911.11367} \BibitemShut {NoStop}%
\bibitem [{\citenamefont {{Koshino}}\ \emph {et~al.}(2018)\citenamefont
  {{Koshino}}, \citenamefont {{Yuan}}, \citenamefont {{Koretsune}},
  \citenamefont {{Ochi}}, \citenamefont {{Kuroki}},\ and\ \citenamefont
  {{Fu}}}]{Koshino2018}%
  \BibitemOpen
  \bibfield  {author} {\bibinfo {author} {\bibfnamefont {M.}~\bibnamefont
  {{Koshino}}}, \bibinfo {author} {\bibfnamefont {N.~F.~Q.}\ \bibnamefont
  {{Yuan}}}, \bibinfo {author} {\bibfnamefont {T.}~\bibnamefont {{Koretsune}}},
  \bibinfo {author} {\bibfnamefont {M.}~\bibnamefont {{Ochi}}}, \bibinfo
  {author} {\bibfnamefont {K.}~\bibnamefont {{Kuroki}}},\ and\ \bibinfo
  {author} {\bibfnamefont {L.}~\bibnamefont {{Fu}}},\ }\bibfield  {title}
  {\bibinfo {title} {{Maximally Localized Wannier Orbitals and the Extended
  Hubbard Model for Twisted Bilayer Graphene}},\ }\href
  {https://doi.org/10.1103/PhysRevX.8.031087} {\bibfield  {journal} {\bibinfo
  {journal} {Phys. Rev. X}\ }\textbf {\bibinfo {volume} {8}},\ \bibinfo {eid}
  {031087} (\bibinfo {year} {2018})},\ \Eprint
  {https://arxiv.org/abs/1805.06819} {arXiv:1805.06819 [cond-mat.mes-hall]}
  \BibitemShut {NoStop}%
\bibitem [{\citenamefont {Nam}\ and\ \citenamefont {Koshino}(2017)}]{Nam2017}%
  \BibitemOpen
  \bibfield  {author} {\bibinfo {author} {\bibfnamefont {N.~N.~T.}\
  \bibnamefont {Nam}}\ and\ \bibinfo {author} {\bibfnamefont {M.}~\bibnamefont
  {Koshino}},\ }\bibfield  {title} {\bibinfo {title} {Lattice relaxation and
  energy band modulation in twisted bilayer graphene},\ }\href
  {https://doi.org/10.1103/PhysRevB.96.075311} {\bibfield  {journal} {\bibinfo
  {journal} {Phys. Rev. B}\ }\textbf {\bibinfo {volume} {96}},\ \bibinfo
  {pages} {075311} (\bibinfo {year} {2017})},\ \Eprint
  {https://arxiv.org/abs/1706.03908} {1706.03908} \BibitemShut {NoStop}%
\bibitem [{\citenamefont {Tinkham}(2003)}]{Tinkham2003}%
  \BibitemOpen
  \bibfield  {author} {\bibinfo {author} {\bibfnamefont {M.}~\bibnamefont
  {Tinkham}},\ }\href@noop {} {\emph {\bibinfo {title} {{Group Theory and
  Quantum Mechanics}}}}\ (\bibinfo  {publisher} {Dover Publications},\ \bibinfo
  {year} {2003})\BibitemShut {NoStop}%
\end{thebibliography}%
\end{document}